\pdfoutput=1
\documentclass[12pt,a4paper]{article}    
\usepackage{jheppub}
\makeatletter
\renewcommand\@fpheader{\color{white}'}
\usepackage[T1]{fontenc}
\usepackage[utf8]{inputenc}
\usepackage[british]{babel}
\usepackage{amssymb,amstext,amsthm,
	bbm,mathrsfs,amscd,bbold,pifont, tensor}
\usepackage{array}
\usepackage[svgnames,table]{xcolor}
\usepackage[T1]{fontenc}
\usepackage[utf8]{inputenc}
\usepackage{enumitem}
\usepackage{booktabs,caption}
\usepackage[normalem]{ulem}
\usepackage{verbatim}
\usepackage{braket}
\usepackage{mdframed}
\usepackage{tikz,pgfplots}
\usetikzlibrary{calc,arrows,cd,arrows.meta,decorations.markings}
\hypersetup{%
	pdftitle   = {Fluids on null hypersurfaces and Carrollian hydrodynamics},
	pdfkeywords = {null fluids, Carrollian symmetries, black holes, flat space holography},
	pdfauthor  = {Emil Have},
}
\theoremstyle{plain}

\theoremstyle{definition}

\renewcommand{\c}{\mathfrak{c}}

\newcommand{\s}{\mathfrak{s}}

\newcommand{\f}{\mathfrak{f}}

\newcommand{\x}{\boldsymbol{x}}
\newcommand{\y}{\boldsymbol{y}}

\ifdefined\C
\renewcommand{\C}{\boldsymbol{C}}
\else
\newcommand{\C}{\boldsymbol{C}}
\fi

\newcommand{\xx}{\mathbb{x}}

\newcommand{\yy}{\mathbb{y}}

\newcommand{\scri}{\mathscr{I}}

\newcommand{\D}{\partial}

\newcommand{\pd}[2]{\frac{\partial #1}{\partial #2}}
\definecolor{gris}{rgb}{0.5,0.5,0.5}
\definecolor{darkgreen}{rgb}{0.0,0.5,0.0}
\newcommand{\zero}{{\color{gris}0}}
\allowdisplaybreaks[0]
\numberwithin{equation}{section}
\usetikzlibrary{shapes,decorations}
\pgfdeclarelayer{nodelayer}
\pgfdeclarelayer{edgelayer}
\pgfsetlayers{edgelayer,nodelayer,main}
\tikzstyle{ghost}=[fill=none, draw=none, shape=circle]
\tikzstyle{cross}=[fill=none, draw=black, shape=circle]
\tikzstyle{grey line}=[-, draw={rgb,255: red,128; green,128; blue,128}]
\tikzstyle{blue line}=[-, draw=blue, fill={rgb,255: red,162; green,246; blue,255}]
\tikzstyle{dash blue}=[-, draw=blue, dashed]
\tikzstyle{dash grey}=[-, draw={rgb,255: red,128; green,128; blue,128}, dashed]
\tikzstyle{thick black line}=[-, thick]
\tikzstyle{blue fill}=[-, draw=none, fill={rgb,255: red,126; green,214; blue,255}]
\tikzstyle{black line}=[-]
\tikzstyle{red fill}=[-, fill={rgb,255: red,255; green,162; blue,164}, draw={rgb,255: red,255; green,0; blue,4}]
\tikzstyle{new edge style 0}=[-, draw=none, fill={rgb,255: red,255; green,162; blue,164}]
\begin{document}
	
	\title{\bf{Fluids on null hypersurfaces and  Carrollian hydrodynamics}}

\author[1,2]{Emil Have}

\affiliation[1]{The Mathematical Institute, University of Oxford,\\ Woodstock Road, Oxford OX2 6GG, United Kingdom}
\affiliation[2]{Center of Gravity, Niels Bohr Institute, University of Copenhagen,\\
Blegdamsvej 17, DK-2100 Copenhagen Ø, Denmark}


\emailAdd{emil.have@maths.ox.ac.uk}
	
	\abstract{We study intrinsic ideal fluids on null hypersurfaces and their realisations by causal fluids living in the ambient spacetime. An embedded null hypersurface is equipped with an induced Carrollian structure with a temporal Weyl symmetry inherited from rescalings of the null generator, and we develop the corresponding null submanifold and variational calculus. The intrinsic fluid phases are classified according to their realisation of Carroll boosts and temporal Weyl symmetry. If we demand that an ambient fluid be confined to a null hypersurface, the ambient fluid velocity must be null, which realises an intrinsic zero-temperature Carrollian phase with unbroken Carroll boosts. In general, timelike or null ambient fluids will cross a null hypersurface and lead to an open Carrollian system. On a null foliation, the sources correspond to transport between null leaves. We then show that there exists a decoupling limit of a timelike ambient fluid which gives rise to an autonomous Carrollian fluid with spontaneously broken Carroll boosts and no source terms. This decoupling limit corresponds to an infinite boost while keeping the enthalpy finite and nonzero, which leads to a divergent ambient energy-momentum tensor. Nevertheless, the induced Carrollian responses do not see the divergent term and remain finite, and an independent Carroll boost Goldstone emerges from a subleading component of the ambient fluid velocity in this limit.}
	
	\maketitle
	
\section{Introduction}
\label{sec:introduction}

What does it mean for a fluid to live on a null hypersurface? Null hypersurfaces arise naturally in Lorentzian geometry, with important examples including black hole horizons~\cite{Hawking:1973uf,Gourgoulhon:2005ng} and (after conformal compactification) null infinity~\cite{Penrose:1962ij,Penrose:1965am}. Their induced geometry is Carrollian~\cite{Duval:2014uoa,Duval:2014uva,Hartong:2015xda,Penna:2018gfx,Donnay:2019jiz,Ciambelli:2019lap}, which suggests that Carrollian hydrodynamics~\cite{deBoer:2017ing,Ciambelli:2018wre,Ciambelli:2018xat,Petkou:2022bmz,deBoer:2021jej,Freidel:2022bai,Freidel:2022vjq,Redondo-Yuste:2022czg,deBoer:2023fnj,Armas:2023dcz,Chabirand:2026sxe} provides the appropriate framework for describing fluid dynamics on null hypersurfaces. However, a causal fluid velocity tangent to a null hypersurface must necessarily be null and aligned with its generators. At the same time, unbroken Carroll boost symmetry imposes vanishing enthalpy at ideal order, $\mathcal{E} + P = 0$~\cite{deBoer:2017ing,deBoer:2021jej} and hence $sT = 0$. A thermal Carrollian fluid requires spontaneously broken Carroll boosts with an associated Goldstone field~\cite{Armas:2023dcz}. It is therefore not clear which intrinsic Carrollian fluid phases can arise from causal fluids in an ambient Lorentzian spacetime. We address this question by first constructing fluids intrinsically on null hypersurfaces and subsequently investigating whether and how they can be realised by ambient causal fluids by restricting to embedded null hypersurfaces.

Carroll symmetry was originally introduced by L\'evy-Leblond and, independently, by Sen Gupta by performing a contraction of the Poincar\'e group in the limit $c\to 0$, where $c$ is the speed of light~\cite{Levy1965,SenGupta1966OnAA}. The corresponding intrinsic geometry was subsequently studied by Henneaux~\cite{Henneaux:1979vn}. A null hypersurface embedded in a Lorentzian spacetime carries precisely such a structure: its induced metric is degenerate, with a one-dimensional kernel spanned by a null vector field. In addition, the freedom to rescale the null generator (``null boosts'') gives rise to a temporal Weyl symmetry of the induced Carrollian geometry, which is not present in the Carrollian geometry that appears in the $c \to 0$ limit of a Lorentzian geometry. It is also important to emphasise that the Carrollian geometry that appears in the limit has the same dimension as its parent, while the one induced on a null hypersurface requires no limit and is codimension-$1$.

The differential geometry of null hypersurfaces, particularly using auxiliary vectors known as riggings, is well developed~\cite{Katsuno_1980,Katsuno_1981,Mars:1993mj,Guti_rrez_2015}, and its relation to Carrollian geometry has been explored extensively in recent years~\cite{Freidel:2022bai,Freidel:2022vjq,Freidel:2024emv,Ciambelli:2025unn}. Here, we develop a null submanifold calculus in the spirit of the embedding formalism of~\cite{Carter:1997pb,Armas:2013hsa,Armas:2013goa,Armas:2017pvj,Armas:2019gnb}. The advantage of this approach is that it provides a systematic framework for relating ambient and intrinsic tensor fields, while keeping track of their transformation laws, and we use this formalism to develop a variational calculus for fixed null embeddings, which allows us to identify how the intrinsic fluid responses on a null hypersurface are related to the ambient energy-momentum tensor. An important (and well-known) feature is that the intrinsic momentum-stress tensor is defined only up to terms proportional to the square of the Carrollian vector field, and, as we shall see, this ambiguity plays a crucial role in relating the intrinsic and ambient hydrodynamics. Throughout this work, we restrict attention to ideal hydrodynamics.

Flat space holography\footnote{There are two complementary approaches to flat space holography: the celestial approach~\cite{Kapec:2016jld,Pasterski:2016qvg,Pasterski:2017kqt,Pasterski:2021raf},	which seeks to recast four-dimensional gravitational $S$-matrix elements as correlators of a 2d CFT on the celestial sphere (for reviews, see~\cite{Raclariu:2021zjz,Pasterski:2021rjz,McLoughlin:2022ljp,Donnay:2023mrd,Zhu:2026ofh}), and the Carrollian approach, where the dual theory lives on null infinity $\scri$~\cite{Hartong:2015usd,Ciambelli:2018wre,Fiorucci:2025twa,Hartong:2025jpp,Hartong:2026rbr,Fiorucci:2026nyg,Hartong:2026bvh} (recent reviews include~\cite{Bagchi:2025vri,Nguyen:2025zhg,Ruzziconi:2026bix}). Links between the two have been explored in~\cite{Donnay:2022aba,Bagchi:2022emh,Donnay:2022wvx}.} provides an important motivation for studying fluids on null hypersurfaces. As mentioned above, in Penrose's conformal compactification~\cite{Penrose:1962ij,Penrose:1965am}, null infinity $\scri$ of an asymptotically flat spacetime is a null hypersurface in the conformally completed spacetime.  From the perspective of a chosen conformal completion, the description of fluids intrinsic to null hypersurfaces is thus directly relevant to the geometry of null infinity.\footnote{See also~\cite{Armas:2023dcz} for a fluid with conformal Carrollian symmetry, which is the relevant symmetry for $\scri$.} The submanifold framework we develop in this work allows us to investigate how such intrinsic fluid dynamics can arise from fields in the ambient spacetime, and, furthermore, how ambient energy-momentum transport gives rise to sources in the intrinsic conservation equations. This latter  observation is particularly interesting in the context of flat space holography, where the conservation equations are sourced by gravitational radiation~\cite{Donnay:2022aba,Donnay:2022wvx}.

The relation between lightlike fluid motion and Carrollian hydrodynamics has previously been emphasised in~\cite{Bagchi:2023ysc,Bagchi:2023rwd}. It is important, however, to distinguish between a fluid moving at the speed of light~\cite{Armas:2025uyv,Armas:2026nwa}, a fluid supported on a null hypersurface, and a fluid obtained as the $c\to 0$ limit of relativistic hydrodynamics. These are distinct notions, and, in particular, a null fluid velocity need not be hypersurface-orthogonal and therefore it does not, in general, define a foliation by null hypersurfaces. Furthermore, the Carrollian structure induced on a null hypersurface does not require taking a $c\to 0$ limit; this limit instead gives rise to a particular phase of Carrollian hydrodynamics with spontaneously broken boosts~\cite{Armas:2023dcz}. As we will show, this phase can be constructed intrinsically on a null hypersurface, provided that the temporal Weyl symmetry is compensated by a Stueckelberg field (if instead it is explicitly broken one finds a more general phase that depends on an additional scale). 
\begin{figure}[t]
	\centering
	\resizebox{\textwidth}{!}{%
		\begin{tikzpicture}[
			>={Stealth[length=2.3mm,width=1.7mm]},
			cell/.style={
				draw=blue!45!black,
				fill=blue!5,
				rounded corners=3pt,
				text width=2.85cm,
				minimum height=1.95cm,
				align=flush center,
				font=\small,
				inner sep=4pt
			},
			amb/.style={
				draw=orange!75!black,
				fill=orange!7,
				rounded corners=3pt,
				align=flush center,
				font=\small,
				inner sep=5pt
			},
			head/.style={align=flush center,font=\small},
			lab/.style={align=flush center,font=\footnotesize,text=orange!70!black},
			realise/.style={->,thick,orange!85!black}
			]
			\node[head] at (2.6,4.5)
			{Temporal Weyl\\unbroken};
			\node[head] at (5.9,4.5)
			{Explicitly broken\\by $\kappa$};
			\node[head] at (9.2,4.5)
			{Compensated\\by $\Omega$};
			\node[head,anchor=east] at (0.95,2.95)
			{Carroll boosts\\broken\\thermal, $\theta^a$};
			\node[head,anchor=east] at (0.95,0.45)
			{Carroll boosts\\unbroken\\zero temperature};
			\node[cell] (T1) at (2.6,2.95)
			{$P\neq0$ branch\\
				$P=s_0T$\\
				$T^a=0$ on shell};
			\node[cell] (T2) at (5.9,2.95)
			{$P_{\mathrm{b,e}}(T,\vec u^2,\kappa)$\\
				no Weyl constraint};
			\node[cell] (T3) at (9.2,2.95)
			{$\mathscr P(T,\vec u^2,\Omega)$\\
				background or dynamical};
			\node[
			cell,
			draw=black!35,
			fill=black!5,
			text=black!65
			] (Z1) at (2.6,0.45)
			{No nontrivial\\
				ideal phase\\
				without a scale};
			\node[cell] (Z2) at (5.9,0.45)
			{$P_{\mathrm e}(\vec k^2,\kappa)$};
			\node[cell] (Z3) at (9.2,0.45)
			{$\Omega^{-1}P_{\mathrm s}(\vec k^2)$};
			\node[
			amb,
			text width=6.3cm,
			minimum height=1.15cm
			] (NF) at (7.55,-2.15)
			{Confined aligned null fluids\\
				zero-temperature intrinsic phases};
			\draw[realise]
			(Z2.south|-NF.north) -- (Z2.south);
			\draw[realise]
			(Z3.south|-NF.north) -- (Z3.south);
			\node[lab,anchor=east] at (5.7,-1.05)
			{explicit $\kappa$};			
			\node[lab,anchor=west] at (9.4,-1.05)
			{background $\Omega$};
			\node[
			amb,
			text width=2.5cm,
			minimum height=1.95cm
			] (CF) at (14.55,2.95)
			{Crossing timelike fluid\\
				open system\\
				$B\to0$};
			\draw[realise]
			(CF.west) --
			node[lab,above=2pt] {thermal limit}
			(T3.east);
			\draw[realise,densely dashed,rounded corners=4pt]
			(CF.south) --
			(CF.south|-NF.east) --
			node[lab,above=3pt] {null-fluid limit}
			(NF.east);
		\end{tikzpicture}
	}
	
	\caption{Intrinsic ideal Carrollian fluid phases and their realisations by ambient causal fluids. The blue boxes classify the intrinsic phases according to the realisation of Carroll boosts (rows) and temporal Weyl symmetry (columns). Spontaneously broken Carroll boosts give rise to thermal phases with a Carroll boost Goldstone, while unbroken Carroll boosts lead to zero-temperature phases with vanishing enthalpy. The grey box indicates the absence of a nontrivial ideal phase when there is no additional scale. In the compensated column, a compensator field $\Omega$ that is either a background Stueckelberg field or a dynamical Goldstone gives rise to different phases, with the latter subject to the Goldstone equation of motion $\mathcal O_\Omega = 0$. The orange boxes and arrows indicate the ambient fluid phases. Causal fluids confined to a null hypersurface must have a null fluid velocity aligned with the generator $v^\mu$ of the null hypersurface, which then realise the Carrollian zero-temperature intrinsic phases. General ambient fluids cross a null hypersurface and induce open Carrollian systems. An infinite boost of a crossing timelike parent, with $B = -v_\mu U^\mu \to 0$, where $U^\mu$ is the ambient fluid velocity, admits two distinct scalings of the enthalpy of the timelike fluid $w_{\text{T}}$. Taking $w_{\text{T}} = \mathcal O(B^2)$ produces a finite ambient null fluid (dashed arrow), whereas keeping $w_{\text{T}}$ finite and nonzero produces an autonomous thermal Carrollian fluid with an emergent boost Goldstone (solid arrow).}
	\label{fig:phase-map}
\end{figure}
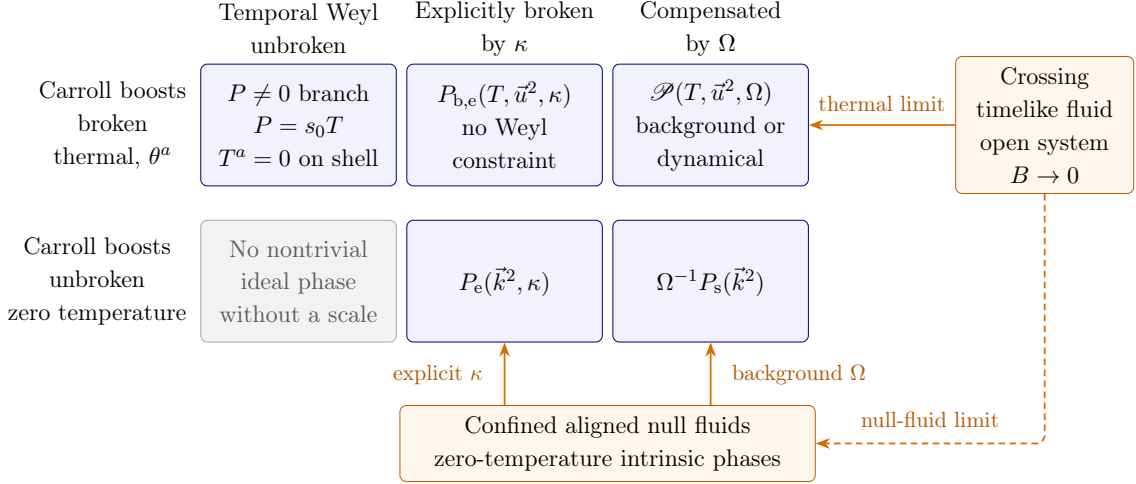

To make these distinctions precise, we first construct the intrinsic ideal fluids that can live on a null hypersurface. Building on the framework of~\cite{Armas:2023dcz}, we classify these phases according to how they realise Carroll boosts and temporal Weyl symmetry. Spontaneously broken Carroll boosts give rise to thermal fluids with an independent boost Goldstone as in~\cite{Armas:2023dcz}, while unbroken boosts characterise zero-temperature phases in analogy with~\cite{Armas:2025uyv,Armas:2026nwa}. The temporal Weyl symmetry can remain unbroken, be explicitly broken or be realised through a compensator field, and this compensator can be treated as either a background Stueckelberg field or a dynamical Goldstone field, with these two choices leading to distinct phases.  We have summarised all of these phases in Figure~\ref{fig:phase-map}.

We subsequently investigate how these intrinsic phases can arise from ambient causal fluids (cf.~Figure~\ref{fig:phase-map}), starting with fluids that move at the speed of light. The covariant effective theory of such null fluids was developed in~\cite{Armas:2025uyv}, and their gravitational realisations in terms of asymptotically AdS and asymptotically flat pp-wave spacetimes were constructed in~\cite{Armas:2026nwa}. These are effective theories of null matter in an ambient Lorentzian spacetime and do not, in general, describe fluids supported on null hypersurfaces. For an ambient null fluid to be confined to a null hypersurface, its velocity must be null and aligned with the null generators, and we show that the corresponding intrinsic theory is a zero-temperature Carrollian fluid phase with unbroken boosts. More generally, ambient null or timelike fluids may cross a null hypersurface, in which case the induced Carrollian system becomes open. By considering a foliation by null hypersurfaces, we show that these sources can be described in terms of transverse transport between neighbouring leaves of the null foliation.

Finally, we demonstrate that a decoupling limit of a timelike ambient fluid leads to an autonomous thermal Carrollian fluid with spontaneously broken boosts and no source terms. This decoupling limit is implemented by infinitely boosting the ambient fluid velocity, similarly to the null limit in~\cite{Armas:2025uyv,Armas:2026nwa}, but rather than sending the ambient enthalpy to zero, the decoupling limit keeps the enthalpy finite and nonzero. While the ambient energy-momentum tensor diverges in the decoupling limit, the induced Carrollian energy-momentum tensor remains finite and satisfies unsourced conservation equations. In this sense, the resulting intrinsic fluid is autonomous, and in particular the subleading component of the ambient fluid velocity becomes an independent Carroll boost Goldstone with an emergent timelike Stueckelberg symmetry in the decoupling limit.

\paragraph{Outline.}
This paper is organised as follows: we begin in Section~\ref{sec:submanifold-calculus} by developing the submanifold and variational calculi for null hypersurfaces. In Section~\ref{sec:ind-carrollian-structure}, we demonstrate how the submanifold calculus gives rise to the induced Carrollian structure and its symmetries, and how these symmetries act on the projectors. This is followed by Section~\ref{sec:variational-calculus}, where a variational calculus for fixed null hypersurfaces is developed, which allows us to translate between an ambient energy-momentum tensor and induced Carrollian responses on a null hypersurface along with their Ward identities. Then, in Section~\ref{sec:the-M-vector}, we identify the $M$-vector field introduced in~\cite{Hartong:2015xda} as the difference between the auxiliary null vector and a chosen reference vector that defines a transverse splitting.

Then, in Section~\ref{sec:Carroll-hydro}, we construct the intrinsic ideal fluid phases that may be realised on a null hypersurface. We begin by considering spontaneously broken Carroll boosts in Section~\ref{sec:spont-break-boosts}, where the unbroken temporal Weyl symmetry severely restricts the thermodynamics. Then, in Section~\ref{sec:Carroll-zero-temperature} we investigate phases where the Carroll boosts remain unbroken, in which case there is no temperature. A nontrivial fluid phase then requires that the temporal Weyl symmetry be broken. Finally, we discuss the scenario where both the Carroll boosts and the temporal Weyl symmetry are broken in Section~\ref{sec:breaking-both-syms}. 

This is followed by Section~\ref{sec:null-fluids-on-surfs}, where we discuss the relation between ambient lightlike fluids and fluids on a null hypersurface. In Section~\ref{sec:causal-confinement} we demonstrate that in order for an ambient fluid to be confined to a null hypersurface, its fluid velocity must be null and aligned with the generator of the null hypersurface, i.e., it must be a null fluid as constructed in~\cite{Armas:2025uyv,Armas:2026nwa}, which we briefly review in Section~\ref{sec:null-fluids-review}. We then show in Section~\ref{sec:twistless-null-fluids} how such aligned ambient null fluids give rise to zero-temperature Carrollian fluids with unbroken Carroll boosts, which we constructed from the intrinsic perspective in Section~\ref{sec:Carroll-hydro}. 

Finally, in Section~\ref{sec:open-Carroll-fluids}, we study causal ambient fluids crossing the leaves of a null foliation. We first derive the sourced intrinsic Ward identities in Section~\ref{sec:foliation-balance}, showing that the sources describe transverse transport between neighbouring leaves. In Section~\ref{sec:timelike-Carroll}, we consider crossing timelike and null fluids and identify the associated Carroll boost field. We then introduce the singular decoupling limit in Section~\ref{sec:decouling-limit}, and demonstrate in Section~\ref{sec:Carroll-recovery} how it produces an autonomous thermal Carrollian fluid with an independent boost Goldstone.

We conclude in Section~\ref{sec:discussion} with a brief discussion of future avenues for research opened by this work.

\section{Submanifold calculus for null hypersurfaces}
\label{sec:submanifold-calculus}

In this section, we develop a submanifold calculus for null hypersurfaces along the lines of the embedding formalism of~\cite{Carter:1997pb,Armas:2013hsa,Armas:2013goa,Armas:2017pvj,Armas:2019gnb}, and show how Carrollian symmetries act on the embedding data. Throughout this work, the embedding remains non-dynamical, that is, we do not consider variations of the embedding maps. Such variations are natural in a fully dynamical null submanifold theory but are not necessary for the fluid applications considered in this paper, and we leave them to future work.

\subsection{Induced Carrollian geometry}
\label{sec:ind-carrollian-structure}

\begin{figure}[h]
\centering
\begin{tikzpicture}[scale=1.2,
	>={Stealth[length=2.3mm,width=1.7mm]},
	vec/.style={->,very thick,line cap=round},
	tvec/.style={->,thick,line cap=round,black!65},
	lab/.style={font=\small},
	cone/.style={gray!75},
	gen/.style={blue!22},
	rig/.style={orange!85!black},
	midarrow/.style={postaction={decorate},decoration={markings,
			mark=at position 0.55 with {\arrow{Stealth[length=1.6mm,width=1.3mm]}}}},
	]
	\begin{scope}[shift={(-7.25,0.45)}]
		\def\blob{(-2.0,-1.35) (0.3,-1.6) (2.05,-1.15) (2.0,1.25) (1.25,2.45) (-0.7,2.4) (-2.15,1.55) (-2.25,0.1)}
		\filldraw[fill=blue!5, draw=blue!40!black!55]
		plot[smooth cycle, tension=0.7] coordinates {(-2.0,-1.35) (0.3,-1.6) (2.05,-1.15) (2.0,1.25) (1.25,2.45) (-0.7,2.4) (-2.15,1.55) (-2.25,0.1)};
		\begin{scope}
			\clip plot[smooth cycle, tension=0.7] coordinates {(-2.0,-1.35) (0.3,-1.6) (2.05,-1.15) (2.0,1.25) (1.25,2.45) (-0.7,2.4) (-2.15,1.55) (-2.25,0.1)};
			\foreach \xx in {-1.7,-1.1,-0.55,0.55,1.1,1.7}
			\draw[gen, midarrow] (\xx,-2) -- (\xx,3);        
			\draw[black!35] (-1.4,-1.96) -- (1.6,2.24);         
			\draw[black!35] (2.5,-1.0) -- (-2.5,1.0);           
			\draw[rig, thick] (-2.6,0) -- (2.6,0);              
			\draw[rig, thick, densely dashed] (-2.6,-1.56) -- (2.6,1.56); 
		\end{scope}
		\draw[rig, ->] (1.0,0) arc (0:29:1.0);
		\node[lab, rig, align=left, anchor=west] at (0.65,-0.70) {Carroll\\[-2pt]boost};
		\node[lab, rig] at (-1.55,-0.22) {$\ker\tau_a$};
		\node[lab, rig] at (-1.3,-1.1) {$\ker\tau'_a$};
		\draw[tvec] (0,0) -- (0.8,1.12) node[lab, right=0pt] {$\partial_0$};
		\draw[tvec] (0,0) -- (-0.8,0.32) node[lab, above left=-1pt] {$\partial_1$};
		\draw[vec,blue!65!black] (0,0) -- (0,1.75) node[lab, above] {$v^a\partial_a$};
		\fill (0,0) circle (1pt) node[lab, below right=0pt] {$p$};
		\node[lab] at (-1.8,2.45) {$(N,v,h)$};
	\end{scope}
	
	\draw[->, thick, black!75] (-5.0,1.35) to[bend left=18] node[lab, above] {$X$} (-2.8,1.35);
	\draw[->, thick, black!75] (-2.8,-0.05) to[bend left=18] node[lab, below] {$X^*$} (-5.0,-0.05);

	\begin{scope}[x={(0.680cm,0.3316cm)}, y={(-1.2804cm,0.1762cm)}, z={(0cm,1.4007cm)}]
		\def\R{3.2}
		\def\aL{133.53} \def\aR{282.41}  
		\def\NTconebody#1#2#3#4{%
			\begin{scope}[shift={({#1},{#2},{#3})}]
				\shade[left color=black!3, right color=black!24, opacity=0.4]
				(0,0,0) -- plot[domain=\aL:\aR, samples=48] ({#4*cos(\x)},{#4*sin(\x)},#4) -- cycle;
				\fill[black!5, opacity=0.45]
				plot[domain=0:360, samples=80] ({#4*cos(\x)},{#4*sin(\x)},#4) -- cycle;
				\draw[cone, densely dashed]
				plot[domain=\aR-360:\aL, samples=60] ({#4*cos(\x)},{#4*sin(\x)},#4);
				\draw[cone] plot[domain=\aL:\aR, samples=48] ({#4*cos(\x)},{#4*sin(\x)},#4);
				\draw[cone] (0,0,0) -- ({#4*cos(\aL)},{#4*sin(\aL)},#4);
				\draw[cone] (0,0,0) -- ({#4*cos(\aR)},{#4*sin(\aR)},#4);
		\end{scope}}

		\filldraw[fill=blue!5, draw=blue!40!black!55, line join=round]
		plot[domain=2.1:5.0, samples=2]   ({-\R+\x*cos(-0.5 r)},{\x*sin(-0.5 r)},{-\R+\x})
		-- plot[domain=-0.5:0.55, samples=40] ({-\R+5.0*cos(\x r)},{5.0*sin(\x r)},{-\R+5.0})
		-- plot[domain=5.0:2.1, samples=2]   ({-\R+\x*cos(0.55 r)},{\x*sin(0.55 r)},{-\R+\x})
		-- plot[domain=0.55:-0.5, samples=40] ({-\R+2.1*cos(\x r)},{2.1*sin(\x r)},{-\R+2.1})
		-- cycle;
		\foreach \f in {-0.36,-0.22,-0.09,0.13,0.27,0.41}
		\draw[gen] ({-\R+2.1*cos(\f r)},{2.1*sin(\f r)},{-\R+2.1})
		-- ({-\R+5.0*cos(\f r)},{5.0*sin(\f r)},{-\R+5.0});
		\foreach \s in {3.9,4.6}
		\draw[blue!32, densely dotted] plot[domain=-0.5:0.55, samples=40]
		({-\R+\s*cos(\x r)},{\s*sin(\x r)},{-\R+\s});
		\node[lab] at ({-\R+2.4*cos(0.45 r)},{2.4*sin(0.45 r)},{-\R+2.4}) {$X(N)$};

		\draw[black!35] plot[domain=-1.1:1.3, samples=30]
		({-\R+(\R+0.7*\x)*cos(-\x/\R r)},{(\R+0.7*\x)*sin(-\x/\R r)},{0.7*\x});
		\draw[black!35] plot[domain=-1.3:1.35, samples=30]
		({-\R+(\R+0.2*\x)*cos(\x/\R r)},{(\R+0.2*\x)*sin(\x/\R r)},{0.2*\x});
		\draw[tvec] (0,0,0) -- (0.7,-1.0,0.7) node[lab, right=0pt] {$u^\mu_0$};
		\draw[tvec] (0,0,0) -- (0.2,1.0,0.2) node[lab, below left=-2pt] {$u^\mu_1$};

		\draw[vec,blue!65!black] (0,0,0) -- (1.78,0,1.78) node[lab, above] {$v^\mu$};
		\NTconebody{0}{0}{0}{1.5}

		\draw[rig, thin, densely dotted] plot[domain=0.43:0.7, samples=20]
		({(\x*\x-1)},{2*\x},{(1+\x*\x)});                        
		\draw[rig, thick, densely dotted] plot[domain=-0.7:0.43, samples=40]
		({(\x*\x-1)},{2*\x},{(1+\x*\x)});
		\draw[rig, ->, thick] plot[domain=-0.1:-0.68, samples=20]
		({(\x*\x-1)},{2*\x},{(1+\x*\x)});
		\node[lab, rig, align=left, anchor=west] at (-0.54,-1.46,1.5) {Carroll\\[-2pt]boost};
		\draw[vec,rig] (0,0,0) -- (-1,0,1) node[lab, below left=0pt] {$\tau^\mu$};
		\fill (0,0,0) circle (1pt) node[lab, below=2pt] {$X(p)$};
	\end{scope}
	\node[lab] at (3.9,3.3) {$(M,g)$};
\end{tikzpicture}
\caption{Schematic representation of a null embedding $X : N \hookrightarrow M$. The induced metric $h_{ab}$ is degenerate, and its kernel is spanned by $v^a$, whose pushforward is the tangent null normal $v^\mu$. Together, the pair $(v^a,h_{ab})$ forms the induced Carrollian structure. The tangent map is $u^\mu_a = \D_a X^\mu$. Choosing an auxiliary null vector $\tau^\mu$ satisfying $v_\mu\tau^\mu=-1$ defines the intrinsic clock one-form $\tau_a = u^\mu_a \tau_\mu$. The set of all such allowed auxiliary null vectors $\tau^\mu$ form a parabola on the light cone at $X(p)$, and moving along it by performing a null rotation induces a Carroll boost on $N$, which changes the spatial subspace $\ker\tau_a$ while preserving the Carrollian vector field. Null boosts acting on the pair $(v^\mu,\tau^\mu)$ induce temporal Weyl transformations.}
\label{fig:null-embedding}
\end{figure}
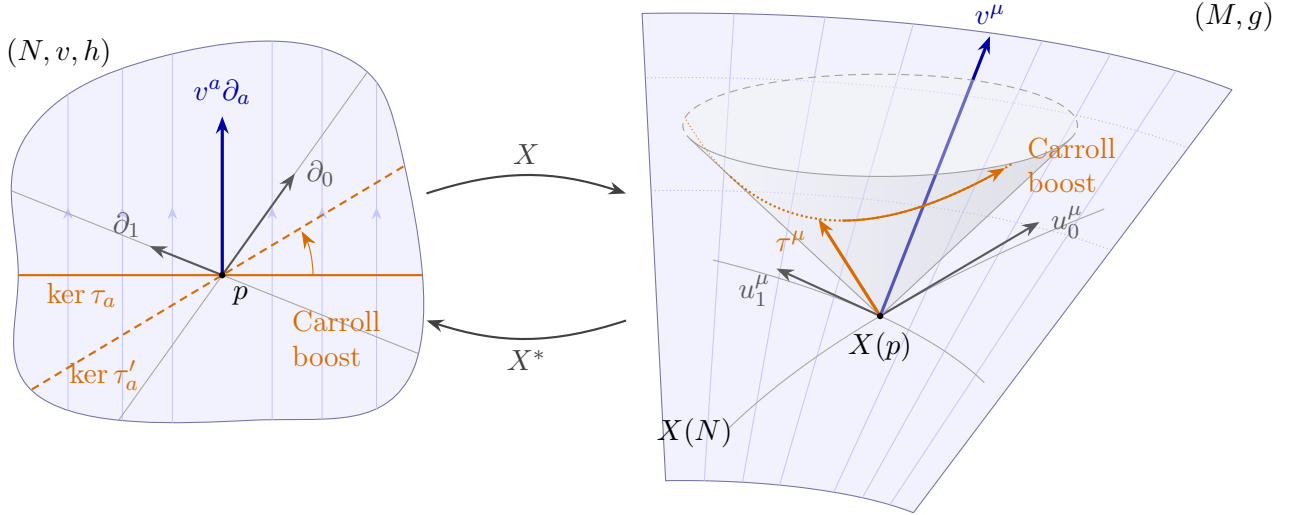

Let $(M,g)$ be an oriented $(d+2)$-dimensional Lorentzian spacetime with local coordinates $x^\mu$, where $\mu = 0,\dots,d+1$. We consider an oriented null hypersurface
\begin{equation}
N\overset{X}{\hookrightarrow} M\,, \qquad \sigma^a \mapsto X^\mu(\sigma^a)\,,
\end{equation}
embedded into $M$ with embedding maps $X^\mu(\sigma^a)$, where $\sigma^a$ with $a=0,\dots,d$ are coordinates on $N$. The embedding maps determine the location of the hypersurface via $x^\mu = X^\mu(\sigma^a)$ (cf.~Figure~\ref{fig:null-embedding}). The corresponding tangent maps are given by
\begin{equation}
\label{eq:pullback-maps}
    u^\mu_a = \D_a X^\mu\,,
\end{equation}
which are the components of the pushforward $X_*(\D_a) = u^\mu_a\D_\mu$ of the coordinate basis vectors $\D_a \in T_p N$. These tangent maps allow us to pull back covariant ambient tensors to $N$; in particular, the pullback of the metric gives rise to the induced metric on $N$
\begin{equation}
\label{eq:induced-metric}
    h_{ab} := u^\mu_a u^\nu_b g_{\mu\nu}\,.
\end{equation}
Let $v_\mu$ be a normal $1$-form to $N$
\begin{equation}
\label{eq:v-u-zero}
    v_\mu u^\mu_a = 0\,,
\end{equation}
and since $N$ is null, it holds on $N$ that
\begin{equation}
v^\mu v_\mu = 0\,,\qquad v^\mu = g^{\mu\nu} v_\nu\,,
\end{equation}
and in particular the vector $v^\mu$ is also tangent to $N$. Since the $u^\mu_a$ span the tangent space of the embedded hypersurface $X_*(T_pN)\subset T_{X(p)}M$, there exists a unique tangent vector $v^a\D_a \in T_p N$ such that
\begin{equation}
\label{eq:unique-decomp}
    v^\mu = v^a u^\mu_a\,.
\end{equation}
We may thus view $v^\mu$ as the components of the pushforward of $v^a\D_a \in T_pN$, which in turn satisfies 
\begin{equation}
    v^a h_{ab} = v_\mu u^\mu_b = 0\,.
\end{equation}
Thus, the induced metric has a one-dimensional kernel spanned $v^a$, and the pair
\begin{equation}
\label{eq:ind-carr-struct}
    (v^a,h_{ab})\,,
\end{equation}
defines the induced Carrollian structure on $N$, whose symmetries we will explore below.

We next introduce an auxiliary transverse null vector $\tau^\mu$, or null rigging, satisfying
\begin{equation}
\label{eq:tau-props}
    v_\mu \tau^\mu = -1\,,\qquad g(\tau,\tau)\big\vert_N = 0\,,
\end{equation}
where the first condition implies that $\tau^\mu\D_\mu \notin X_*(T_pN)$. Such a vector field always exists locally, though it is not canonical and the freedom in this choice will become the Carroll boost symmetry of the induced geometry. We may then decompose the ambient metric $g_{\mu\nu}$ along $N$ according to
\begin{equation}
\label{eq:ambient-metric-decomp}
    g_{\mu\nu} = h_{\mu\nu} - \tau_\mu v_\nu - \tau_\nu v_\mu\,,
\end{equation}
where 
\begin{equation}
h_{\mu\nu} v^\mu = 0 = h_{\mu\nu}\tau^\nu\,.    
\end{equation}
It is useful to introduce spatial vielbeine $e^A_\mu$ with $A = 1,\dots,d$ satisfying
\begin{equation}
    \label{eq:vielbein-amb}
    \delta_{AB} e^A_\mu e^B_\nu = h_{\mu\nu}\,,\qquad e_\mu^A v^\mu = 0 = e_\mu^A \tau^\mu\,,\qquad g^{\mu\nu}e_\mu^A e_\nu^B =\delta^{AB}\,.
\end{equation}
The property~\eqref{eq:v-u-zero} implies that 
\begin{equation}
    h_{ab} = u^\mu_a u^\nu_b h_{\mu\nu}\,,
\end{equation}
where $h_{ab}$ is the degenerate induced metric defined in~\eqref{eq:induced-metric}. Their pullbacks
\begin{equation}
e^A_a := u_a^\mu e^A_\mu\,,
\end{equation}
satisfy the relations
\begin{equation}
    h_{ab} = \delta_{AB}e^A_a e^B_b\,,\qquad e^A_a v^a = e^A_\mu u_a^\mu v^a = e_\mu^A v^\mu = 0\,.
\end{equation}
The components of the pullback of $\tau_\mu$ to $N$ are given by $\tau_a = u^\mu_a \tau_\mu$ and satisfy, by the same argument as above,
\begin{equation}
    \tau_a v^a = \tau_\mu v^\mu = -1\,,
\end{equation}
and hence this determines a splitting
\begin{equation}
    T_p N = \braket{v}\oplus \ker\tau\,,
\end{equation}
which allows us to introduce the spatial projector
\begin{equation}
\label{eq:h-projector}
   h^a_b := \delta^a_b + v^a\tau_b\,.
\end{equation}
We use this to define the spatial inverse $h^{ab}$ on $N$ via
\begin{equation}
    h^{ac}h_{cb} = h^a_b\,,\qquad h^{ab}\tau_a = 0\,.
\end{equation}
We emphasise that $h^{ab}$ depends on the choice of $\tau_a$, i.e., $h_{ab}$ has no preferred inverse, which is nothing but the statement that $h^{ab}$ transforms under Carroll boosts.

Using $h^{ab}$, we define the inverse projection
\begin{equation}
\label{eq:explicit-inverse}
    u_\mu^a := h^{ab}u^\nu_b g_{\mu\nu} - v^a\tau_\mu = h^{ab}u^\nu_b h_{\mu\nu} - v^a\tau_\mu\,,
\end{equation}
which is constructed such that
\begin{equation}
    v^a = u_\mu^a v^\mu\,,\qquad h^{ab} = u^a_\mu u^b_\nu g^{\mu\nu}\,,\qquad \tau^\mu u_\mu^a = 0\,.
\end{equation}
We furthermore have the relations
\begin{equation}
\label{eq:completeness-rels}
    u^a_\mu u^\mu_b = \delta^a_b\,,\qquad u^\mu_a u^a_\nu = \delta^\mu_\nu + \tau^\mu v_\nu\,,\qquad \delta^a_b = h^{ac}h_{cb} - v^a\tau_b\,.
\end{equation}
The second equation above follows from the observation that $(u^\mu_a,-\tau^\mu)$ form a basis for $T_{X(p)}M$ for $p\in N$, with $(u_\mu^a,v_\mu)$ the corresponding dual basis for $T^*_{X(p)}M$. For any chosen representatives $(\tau^\mu,h^{\mu\nu})$, the object~\eqref{eq:explicit-inverse} is the unique Carroll-covariant combination that satisfies the above relations.

\paragraph{Temporal Carroll--Weyl symmetry.}

We next determine how the transformations of the induced Carrollian geometry on $N$ are inherited from local Lorentz transformations of the ambient null frame. On $N$, where the metric decomposes as in~\eqref{eq:ambient-metric-decomp}, the ambient data $(v^\mu,\tau_\mu,h_{\mu\nu})$ transform under infinitesimal null boosts and null rotations 
\begin{equation}
\label{eq:trafos}
    \delta v^\mu = \alpha v^\mu\,,\qquad \delta\tau_\mu = - \alpha\tau_\mu + \lambda_\mu\,,\qquad \delta h_{\mu\nu} = 2 \lambda_{(\mu}v_{\nu)}\,,
\end{equation}
where $\alpha$ is the parameter of an infinitesimal null boost, while $\lambda_\mu$ is the parameter of an infinitesimal null rotation satisfying 
\begin{equation}
\label{eq:null-rot-param-props}
\lambda_\mu v^\mu = 0 = \lambda_\mu \tau^\mu\,,    
\end{equation}
implying that
\begin{equation}
    \lambda_\mu = e_\mu^A\lambda_A\,.
\end{equation}
The spatial vielbeine $e^A_\mu$ introduced in~\eqref{eq:vielbein-amb} transform as
\begin{equation}
\label{eq:amb-vielbein-trafo}
    \delta e^A_\mu = v_\mu \lambda^\nu e^A_\nu + R^A{_B}e^B_\mu\,,
\end{equation}
where $R^A{_B}$ is the parameter of an infinitesimal $SO(d)$ rotation. 

The embedding maps $X^\mu(\sigma^a)$ are inert under both null boosts and null rotations, and consequently the tangent vectors $u^\mu_a$ remain invariant
\begin{equation}
    \delta X^\mu = 0\,,\qquad \delta u^\mu_a = 0\,.
\end{equation}
Hence, the induced fields transform as
\begin{equation}
\label{eq:inuced-Carroll-trafos}
    \delta v^a = \alpha v^a\,,\qquad \delta \tau_a = -\alpha\tau_a + \lambda_a\,,\qquad \delta h_{ab} = 0\,,
\end{equation}
where
\begin{equation}
    \lambda_a := u^\mu_a \lambda_\mu\,,\qquad v^a \lambda_a = 0\,.
\end{equation}
The ambient null rotations thus become Carroll boosts on $N$, while null boosts rescale the Carrollian vector while leaving the spatial ruler unchanged. We will refer to this latter transformation  as a temporal Weyl transformation, and to the full symmetry as temporal Carroll--Weyl symmetry, which is the anisotropic Carroll--Lifshitz transformations discussed in Section~4 of~\cite{Hartong:2015xda} but with $z=\infty$. This symmetry will have drastic implications for the intrinsic fluid dynamics on null hypersurfaces we develop in the next section.

Using~\eqref{eq:h-projector}, we may derive the transformation properties of $h^{ab}$ using those of $\tau_a$, leading to
\begin{equation}
    \delta h^{ab} = 2v^{(a}\lambda^{b)}\,,
\end{equation}
with $\lambda^a := h^{ab}\lambda_b$. Note furthermore that $u_\mu^a$ as defined in~\eqref{eq:explicit-inverse} transforms under Carroll boosts as
\begin{equation}
\label{eq:projector-trafo}
    \delta u_\mu^a = \lambda^a v_\mu\,,
\end{equation}
which follows from the identity $\lambda^b u^\nu_b h_{\mu\nu} = \lambda_\mu$, which uses the properties of the null rotation parameter~\eqref{eq:null-rot-param-props}. Finally, we may also express $h^{ab}$ in terms of the induced inverse vielbein $e_A^\mu$ introduced in~\eqref{eq:vielbein-amb}
\begin{equation}
\label{eq:inv-vielbein-def}
  h^{ab} = \delta^{AB}e_A^a e_B^b\,,\qquad  e_A^a := u_\mu^a e^\mu_A\,,
\end{equation}
which satisfies $e^a_A\tau_a = 0$ and transforms as
\begin{equation}
    \delta e^a_A = v^a \lambda_A + R_A{^B}e_B^a\,,
\end{equation}
where we used the transformation law~\eqref{eq:amb-vielbein-trafo} and the relation $v_\mu e^\mu_A = 0$.

\paragraph{Volume form.} 

We now discuss the volume form induced on $N$, which is an important ingredient for constructing actions on $N$. Let $\boldsymbol{\epsilon}_M$ denote the ambient volume form. The auxiliary null vector $\tau^\mu$ defines a $(d+1)$-form along $N$ by
\begin{equation}
\label{eq:ind-vol}
    \tilde{\boldsymbol{\epsilon}}_N := -\tau\lrcorner \,\boldsymbol{\epsilon}_M\,,
\end{equation}
where ``$\lrcorner$'' denotes contraction. The pullback of $\tilde{\boldsymbol{\epsilon}}_N$ gives the induced volume form on $N$
\begin{equation}
    \boldsymbol{\epsilon}_N := X^*\tilde{\boldsymbol{\epsilon}}_N\,,
\end{equation}
where the sign in~\eqref{eq:ind-vol} follows from $v^\mu\tau_\mu = -1$ when requiring that
\begin{equation}
\label{eq:restriction-of-volume-form}
    \boldsymbol{\epsilon}_M\big\vert_N = v\wedge\tilde{\boldsymbol{\epsilon}}_N\,,
\end{equation}
where, in particular, the right-hand side is not a pullback since the ambient top-form pull back to zero for dimensional reasons. In terms of the coframe $(\tau,e^A)$ on $T_p^*N$, we explicitly have
\begin{equation}
\label{eq:induced-volume-form}
    \boldsymbol{\epsilon}_N = \tau \wedge e^1\wedge\cdots\wedge e^d = e\,d^{d+1}\sigma\,,
\end{equation}
where
\begin{equation}
    e = \det(\tau_a,e^A_a)\,.
\end{equation}
In particular, under Carroll boosts $\delta_\lambda e^A_a = 0$ and $\delta_\lambda\tau_a = \lambda_a$, where $\lambda_a$ is spatial, so the volume form is invariant. Under temporal Weyl transformations it has weight $-1$
\begin{equation}
    \delta_\alpha \boldsymbol{\epsilon}_N = -\alpha \boldsymbol{\epsilon}_N\,,
\end{equation}
which comes from the single factor of $\tau$ in~\eqref{eq:induced-volume-form}.

\paragraph{Induced connection.}

We now discuss the connection $D$ induced on $N$ by the ambient Levi--Civita connection $\nabla$ using the formalism developed above. For a vector $X^a$ and a $1$-form $Y_a$ on $N$, we define
\begin{equation}
\label{eq:ind-conn-on-vecs-and-1-forms}
    D_a X^b = u^\mu_a u^b_\nu\nabla_\mu (u^\nu_c X^c)\,,\qquad D_a Y_b = u^\mu_a u^\nu_b\nabla_\mu (u^c_\nu Y_c)\,.
\end{equation}
The corresponding connection coefficients $\gamma^c_{ab}$ are given by
\begin{equation}
\label{eq:ind-connection-def}
    \gamma_{ab}^c = u^c_\mu\D_a u^\mu_b + u^\mu_a u^\nu_b u^c_\rho \Gamma^\rho_{\mu\nu} = u^c_\nu u^\mu_a\nabla_\mu u^\nu_b\,, 
\end{equation}
Since the Levi--Civita connection $\nabla$ is torsion-free, and $u^\mu_b = \D_b X^\mu$ so that
\begin{equation}
\label{eq:integrability}
    \D_{[a}u^\mu_{b]} = 0\,,
\end{equation}
it follows that $D$ is also torsion-free
\begin{equation}
    \gamma_{[ab]}^c = 0\,,
\end{equation}
though, as we shall see, it is not adapted to the induced Carrollian structure. 

The analogue of the Weingarten identity is
\begin{equation}
\label{eq:Weingarten}
    u^\nu_a\nabla_\nu u^\mu_b = u^\mu_c \gamma^c_{ab} - \tau^\mu{K}_{ab}\,,
\end{equation}
where we defined the extrinsic curvature of the null surface as
\begin{equation}
\label{eq:intrinsic-torsion}
    {K}_{ab} = -u^\mu_a u^\nu_b\nabla_{(\mu} v_{\nu)} = - \frac{1}{2}u^\mu_a u^\nu_b\pounds_v g_{\mu\nu}\,.
\end{equation}
Now, taking the derivative $\D_a$ of $v_\mu u^\mu_b = 0$ and antisymmetrising in $a,b$, the condition~\eqref{eq:integrability} implies that the pullback of $dv$ vanishes. Thus, for any smooth extension of the normal $1$-form away from $N$, we have
\begin{equation}
\label{eq:HSO-induced-conn}
    v \wedge dv\big\vert_N = 0\,,
\end{equation}
implying that along $N$ there exists a $1$-form $\alpha$ such that $dv = v\wedge\alpha$, or, in components,
\begin{equation}
    \nabla_{[\mu}v_{\nu]} = v_{[\mu}\alpha_{\nu]}\,.
\end{equation}
Contracting this with $v^\nu$ gives
\begin{equation}
    v^\nu\nabla_\nu v_\mu = \frac{1}{2}\D_\mu(v^2) - v_\mu v^\nu\alpha_\nu\,,
\end{equation}
along $N$. Since $v^2$ vanishes on $N$, its derivative is normal to $N$
\begin{equation}
	\D_\mu (v^2) = - \tau^\nu \D_\nu(v^2)v_\mu \,,
\end{equation}
and hence the null normal is geodesic
\begin{equation}
\label{eq:geodesic}
    v^\nu\nabla_\nu v_\mu = c_v v_\mu\,,\qquad c_v = -v^\mu\alpha_\mu - \frac{1}{2}\tau^\mu\D_\mu(v^2)\,,
\end{equation}
where $c_v$ is the inaffinity. If the extension of $v$ is null in a neighbourhood of $N$, the last term vanishes and $c_v = -v^\mu\alpha_\mu$. This in turn implies that $K_{ab}$ defined in~\eqref{eq:intrinsic-torsion} is spatial, i.e., its contraction with $v^a$ vanishes
\begin{equation}
    v^a K_{ab} = -\frac{1}{2}u^\nu_b\left(v^\mu\nabla_\mu v_\nu + \frac{1}{2}\D_\nu(v^2)\right) = 0\,,
\end{equation}
which is independent of the extension of $v$. The vanishing pullback of $dv$ also implies that the symmetrisation in the first equality in~\eqref{eq:intrinsic-torsion} is redundant.

The extrinsic curvature $K_{ab}$ also represents the intrinsic torsion of the Carrollian geometry induced on $N$~\cite{Figueroa-OFarrill:2020gpr}.\footnote{
	Strictly speaking, describing $N$ as a $G$-structure requires adding the temporal Weyl symmetry to the cases considered in~\cite{Figueroa-OFarrill:2020gpr}. The additional generator shifts torsion components along $v^a$, which may be removed by Carrolls boosts and so does not enlarge the image of the Spencer differential. The intrinsic torsion is therefore still $K_{ab}$, but now with temporal Weyl weight $+1$ since $\delta_\alpha K_{ab} = \alpha K_{ab}$. I am grateful to Jos\'e Figueroa-O'Farrill for useful discussions on this point. 
	} This follows from the fact that the pullback commutes with the Lie derivative along a vector whose flow preserves $N$, i.e.,
\begin{equation}
\label{eq:has-intrinsic-Lie-derivative}
    X^*\pounds_v g = \pounds_{v}^N(X^*g) \Rightarrow K_{ab} = -\frac{1}{2}\pounds^N_{v}h_{ab}\,,
\end{equation}
where ``$\pounds^N_{v}$'' denotes the Lie derivative on $N$ along $v^a$. Next, we consider the action of $D$ on the Carrollian data on $N$. First, observe that
\begin{equation}
\label{eq:D-on-h-lower}
        D_a h_{bc} = u^\rho_a u^\sigma_b u^\lambda_c\nabla_\rho( u^d_\sigma u^e_\lambda h_{de}) = u^\rho_a u^\sigma_b u^\lambda_c\nabla_\rho(g_{\mu\nu}u^\mu_d u^\nu_e u^d_\sigma u^e_\lambda ) = -2{K}_{a(b}\tau_{c)}\,,
\end{equation}
where we used the completeness relations~\eqref{eq:completeness-rels}, while
\begin{equation}
    D_a v^b = -h^{bc}K_{ac} + v^b \omega_a\,,
\end{equation}
where we introduced the H\'aj\'i\v cek $1$-form~\cite{Hajicek:1973hof}
\begin{equation}
\label{eq:Hajicek}
    \omega_a := u^\mu_a v^\nu \nabla_\mu \tau_\nu = v^b D_a \tau_b\,.
\end{equation}
The H\'aj\'i\v cek $1$-form transforms as
\begin{equation}
    \delta \omega_a = \lambda^b K_{ab} + \D_a\alpha\,,
\end{equation}
i.e., it acts as a Weyl connection for the null boost, but it picks up a piece proportional to the extrinsic curvature under Carroll boosts.
We furthermore get
\begin{subequations}
\begin{align}
    D_a \tau_b &= G_{ab} + F_{ab} + a_{[a} \tau_{b]}  \,, \label{eq:D-tau}
    \\
    D_a h^{bc} &= 2v^{(b} h^{c)d}\left(G_{ad} + F_{ad} - \frac{1}{2} a_d \tau_a\right)\,,
\end{align}
\end{subequations}
where we defined
\begin{equation}
        F_{ab} = h^c_{[a} h^d_{b]}D_c\tau_d\,,\qquad  a_a = \pounds^N_v \tau_a = u^\mu_a\pounds_v\tau_\mu = v^b D_b\tau_a - \omega_a\,,\qquad G_{ab} = \frac{1}{2}u^\mu_a u^\nu_b \pounds_\tau g_{\mu\nu}\,,
\end{equation}
The objects $F_{ab}$ and $a_a$ are both spatial, i.e.,
\begin{equation}
    v^a F_{ab} = 0 = v^a a_a\,,
\end{equation}
while $G_{ab}$ is like an extrinsic curvature for the auxiliary null vector.\\

Next, we work out the explicit expression for the connection coefficients $\gamma^c_{ab}$. Using~\eqref{eq:D-tau}, we find that $D_{(a}\tau_{b)} = G_{ab}$, implying that 
\begin{equation}
\label{eq:tau-proj-gamma}
    \gamma^c_{ab}\tau_c = \D_{(a}\tau_{b)} - G_{ab}\,.
\end{equation}
Similarly, using~\eqref{eq:D-on-h-lower} and writing out the explicit expression for the covariant derivative, we add the standard cyclic combination of this expression to obtain
\begin{equation}
\label{eq:h-proj-gamma}
    \gamma^d_{ab} h_{dc} = \frac{1}{2}\left( \D_a h_{bc} + \D_b h_{ac} - \D_c h_{ab} \right) + K_{ab}\tau_c\,,
\end{equation}
where the last term vanishes upon contraction with $h^{cd}$. Using the last completeness relation in~\eqref{eq:completeness-rels}, we may therefore combine~\eqref{eq:tau-proj-gamma} and~\eqref{eq:h-proj-gamma} to obtain
\begin{equation}
\begin{split}
    \gamma^c_{ab} &= h^c_d \gamma^d_{ab} - v^c\tau_d\gamma^d_{ab}\\
    &= -v^c \D_{(a}\tau_{b)} + \frac{1}{2} h^{cd}\left( \D_a h_{bd} + \D_b h_{ad} - \D_d h_{ab} \right) + v^c G_{ab}\,,
\end{split}
\end{equation}
which in particular involves $G_{ab}$ which is not intrinsic to the surface. In general, connections on Carrollian geometries contain an undetermined symmetric tensor, but in the embedded geometry this piece is precisely $G_{ab}$. In this way, the ambient Levi--Civita connection induces a canonical connection on the embedded null hypersurface for the chosen auxiliary null vector.

The induced connection is invariant under temporal Weyl rescalings but not under Carroll boosts. In particular, the transformation law for $u^a_\mu$ in~\eqref{eq:projector-trafo} combined with the definition of the components of the induced connection in~\eqref{eq:ind-connection-def} implies that 
\begin{equation}
    \delta_\lambda \gamma^c_{ab} = \lambda^c K_{ab}\,,
\end{equation}
where we used the Weingarten identity~\eqref{eq:Weingarten} along with the relation $v_\mu u^\mu_a = 0$. 

\subsection{Variational calculus and induced response identities}
\label{sec:variational-calculus}

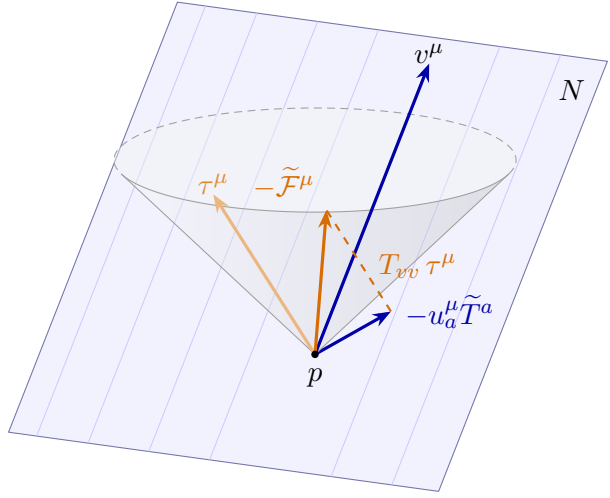
\begin{figure}
	\centering
	\begin{tikzpicture}[
		scale=1.4,
		x={(0.469cm,0.2287cm)},
		y={(-0.883cm,0.1215cm)},
		z={(0cm,0.966cm)},
		>={Stealth[length=2.4mm,width=1.8mm]},
		vec/.style={->,very thick,line cap=round},
		lab/.style={font=\small},
		cone/.style={gray!75},
		ray/.style={dashed, orange!85!black},
		]
		\def\aL{133.53}
		\def\aR{282.41}
		\def\NTcone#1#2#3#4{%
			\begin{scope}[shift={(#1,#2,#3)}]
				\draw[blue!45!black, thick] (0,0,0) -- (#4,0,#4);
				\shade[left color=black!3, right color=black!24, opacity=0.4]
				(0,0,0) -- plot[domain=\aL:\aR, samples=48] ({#4*cos(\x)},{#4*sin(\x)},#4) -- cycle;
				\fill[black!5, opacity=0.45]
				plot[domain=0:360, samples=80] ({#4*cos(\x)},{#4*sin(\x)},#4) -- cycle;
				\draw[cone, densely dashed]
				plot[domain=\aR-360:\aL, samples=60] ({#4*cos(\x)},{#4*sin(\x)},#4);
				\draw[cone] plot[domain=\aL:\aR, samples=48] ({#4*cos(\x)},{#4*sin(\x)},#4);
				\draw[cone] (0,0,0) -- ({#4*cos(\aL)},{#4*sin(\aL)},#4);
				\draw[cone] (0,0,0) -- ({#4*cos(\aR)},{#4*sin(\aR)},#4);
		\end{scope}}
		\filldraw[fill=blue!5, draw=blue!40!black!55, line join=round]
		(-0.9,2.8,-0.9) -- (-0.9,-1.8,-0.9) -- (2.5,-1.8,2.5) -- (2.5,2.8,2.5) -- cycle;
		\foreach \yy in {-1.3,-0.65,0.65,1.3,1.95,2.5}
		\draw[blue!20] (-0.9,\yy,-0.9) -- (2.5,\yy,2.5);
		\node[lab] at (2.25,-1.55,2.25) {$N$};
		\draw[vec, blue!65!black] (0,0,0) -- (0.4,-0.6,0.4)            
		node[lab, right=1pt] {$-u^\mu_a\widetilde T^{a}$};
		\NTcone{0}{0}{0}{1.9}
		\draw[vec, blue!65!black] (0,0,0) -- (2.3,0,2.3);
		\node[lab] at (2.3,0,2.4) {$v^\mu$};
		\draw[vec, orange!85!black!50] (0,0,0) -- (-2.05,0,2.05);
		\node[lab, orange!85!black!80] at (-2.05,0,2.15) {$\tau^\mu$};
		\draw[ray, thick, -] (0.4,-0.6,0.4) -- (-0.9,-0.6,1.7)
		node[lab, pos=0.45, right=2pt] {$T_{vv}\,\tau^\mu$};
		\draw[vec, orange!85!black] (0,0,0) -- (-0.9,-0.6,1.7)            
		node[lab, above left=0pt] {$-\widetilde{\mathcal F}^{\mu}$};
		\fill (0,0,0) circle (1pt) node[lab, below=1pt] {$p$};
	\end{tikzpicture}
\caption{Decomposition of the ambient null traction at a point $p$ of a null hypersurface $N$. The null normal $v^\mu$ is tangent to $N$, while the auxiliary null vector $\tau^\mu$, satisfying $v_\mu\tau^\mu=-1$, defines the tangential projector via $u^\mu_a u^a_\nu = \delta^\mu_\nu + \tau^\mu v_\nu$. Applying this projector to the null traction $-\widetilde{\mathcal F}^\mu = -v_\nu T^{\mu\nu}$ gives the tangential vector $-u^\mu_a\widetilde T^a$, where $\widetilde T^a = u^a_\mu\widetilde{\mathcal F}^\mu$ is the induced energy current. Note that the figure illustrates a causal flux with $T_{vv} := v_\mu v_\nu T^{\mu\nu} > 0$ so that $-\widetilde{\mathcal F}^\mu$ is future-directed for future-directed $v^\mu$, since $v_\mu(-\widetilde{\mathcal F}^\mu) = -T_{vv} < 0$. The dashed orange line represents the transverse contribution $T_{vv}\tau^\mu$ appearing in the decomposition $-\widetilde{\mathcal F}^\mu = -u^\mu_a\widetilde T^a + T_{vv}\tau^\mu$. Note that while the tangential projection depends on the choice of auxiliary null vector, the full ambient null traction is independent of this choice.}
\label{fig:null-traction}
\end{figure}

We now develop the variational calculus of embedded null hypersurfaces at fixed embedding, i.e.,
\begin{equation}
    \delta X^\mu = 0 \implies \delta u^\mu_a = 0\,.
\end{equation}
Later, in Sections~\ref{sec:null-fluids-on-surfs} and~\ref{sec:open-Carroll-fluids}, we will investigate the physics of ambient fluids restricted to null hypersurfaces. To that end, we consider the variation of the ambient metric along $N$. Since $(u^\mu_a,-\tau^\mu)$ is a basis for $T_{X(p)}M$, there are three natural projections of $\delta g_{\mu\nu}$ at $N$, namely
\begin{equation}
    u^\mu_a u^\nu_b \delta g_{\mu\nu} = \delta h_{ab}\,,\qquad -\tau^\mu u^\nu_a \delta g_{\mu\nu}\,,\qquad \tau^\mu\tau^\nu\delta g_{\mu\nu}\,.
\end{equation}
and using completeness~\eqref{eq:completeness-rels}, we find that
\begin{equation}
    \delta g_{\mu\nu}\big\vert_{N} = u^a_\mu u^b_\nu \delta h_{ab} - 2v_{(\mu}u_{\nu)}^a \tau^\rho u^\sigma_a \delta g_{\rho\sigma} + v_\mu v_\nu \tau^\rho\tau^\sigma \delta g_{\rho\sigma}\,.
\end{equation}
Since the embedding remains fixed, we have
\begin{equation}
    -\tau^\rho u^\sigma_a \delta g_{\rho\sigma} = -\delta\tau_a + u_{\rho a}\delta \tau^\rho\,,
\end{equation}
where $u_{\rho a} := g_{\rho\sigma}u^\sigma_a$. Furthermore, since $\tau^\mu$ remains null under variations, 
\begin{equation}
    0 = \delta(\tau^\rho\tau^\sigma g_{\rho\sigma}) \implies \tau^\rho \tau^\sigma\delta g_{\rho\sigma} = -2\tau_\rho\delta\tau^\rho\,,
\end{equation}
which means that we write
\begin{equation}
\label{eq:delta-g-on-N}
    \delta g_{\mu\nu}\big\vert_{N} = u^a_\mu u^b_\nu \delta h_{ab} - 2v_{(\mu}u_{\nu)}^a\delta\tau_a  + 2v_{(\mu}g_{\nu)\rho}\delta\tau^\rho\,,
\end{equation}
where we used completeness in the form $u^a_\nu u_{\mu a} = g_{\nu\mu} + v_\nu\tau_\mu$ to obtain the last term. In particular, the decomposition of the variation of the ambient metric is parametrised by both the intrinsic (induced) Carrollian data $(\tau_a,h_{ab})$ along with the variations of the auxiliary transverse null vector $\tau^\mu$. As we will see below, its conjugate response is the energy-momentum flux through the null surface, which we call the null traction. We have illustrated this in Figure~\ref{fig:null-traction}.

We now derive the induced responses, which arise as projections of the ambient energy-momentum tensor restricted to $N$. The ambient energy-momentum tensor $T^{\mu\nu}$ is the response to varying the metric, so restricting $T^{\mu\nu}\delta g_{\mu\nu}$ to $N$ using our decomposition~\eqref{eq:delta-g-on-N} will give precisely these projections, i.e.,
\begin{equation}
\label{eq:variation-decomp}
    \frac{1}{2}\left(T^{\mu\nu}\delta g_{\mu\nu}\right)\big\vert_N = -\widetilde T^a\delta \tau_a + \frac{1}{2}\widetilde{\mathcal{T}}^{ab}\delta h_{ab} + \widetilde{\mathcal{F}}_\mu \delta \tau^\mu\,,
\end{equation}
where we defined
\begin{equation}
\label{eq:currents-on-surface}
    \widetilde T^a := v_\mu u^a_\nu T^{\mu\nu}\,,\qquad \widetilde{\mathcal{T}}^{ab} := u_\mu^a u_\nu^b T^{\mu\nu}\,,\qquad \widetilde{\mathcal{F}}_\mu := v_\nu T^\nu{_\mu}\,.
\end{equation}
The first two quantities are similar to the Carrollian and the momentum-stress tensor, respectively, while $\widetilde{\mathcal{F}}_\mu$ is the null traction discussed above. This null traction may also be viewed as the obstruction to describing the ambient response intrinsically on $N$ purely in terms of Carrollian currents. The tildes emphasise that these objects are universal projections of the ambient energy-momentum tensor, with the precise relation to the intrinsic Carrollian responses on $N$ depending on the physical realisation, and in particular on the normalisation of the hypersurface volume form, i.e., how precisely $\boldsymbol{\epsilon}_N$ is related to $\boldsymbol{\epsilon}_M$. Indeed, it is important to emphasise that~\eqref{eq:variation-decomp} is a local identity on $N$, and the ambient action itself will not in general restrict to a lower-dimensional action. We will return to this in Sections~\ref{sec:null-fluids-on-surfs} and~\ref{sec:open-Carroll-fluids}. In a more general treatment, where the embedding itself is allowed to vary, there would be additional responses corresponding to varying the extrinsic geometry which would enter the shape equation. We leave the corresponding dynamical submanifold calculus to future work. 

Finally, the variations must be such that $N$ remains null. Since we keep the null hypersurface fixed, the normal $1$-form $v_\mu$ can only vary by a rescaling, implying that
\begin{equation}
   0 = \delta (g^{\mu\nu} v_\mu v_\nu ) = - v^\mu v^\nu \delta g_{\mu\nu}\,.
\end{equation}
Using~\eqref{eq:unique-decomp}, this is equivalent to the standard Carrollian constraint which holds regardless of whether the embedding is kept fixed or not
\begin{equation}
    v^a v^b\delta h_{ab} = 0\,,
\end{equation}
which in turn implies that
\begin{equation}
\label{eq:intrinsic-ambiguity}
    \widetilde{\mathcal{T}}^{ab} \sim \widetilde{\mathcal{T}}^{ab} + f v^a v^b\,,
\end{equation}
where $f$ is an arbitrary function on $N$. In other words, the intrinsic momentum-stress response $\widetilde{\mathcal{T}}^{ab}$ is only defined up to terms proportional to $v^a v^b$, which is an intrinsic ambiguity of $\widetilde{\mathcal{T}}^{ab}$.

\paragraph{Identities for responses and Ward identities.}

The projected responses in~\eqref{eq:currents-on-surface} are not independent and can be related in a way that corresponds to the Ward identities for Carroll boosts and temporal Weyl transformations from the intrinsic perspective of $N$. To see this, note that raising the index in~\eqref{eq:explicit-inverse} gives
\begin{equation}
\label{eq:raised-inverse-projection}
    u^a_\nu g^{\nu\mu}=h^{ab}u^\mu_b-v^a\tau^\mu\,,
\end{equation}
which, using~\eqref{eq:currents-on-surface}, implies that
\begin{equation}
\label{eq:embedded-response-identity}
    \widetilde T^a=h^{ab}\widetilde{\mathcal F}_b - v^a\tau^\mu\widetilde{\mathcal F}_\mu\,,\qquad \widetilde{\mathcal F}_a:=u^\mu_a\widetilde{\mathcal F}_\mu\,.
\end{equation}
Its temporal and spatial projections are therefore
\begin{equation}
\label{eq:embedded-Ward-identities}
    \tau_a\widetilde T^a=\tau^\mu\widetilde{\mathcal F}_\mu\,,\qquad \widetilde T^b h_{ba}=h^b_a\widetilde{\mathcal F}_b\,.
\end{equation}
Formally, these same identities arise as Ward identities by requiring that~\eqref{eq:variation-decomp} is invariant under the Carroll boosts and temporal Weyl transformations in~\eqref{eq:inuced-Carroll-trafos} induced from the ambient transformations in~\eqref{eq:trafos}, which act on $\tau^\mu$ as
\begin{equation}
	\delta \tau^\mu = -\alpha \tau^\mu + u^\mu_a\lambda^a\,,
\end{equation} 
so that invariance of~\eqref{eq:variation-decomp} becomes
\begin{equation}
    0=\alpha\left(\tau_a\widetilde T^a-\tau^\mu\widetilde{\mathcal F}_\mu\right)
      +\lambda_a\left(h^{ab}\widetilde{\mathcal F}_b-\widetilde T^a\right)\,.
\end{equation}
Since $\alpha$ and the spatial parameter $\lambda_a$ are arbitrary, this precisely reproduces~\eqref{eq:embedded-Ward-identities}. Notably, these Ward identities have extra terms compared to what one would find in an intrinsic Carrollian fluid on a null hypersurface: these extra terms involve the null traction and arise because the source multiplet also contains $\tau^\mu$ in addition to the intrinsic Carrollian structure. In this way, the Carroll boost and temporal-Weyl Ward identities are encoded in the submanifold calculus as identities among the projected ambient responses even before imposing ambient conservation or specifying an intrinsic functional. We again emphasise that their precise relation to intrinsic Ward identities depends on the physical realisation of the theory on the hypersurface and the choice of measure, as we will see in Sections~\ref{sec:null-fluids-on-surfs} and~\ref{sec:open-Carroll-fluids}. 

\paragraph{The diffeomorphism Ward identity.}

The ambient diffeomorphism Ward identity encodes conservation of the energy-momentum tensor
\begin{equation}
	\label{eq:ambient-cons}
	\nabla_\mu T^\mu{_\nu} = 0\,,
\end{equation}
which splits into its tangential $u^\mu_a$-projection along $N$ and its transverse $\tau^\mu$-projection transverse to it (recall that $(u^\mu_a,-\tau^\mu)$ form a basis for $T_{X(p)}M$, while the null normal $v^\mu = v^a u^\mu_a$ is itself tangent). In this section, we will discuss what these projections correspond to from the intrinsic Carrollian perspective. To proceed, it is useful to define the analogue of the Carrollian energy-momentum tensor for the induced responses
\begin{equation}
\label{eq:embedded-EMT}
    \widetilde T^a{_b}:=-\tau_b\widetilde T^a + \widetilde{\mathcal T}^{ac}h_{cb}\,,
\end{equation}
which, using completeness~\eqref{eq:completeness-rels}, is related to the ambient energy-momentum tensor via
\begin{equation}
\label{eq:embedded-EMT-decomp}
    T^\mu{_\nu}u^\nu_a = (u^\mu_b u^b_\rho - \tau^\mu v_\rho)T^{\rho\sigma}(h_{ac} u^c_\sigma - \tau_a v_\sigma ) = u^\mu_b\widetilde T^b{_a} - \tau^\mu\widetilde{\mathcal F}_a\,,
\end{equation}
where  to rewrite it in terms of the induced responses we used that
\begin{equation}
	\widetilde{\mathcal F}_a =h_{ab}\widetilde{T}^b - \tau_a v_\rho v_\sigma T^{\rho\sigma}\,.
\end{equation}
Before we project the ambient conservation equation~\eqref{eq:ambient-cons}, it is useful to derive the ``intrinsic-looking'' diffeomorphism Ward identity, i.e., the Ward identity one would obtain on the null hypersurface if only the induced Carrollian structure were present. To obtain this, we formally vary the first two terms in~\eqref{eq:variation-decomp} under diffeomorphisms while keeping $\tau^\mu$ fixed. This yields the expression
\begin{equation}
	\label{eq:intrinsic-looking-diffeo-WI}
	\begin{split}
			\widetilde{\mathcal W}_a &:= e^{-1}\D_b(e\widetilde T^b{_a})+\widetilde T^b\D_a\tau_b-\frac12\widetilde{\mathcal T}^{bc}\D_a h_{bc} \\
			&~=D_b\widetilde T^b{_a}-\omega_b\widetilde T^b{_a} + \widetilde T^bD_a\tau_b - \frac{1}{2}\widetilde{\mathcal T}^{bc}D_a h_{bc}\,,
	\end{split}
\end{equation}
where we used that 
\begin{equation}
	D_a e:=\D_a e-\gamma^b_{ab}e = -\omega_a e\,,
\end{equation}
which in turn follows from the identity
\begin{equation}
	e^{-1}D_a e=-v^bD_a\tau_b+\frac12h^{bc}D_a h_{bc}=-\omega_a\,.
\end{equation}
A closed Carrollian system would satisfy $\widetilde{\mathcal{W}}_a = 0$. Note also that $\widetilde{\mathcal W}_a$ is insensitive to the ambiguity~\eqref{eq:intrinsic-ambiguity}, since $v^a v^b D_c h_{ab} = 0$. We next project the ambient Ward identity~\eqref{eq:ambient-cons} along $N$; using the completeness relation for the projectors~\eqref{eq:completeness-rels} once again, we find
\begin{equation}
\label{eq:WI-split}
    u^\nu_a\nabla_\mu T^\mu{_\nu}
       =u^\rho_bu^b_\mu u^\nu_a\nabla_\rho T^\mu{_\nu}
        -\tau^\rho v_\mu u^\nu_a\nabla_\rho T^\mu{_\nu}\,,
\end{equation}
where all ambient quantities are evaluated on $N$. The first term only involves tangential derivatives, and to evaluate it we need the following useful identity
\begin{equation}
\label{eq:auxiliary-expansion}
\begin{split}
    u^b_\mu u^\rho_b\nabla_\rho\tau^\mu
       &=h^{bc}u^\rho_bu^\mu_c\nabla_\rho\tau_\mu
          -v^b u^\rho_b\tau_\mu\nabla_\rho\tau^\mu\\
       &=h^{bc}D_b\tau_c=h^{bc}G_{bc}=:G\,,
\end{split}
\end{equation}
where we used the explicit expression for the inverse projector in~\eqref{eq:explicit-inverse} in the second equality, and $\tau_\mu \nabla_\rho\tau^\mu =0$ as well as~\eqref{eq:ind-conn-on-vecs-and-1-forms} in the third. Then, using this identity along with the Weingarten identity~\eqref{eq:Weingarten} and the decomposition~\eqref{eq:embedded-EMT-decomp}, we can write the first term in~\eqref{eq:WI-split} as
\begin{equation}
\label{eq:tang-proj-WI}
\begin{split}
	u^\rho_bu^b_\mu u^\nu_a\nabla_\rho T^\mu{_\nu} &= D_b\widetilde T^b{_a} + \widetilde{\mathcal T}^{bc}\tau_c K_{ab} - G\widetilde{\mathcal F}_a\\
	&= \widetilde{\mathcal W}_a + \omega_b\widetilde T^b{_a}
	-\widetilde T^bD_a\tau_b-G\widetilde{\mathcal F}_a\,,
\end{split}
\end{equation}
where we used that $D_a h_{bc}=-2K_{a(b}\tau_{c)}$ to write this in terms of $\widetilde{\mathcal W}_a$ defined in~\eqref{eq:intrinsic-looking-diffeo-WI}. Hence, the tangentially projected ambient diffeomorphism Ward identity~\eqref{eq:WI-split} becomes
\begin{equation}
\label{eq:single-surf-sourced-WI}
   0 = u^\nu_a\nabla_\mu T^\mu{_\nu}\big\vert_N \implies \widetilde{\mathcal W}_a = \tau^\rho v_\mu u^\nu_a\nabla_\rho T^\mu{_\nu} - \omega_b\widetilde T^b{_a} + \widetilde T^b D_a\tau_b + G\widetilde{\mathcal F}_a\,.
\end{equation}
Hence, on $N$, ambient conservation gives rise to a sourced Ward identity from the intrinsic perspective, with the right-hand side acting as the source. In general, it cannot be reconstructed from the responses and their tangential derivatives on $N$. In Section~\ref{sec:foliation-balance}, we will show how a null foliation gives rise to a source encoding transport between neighbouring null leaves.

In addition to the tangential projection in~\eqref{eq:WI-split}, we must also consider the transverse projection along $-\tau^\mu$, which we may write as
\begin{equation}
	-\tau^\nu\nabla_\mu T^\mu{_\nu} = -\tau^\nu u^\rho_bu^b_\mu\nabla_\rho T^\mu{_\nu} + \tau^\rho\tau^\nu v_\mu\nabla_\rho T^\mu{_\nu}\,.
\end{equation}
Using the same manipulations as above, we find that the tangential term becomes
\begin{equation}
	-\tau^\nu u^\rho_bu^b_\mu\nabla_\rho T^\mu{_\nu} = -D_b(\widetilde{\mathcal T}^{bc}\tau_c) + \widetilde{\mathcal T}^{bc}G_{bc} + G\tau_b\widetilde T^b\,,
\end{equation}
so that the transverse projection of the ambient diffeomorphism Ward identity on $N$ becomes
\begin{equation}
\label{eq:single-surface-trans-eq}
    -\tau^\nu\nabla_\mu T^\mu{_\nu}\big\vert_N = -D_b(\widetilde{\mathcal T}^{bc}\tau_c) + \widetilde{\mathcal T}^{bc}G_{bc} + G\tau_b\widetilde T^b + \tau^\rho\tau^\nu v_\mu\nabla_\rho T^\mu{_\nu} = 0\,.
\end{equation}
This transverse equation contains information which is not captured by the intrinsic responses on a single null hypersurface. In particular, it describes the transverse evolution of components of the ambient energy-momentum tensor that are invisible from the intrinsic perspective, and hence it generically gives rise to a constraint for the ambient realisation of a Carrollian fluid. In special cases, however, this equation may reduce to one of the intrinsic Carrollian conservation equations, in which case it does not impose an independent constraint. We will see an example of the former for null fluids in Section~\ref{sec:null-fluids-on-surfs}. Another example arises in the thermal decoupling limit of Section~\ref{sec:Carroll-recovery}, where the transverse equation describes additional ambient data without imposing further constraints on the intrinsic Carrollian fluid.

\subsection{Reference splittings and the vector field $M^a$}
\label{sec:the-M-vector}

In many works on Carrollian geometry there appears an additional vector field $M^a$ that shifts under Carroll boosts. Motivated by the analogy with the mass gauge field in Newton--Cartan geometry~\cite{Andringa:2010it}, this field was first introduced in~\cite{Hartong:2015xda} to construct Carroll boost-invariant Carrollian inverse data, and it was subsequently identified there in adapted coordinates for a null hypersurface. Related $M^a$-type fields also emerge in covariant Carrollian expansions~\cite{Hansen:2021fxi}, and in covariant near-boundary expansions of asymptotically flat spacetimes~\cite{Hartong:2015usd,Hartong:2025jpp,Hartong:2026rbr}, as well as in Carrollian string theory~\cite{Bagchi:2024rje,Blair:2025nno}, though their precise role and transformation properties depend on the context.

In this subsection, we shed light on its role for null hypersurfaces. The key additional ingredient is a choice of reference vector $r^\mu$ along $N$ satisfying
\begin{equation}
\label{eq:ref-normalisation}
    v_\mu r^\mu = -1\,.
\end{equation}
This is a choice of transverse splitting, in the sense that the tangent space restricted to $N$ splits as
\begin{equation}
    T_{X(p)}M = X_*(T_pN)\oplus \braket{r}\,.
\end{equation}
To explicitly see this, define for any $Y^\mu\D_\mu \in T_{X(p)}M$ the combination
\begin{equation}
\label{eq:X-par}
    Y^\mu_\parallel := Y^\mu + r^\mu v_\nu Y^\nu\,,
\end{equation}
satisfying
\begin{equation}
    v_\mu Y^\mu_\parallel = 0\,,
\end{equation}
so that $Y^\mu_\parallel\D_\mu \in X_*(T_pN)$, and therefore, by rearranging~\eqref{eq:X-par}
\begin{equation}
    Y^\mu = Y^\mu_\parallel - r^\mu v_\nu Y^\nu\,.
\end{equation}
In particular, $r^\mu$ need not be null, though it can always be made to be so by defining
\begin{equation}
\label{eq:r-naught}
    r_0^\mu = r^\mu + \frac{1}{2} v^\mu r^2 \implies r_0^2 = 0\,.
\end{equation}
Note that $r_0^\mu v_\mu = -1$. In general, since also $v_\mu\tau^\mu = -1$, it follows that the difference between $\tau^\mu$ and $r^\mu$ is tangent to $N$
\begin{equation}
    v_\mu (\tau^\mu - r^\mu) = 0\,,
\end{equation}
implying that we can express this difference in the basis $u^\mu_a$ for $X_*(T_pN)$ as
\begin{equation}
\label{eq:cov-M}
    \tau^\mu - r^\mu = M^a u^\mu_a \implies \tau^\mu = r^\mu + M^a u^\mu_a\,,
\end{equation}
for some $M^a$. This will be the covariant manifestation of the $M^a$ field, and is nothing but the difference between the auxiliary null vector and the fixed choice of reference transverse splitting vector $r^\mu$. Note in particular that the intrinsic Carrollian geometry does not require this additional choice, and $M^a$ arises only when the auxiliary null vector is compared with a chosen reference splitting.

Since $\tau^\mu u_\mu^a = 0$, we have
\begin{equation}
\label{eq:r-a}
    r^a = u_\mu^a r^\mu = - M^a\,.
\end{equation}
The reference vector $r^\mu$ is an additional choice, which we take to be inert under null rotations. The normalisation condition~\eqref{eq:ref-normalisation} then dictates that it has the same weight as $\tau^\mu$ under null boosts, namely
\begin{equation}
    \delta r^\mu = -\alpha r^\mu\,,
\end{equation}
implying that $M^\mu$ transforms as
\begin{equation}
    \delta M^\mu = -\alpha M^\mu + \lambda^\mu\,,
\end{equation}
or, equivalently, that $M^a$ transforms under anisotropic Carroll--Weyl symmetry as
\begin{equation}
    \delta M^a = -\alpha M^a + \lambda^a\,,
\end{equation}
which is consistent with~\eqref{eq:r-a} due to~\eqref{eq:projector-trafo}. Lowering the index of $r^\mu$ and pulling back to $N$, we get
\begin{equation}
\label{eq:hat-tau}
    r_a = u_a^\mu r_\mu = \tau_a - h_{ab}M^b =: \hat\tau_a\,,
\end{equation}
where in the last equality we identified the pullback of the reference $1$-form with the Carroll boost-invariant clock form $\hat\tau_a$ introduced in~\cite{Hartong:2015xda}. Note that unlike the object introduced in that work, $r_a$ transforms under temporal Weyl rescalings as
\begin{equation}
\label{eq:Weyl-trafo-hat-tau}
    \delta_\alpha \hat\tau_a = -\alpha \hat\tau_a\,.
\end{equation}
Equipped with $M^a$, we can now also define an invariant inverse projector 
\begin{equation}
\label{eq:inv-proj}
    \bar u^a_\mu := u^a_\mu - M^a v_\mu\,,\qquad \delta \bar u^a_\mu = 0\,,
\end{equation}
which satisfies 
\begin{equation}
    v^\mu \bar u^a_\mu = v^a\,,\qquad r^\mu \bar u^a_\mu = 0\,,\qquad \tau^\mu \bar u^a_\mu = M^a\,,
\end{equation}
as well as
\begin{equation}
    u_a^\mu \bar u_\mu^b = \delta^a_b\,,\qquad u^\mu_a \bar u^a_\nu = \delta^\mu_\nu  + r^\mu v_\nu\,,
\end{equation}
where we used~\eqref{eq:cov-M}. In this sense, $\bar u^a_\mu$ is the inverse projection adapted to the boost invariant reference splitting vector $r^\mu$ rather than to the auxiliary null vector $\tau^\mu$. Using the invariant projector~\eqref{eq:inv-proj}, we may project the inverse metric to get
\begin{equation}
\label{eq:bar-h}
    \bar h^{ab} := \bar u^a_\mu \bar u^b_\nu g^{\mu\nu} = h^{ab} - 2M^{(a}v^{b)}\,,
\end{equation}
which is another Carroll boost-invariant object introduced in~\cite{Hartong:2015xda}. As in that paper, we can also use the $M$-field to build a boost-invariant version of the inverse spatial vielbein $e^a_A$ in~\eqref{eq:inv-vielbein-def}
\begin{equation}
    \hat e^a_A := e^a_A - v^a M^b e_{b A}\,,
\end{equation}
which in particular satisfies
\begin{equation}
    \hat e^a_A \hat\tau_a = 0\,,\qquad  \hat e^a_A e^B_a = \delta^B_A\,,\qquad \hat e^a_A =\bar h^{ab}e_{bA}\,,
\end{equation}
and which defines the boost-invariant object
\begin{equation}
\label{eq:hat-h}
    \hat h^{ab} := \delta^{AB}\hat e^a_A \hat e^b_B = h^{ab} + v^a v^b(h_{cd}M^c M^d - 2\tau_c M^c) - 2v^{(a}M^{b)}\,.
\end{equation}
Since $\tau^\mu$ is null, the expression for $\tau^\mu$ in terms of $r^\mu$ and $M^a$ in~\eqref{eq:cov-M} implies that
\begin{equation}
    0 = r^2 - h_{ab}M^a M^b + 2\tau_a M^a \implies r^2 = 2\bar\Phi\,,
\end{equation}
where we defined
\begin{equation}
    \bar\Phi := -\tau_a M^a + \frac{1}{2}h_{ab}M^a M^b\,.
\end{equation}
In terms of this scalar, the expression for $\hat h^{ab}$ in~\eqref{eq:hat-h} may be related to $\bar h^{ab}$ in~\eqref{eq:bar-h} via
\begin{equation}
    \hat h^{ab} = \bar h^{ab} + 2v^a v^b\bar\Phi\,.
\end{equation}
Of course, one does not have to choose an additional splitting at all, in which case the field $M^a$ is simply absent. In the paragraph below, we discuss the explicit emergence of $M^a$ in the hypersurface construction of~\cite{Hartong:2015xda}.

\paragraph{Null foliations.}
If the ambient spacetime admits a (local) foliation of null hypersurfaces, there exist adapted coordinates $x^\mu = (u,x^a)$ such that the leaves of the foliation are 
\begin{equation}
\label{eq:null-foliation-sec-2}
    N_u := \{u = \text{constant}\}\,,
\end{equation}
with null normal $1$-form proportional to $du$. With a general lapse $\mathcal N>0$, we may then write\footnote{In order to compare Eq.~(3.9) of~\cite{Hartong:2015xda} with~\eqref{eq:ambient-metric-decomp}, one must make the identification
\begin{equation*}
    U_A \leftrightarrow v_\mu\,,\qquad U^A \leftrightarrow -v^\mu\,,\qquad V^A \leftrightarrow \tau^\mu\,,\qquad V_A \leftrightarrow -\tau_\mu\,,
\end{equation*}
where $A$ is the ambient spacetime index in~\cite{Hartong:2015xda}. With the nullbein conventions of Eqs.~(3.9)--(3.10) there, $g_{AB}U^B=-U_A$ and $g_{AB}V^B=-V_A$, which accounts for the relative signs in this dictionary.
}
\begin{equation}
   v_\mu = \mathcal N\D_\mu u\,,
\end{equation}
where $\mathcal N$ has temporal Weyl weight $+1$. The choice $\mathcal N = 1$ made in~\cite{Hartong:2015xda} fixes the null boost rescaling symmetry of~\eqref{eq:trafos}. Choosing to work in adapted coordinates, which provide a local trivialisation of the foliation, is precisely what selects the reference transverse splitting vector 
\begin{equation}
\label{eq:foliation-r}
	r^\mu = -\mathcal N^{-1}(\D_u)^\mu\,,
\end{equation}
which satisfies
\begin{equation}
    v_\mu r^\mu = -1\,.
\end{equation}
The auxiliary null vector then takes the form
\begin{equation}
\label{eq:null-foliation-M}
	\tau^\mu = -\mathcal N^{-1}(\D_u)^\mu + M^a u^\mu_a\,,
\end{equation}
so that, using~\eqref{eq:foliation-r},
\begin{equation}
    M^\mu = \tau^\mu - r^\mu\,,
\end{equation}
precisely as in~\eqref{eq:cov-M}. The  components of the metric in these coordinates are
\begin{equation}
    \hat\tau_a = -\mathcal N^{-1}g_{ua}\,,\qquad \bar\Phi = \frac{g_{uu}}{2\mathcal N^2}\,,\qquad \bar h^{ab} = g^{ab}\,.
\end{equation}
These reduce to the expressions of~\cite{Hartong:2015xda} for $\mathcal N=1$. In Section~\ref{sec:open-Carroll-fluids}, where we consider a timelike fluid defined on a null foliation, the response to the $M^a$ field (up to a factor of the lapse) is precisely the tangentially projected null traction $\widetilde{\mathcal F}_a$ (cf.~\eqref{eq:currents-on-surface}), which will play the role of a source in the conservation equations. It should in particular not be confused with the Carroll boost field that appears as a certain subleading field of the ambient fluid velocity, from which the Carroll boost Goldstone $\theta^a$ of~\cite{Armas:2023dcz} emerges in the decoupling limit discussed Section~\ref{sec:Carroll-recovery}.

\section{Intrinsic fluids on null hypersurfaces}
\label{sec:Carroll-hydro}

In this section, we develop the theory of Carrollian fluids with a Weyl rescaling symmetry in the time direction, which, as discussed in Section~\ref{sec:ind-carrollian-structure}, is the relevant symmetry for null hypersurfaces. These are the candidate intrinsic theories to be compared with ambient fluids using the formalism developed in Section~\ref{sec:submanifold-calculus}. A general theory of Carrollian fluids was developed in~\cite{Armas:2023dcz}, which demonstrated that thermal Carrollian fluids spontaneously break Carroll boosts, leading to a Carroll boost Goldstone (similar to the dipole Goldstone for fracton fluids~\cite{Armas:2023ouk,Jain:2023nbf}), and this section builds on that work.

\subsection{Spontaneously broken Carroll boosts}
\label{sec:spont-break-boosts}

The background fields defining the $(d+1)$-dimensional geometry consist of the Carrollian structure $(v^a,h_{ab})$ along with the inverses $(\tau_a,h^{ab})$, which transform according to
\begin{equation}
\label{eq:trafo-1}
\begin{split}
    \delta v^a &= \pounds_\xi v^a + \alpha v^a\,,\qquad \delta \tau_a = \pounds_\xi \tau_a -\alpha \tau_a + \lambda_a\,,\\
    \delta h_{ab} &= \pounds_\xi h_{ab}\,,\qquad \delta h^{ab} =\pounds_\xi h^{ab} + 2\lambda^{(a}v^{b)}\,,
\end{split}
\end{equation}
where we, unlike in the above, included the action of infinitesimal diffeomorphisms with parameter $\xi^a$. In addition, due to the spontaneous breaking of Carroll boosts, the field content includes a dynamical Carroll boost Goldstone $\theta^a$, which transforms as
\begin{equation}
\label{eq:trafo-2}
    \delta \theta^a = \pounds_\xi\theta^a - \alpha\theta^a - h^{ab} \lambda_b + \chi v^a\,,
\end{equation}
where $\chi$ is the parameter of a Stueckelberg symmetry that ensures that the time component of the Carroll boost Goldstone is pure gauge. As explained in~\cite{Armas:2023dcz}, we can use the Carroll boost Goldstone to form the Carroll boost and Stueckelberg-invariant combinations
\begin{equation}
\label{eq:check-objets}
    \check \tau_a = \tau_a + h_{ab}\theta^b\,,\qquad \check h^{ab} = h^{ab} + v^{a}v^b(\theta^2 + 2\tau_c \theta^c) + 2v^{(a}\theta^{b)}\,,
\end{equation}
satisfying
\begin{equation}
    \check\tau_a \check h^{ab} = 0 \,,\qquad v^a\check \tau_a = -1\,,\qquad \delta^a_b = -\check\tau_b v^a + \check h^a_b\,,
\end{equation}
where we defined
\begin{equation}
\label{eq:check-h}
    \check h^a_b := \check h^{ac}h_{cb}\,,
\end{equation}
They transform only under anisotropic Weyl rescalings according to
\begin{equation}
    \delta_\alpha \check\tau_a = -\alpha \check\tau_a\,,\qquad \delta_\alpha \check h^{ab} = 0\,,
\end{equation}
and as such make up a Lifshitz geometry with $z = \infty$ (see, e.g.,~\cite{deBoer:2017ing,deBoer:2020xlc,Figueroa-OFarrill:2022kcd}), but with the usual caveat of~\cite{Armas:2023dcz} that these fields are dynamical since they contain the Carroll boost Goldstone. 

The Carroll-boost invariant objects introduced in~\eqref{eq:check-objets} bear a striking resemblance to the Carroll-boost invariant objects constructed using the vector $M^a$ in~\eqref{eq:hat-tau} and~\eqref{eq:hat-h}, but it is important to emphasise that they are physically distinct: whereas $M^a$ arises when a fixed reference splitting is introduced, as explained in Section~\ref{sec:the-M-vector}, the Carroll boost Goldstone is a dynamical field required in the description of Carrollian thermal physics. 

\paragraph{Currents and conservation laws.} Consider a functional $S[\tau_a,h_{ab};\theta^a]$ describing a Carrollian fluid with temporal Weyl symmetry with spontaneously broken Carroll boosts.\footnote{In general, such a functional takes the form of an effective action for the Goldstone, and was shown in~\cite{Armas:2023dcz} to be Ho\v rava--Lifshitz-like (see also~\cite{Hartong:2015zia}).} Its variation reads
\begin{equation}
\label{eq:fluid-functional-variation}
    \delta S = \int d^{d+1}\sigma\,e\left[  -T^a\delta \tau_a + \frac{1}{2}\mathcal{T}^{ab}\delta h_{ab} - K_a \delta \theta^a   \right]\,,
\end{equation}
where $e\,d^{d+1}\sigma$ is the volume form (cf.~\eqref{eq:induced-volume-form}), and where $T^a$ is the energy current, $\mathcal{T}^{ab}$ is the momentum-stress tensor (subject to the ambiguity~\eqref{eq:intrinsic-ambiguity}), and $K_a$ is the response to varying the Goldstone field, i.e., its equation of motion. The measure is given by $e = \det (\tau_a,e_a^A)$, or, equivalently, 
\begin{equation}
    e = \sqrt{\det(\tau_a\tau_b + h_{ab})}\,, 
\end{equation}
which has the same Weyl weight as $\tau_a$, namely
\begin{equation}
\label{eq:measure-Weyl-weight}
    \delta_\alpha e = -\alpha e\,.
\end{equation}
The (off-shell) Ward identities for Carroll boosts, the Stueckelberg symmetry and the temporal Weyl rescalings are
\begin{equation}
    K_a = T^b h_{ba}\,,\qquad v^a K_a = 0\,,\qquad T^a \tau_a + K_a\theta^a = 0\,.
\end{equation}
Combining the first and the last equations, we can write the Ward identity for temporal Weyl rescalings as
\begin{equation}
\label{eq:alternative-Weyl-WI}
    \check \tau_a T^a = 0\,.
\end{equation}
This implies that the ordinary Carroll boost Ward identity
\begin{equation}
    \label{eq:boost-WI}
    T^b h_{ba} = 0\,,
\end{equation}
is now dynamically imposed by the equation of motion for the Carroll boost Goldstone. On-shell, therefore, the temporal Weyl Ward identity combined with~\eqref{eq:boost-WI} implies that
\begin{equation}
    T^ a = 0\qquad \text{(on-shell)}\,.
\end{equation}
Finally, the on-shell diffeomorphism Ward identity takes the form
\begin{equation}
    e^{-1}\D_b(e T^b{_a}) + T^b\D_a\tau_b - \frac{1}{2}\mathcal{T}^{bc}\D_a h_{bc} = 0\,,
\end{equation}
where we defined the Carrollian energy-momentum tensor
\begin{equation}
    T^a{_b} = -\tau_b T^a + \mathcal{T}^{ac}h_{cb}\,.
\end{equation}

\paragraph{Equilibrium partition function.}

We now discuss the equilibrium (or hydrostatic) partition function for Carrollian fluids with a temporal Weyl symmetry. The method of hydrostatic partition functions has a long and illustrious history in the context of (non-)Lorentzian geometry~\cite{PartitionFunction2, Banerjee:2012iz, Armas:2013hsa, Jensen:2014ama, Banerjee:2015uta, deBoer:2017ing,deBoer:2017abi, Armas:2019gnb,deBoer:2020xlc, Armas:2020mpr}. An equilibrium Carrollian background with temporal Weyl symmetry is characterised by the infinitesimal symmetry parameters
\begin{equation}
\label{eq:equil-conf}
\mathbb K = 
\big(k^a,\lambda^{\mathbb K}_a,\chi_{\mathbb K},
\alpha_{\mathbb K}\big)\,,
\end{equation}
consisting of a diffeomorphism generated by the Killing vector $k^a$, together with a Carroll boost parameter $\lambda^{\mathbb K}_a$, a Stueckelberg parameter $\chi_{\mathbb K}$ and a temporal Weyl parameter $\alpha_{\mathbb K}$. The equilibrium conditions are the vanishing of the variations in~\eqref{eq:trafo-1} and~\eqref{eq:trafo-2} of all background fields and the Goldstone, i.e.,
\begin{equation}
\begin{split}
\delta_{\mathbb K} v^a &= \pounds_k v^a + \alpha_{\mathbb K} v^a = 0\,,\qquad \delta_{\mathbb K} \tau_a = \pounds_k \tau_a -\alpha_{\mathbb K} \tau_a +\lambda^{\mathbb K}_a = 0\,,\\
\delta_{\mathbb K}h_{ab} &= \pounds_kh_{ab} = 0\,,\qquad  \delta_{\mathbb K} \theta^a = \pounds_k \theta^a - \alpha_{\mathbb K}\theta^a - h^{ab}\lambda^{\mathbb K}_b + \chi_{\mathbb K} v^a=0\,.
\end{split}
\end{equation}
The boost and Stueckelberg-invariant combinations introduced in~\eqref{eq:check-objets} satisfy
\begin{equation}
\pounds_k\check\tau_a - \alpha_{\mathbb K}\check\tau_a = 0\,, \qquad \pounds_k h_{ab}=0\,, \qquad \pounds_k\check h^{ab}=0\,.
\end{equation}
The Killing vector $k^a$ is identified with the thermal vector and is in particular inert under temporal Weyl transformations, while the Weyl symmetry parameter itself transforms, i.e.,
\begin{equation}
\delta_\alpha k^a=0\,, \qquad \delta_\alpha\alpha_{\mathbb K} = -\pounds_k\alpha\,.
\end{equation}
Following~\cite{Armas:2023dcz}, we define the local temperature and Carrollian fluid velocity as
\begin{equation}
\label{eq:hydrostatic-data}
T := \frac{T_0}{\check\tau_a k^a}\,, \qquad u^a := \frac{k^a}{\check\tau_b k^b}\,, 
\end{equation}
where $T_0$ is a constant. These both have temporal Weyl weight $+1$, and we take $T_0 > 0$ and furthermore  restrict to configurations with $\check\tau_a k^a > 0$ so that the intrinsic temperature is positive. By construction, the Carrollian fluid velocity satisfies
\begin{equation}
    \check\tau_a u^a=1 \implies u^a = -v^a + \vec u^a\,,\qquad \vec u^a := \check h^a_b u^b\,,
\end{equation}
implying that the spatial fluid velocity decomposes as
\begin{equation}
\label{eq:spatial-fluid-vel}
    \vec u^a = v^a\theta^b\vec u_b + h^a_b u^b\,.
\end{equation}
In addition to the temperature defined in~\eqref{eq:hydrostatic-data}, the square of the spatial velocity gives rise to a second hydrostatic scalar
\begin{equation}
\vec u^2 := h_{ab}u^au^b\,.
\end{equation}
The temporal Weyl transformations of the hydrostatic data are
\begin{equation}
\delta_\alpha T=\alpha T\,, \qquad \delta_\alpha u^a=\alpha u^a\,, \qquad \delta_\alpha\vec u^2 = 2\alpha\vec u^2\,.
\end{equation}
Hence, the most general ideal-order hydrostatic partition function is
\begin{equation}
\label{eq:HPF}
S_{\text{HPF}} = \int d^{d+1}\sigma\,eP(T,\vec u^2)\,,
\end{equation}
where $P$ must have Weyl weight $+1$ since the measure has Weyl weight $-1$ (cf.~\eqref{eq:measure-Weyl-weight}), and since this Weyl weight arises through its dependence on $T$ and $\vec u^2$, the pressure must satisfy
\begin{equation}
\label{eq:homogeneity-pde}
\left( T\frac{\partial}{\partial T} + 2\vec u^2 \frac{\partial}{\partial\vec u^2} \right)P = P\,,
\end{equation}
which is solved by
\begin{equation}
P(T,\vec u^2) = Tp\left(\frac{\vec u^2}{T^2}\right)\,,
\end{equation}
for an arbitrary function $p$. The variation of~\eqref{eq:HPF} is now identical to the ideal Carrollian fluid construction of~\cite{Armas:2023dcz}: defining the thermodynamic coefficients by
\begin{equation}
\label{eq:first-law}
dP = s dT + \frac{1}{2}m d\vec u^2\,,\qquad s := \pd{P}{T}\,,\qquad m = 2\pd{P}{\vec u^2}\,,
\end{equation}
and using that the variation of the measure is given by 
\begin{equation}
e^{-1}\delta e = -v^a\delta\tau_a + \frac{1}{2}h^{ab}\delta h_{ab}\,,
\end{equation}
we obtain the currents\footnote{These variations are performed with the symmetry parameters of~\eqref{eq:equil-conf} held fixed.}
\begin{equation}
\label{eq:current-broken-boosts}
\begin{split}
T^a &= P v^a + \left(sT+m\vec u^2\right)u^a\,,\\
\mathcal T^{ab} &= Ph^{ab} + m u^au^b - 2\left(sT+m\vec u^2\right)
u^{(a}\theta^{b)}\,,\\
K_a &= \left(sT+m\vec u^2 \right) h_{ab}u^b\,.
\end{split}
\end{equation}
The Weyl homogeneity condition~\eqref{eq:homogeneity-pde} is equivalent to
\begin{equation}
\label{eq:P-expr}
P = sT + m\vec u^2\,,
\end{equation}
which is again equivalent to the temporal Weyl Ward identity~\eqref{eq:alternative-Weyl-WI}. In particular, the relation~\eqref{eq:P-expr} tells us that the Goldstone equation $K_a = 0$ requires a fluid with nonzero pressure, $P\neq0$, to have vanishing spatial fluid velocity\footnote{As shown in~\cite{Armas:2023dcz}, the Carrollian fluid that emerges from the $c\to 0$ limit of relativistic fluids also has vanishing spatial fluid velocity.}
\begin{equation}
\label{eq:vanishing-spatial-vel}
	\vec u^{\,a} = 0\,,
\end{equation}
in which case $k^a$ is aligned with $v^a$, i.e.,
\begin{equation}
\label{eq:aligned-hydro-conf}
	k^a = \nu v^a\,,
\end{equation}
for some nonzero $\nu$. The condition~\eqref{eq:vanishing-spatial-vel} has drastic consequences: it means that the entropy density $s = p(0) =: s_0$ is constant, and that
\begin{equation}
\label{eq:const-ent}
    P = s_0T\,,\qquad m = 2 T^{-1} p'(0)\,.
\end{equation}
Note in particular that term involving $m$ is $m v^a v^b$, which is of the form~\eqref{eq:intrinsic-ambiguity} and therefore we may discard it. Hence the on-shell currents collapse to
\begin{equation}
\label{eq:collapsed-currents}
        T^a = 0\,,\qquad \mathcal{T}^{ab} = s_0 T (h^{ab} + 2v^{(a}\theta^{b)})\,.
\end{equation}
We note in particular that these fluids do not admit a nontrivial zero-temperature limit since the pressure vanishes in that limit. This suggests the existence of a distinct zero-temperature phase where the temporal Weyl symmetry is broken, but where Carroll boosts remain unbroken. This we discuss below. 

Finally, we note that the intrinsic torsion must vanish for aligned hydrostatic configurations~\eqref{eq:aligned-hydro-conf}, since the condition
\begin{equation}
	\pounds_kh_{ab} = 0\,,	
\end{equation}
implies that
\begin{equation}
	\nu\pounds_vh_{ab} = -2\nu K_{ab} = 0\,,
\end{equation}
and hence $K_{ab} = 0$.

\subsection{Zero-temperature Carrollian fluids with unbroken Carroll boosts}
\label{sec:Carroll-zero-temperature}

When Carroll boosts are not spontaneously broken, there is no well-defined notion of Carrollian temperature. The absence of the associated boost Goldstone therefore naturally characterises a zero-temperature phase. As we shall see in Section~\ref{sec:null-fluids-on-surfs}, this is the natural intrinsic construction of the Carrollian fluids that are induced by null fluids flowing along a null hypersurface.

The equilibrium configuration is still given by~\eqref{eq:equil-conf}, but now the only Carroll boost and temporal Weyl invariant scalar constructed from $k^a$ at zeroth derivative order is
\begin{equation}
\label{eq:k-squared}
    \vec k^2 :=h_{ab}k^ak^b\,.
\end{equation}
The measure, however, has weight $-1$, and hence, in the absence of an additional scale, we cannot construct a hydrostatic partition function that is invariant under temporal Weyl transformations. In other words, a nontrivial zero-temperature phase requires the temporal Weyl symmetry to be either ``spontaneously'' or explicitly broken, in direct analogy with the null fluids constructed in~\cite{Armas:2025uyv,Armas:2026nwa}, which we also review in Section~\ref{sec:null-fluids-review} below.

\paragraph{``Spontaneous'' breaking.}

We first consider the case where the breaking of the temporal Weyl symmetry is mediated by a compensator field $\Omega$ that is either a background field or a dynamical field with an associated equation of motion. Only in the latter case is the compensator Goldstone-like, which is why we write ``spontaneous'' in quotes, though with this qualification now explained we will drop the quotes in what follows. If $\Omega$ instead remains a background field, it becomes a Stueckelberg field for the temporal Weyl symmetry. The compensator field transforms as
\begin{equation}
\label{eq:Omega-Weyl-trafo}
    \delta_\alpha\Omega=-\alpha\Omega\,.
\end{equation}
This now allows us to construct a Weyl-invariant zero-temperature hydrostatic partition function at ideal order
\begin{equation}
\label{eq:spont-zero-T-HPF}
    S^0_{\text{s}} = \int d^{d+1}\sigma\,e\Omega^{-1} P_{\text{s}}(\vec k^2) \,,
\end{equation}
where we used the subscript ``s'' for spontaneous, though $\Omega$ only admits a Goldstone-like interpretation when it is regarded as a dynamical field. The variation of~\eqref{eq:spont-zero-T-HPF} is
\begin{equation}
\label{eq:scale-inv}
    \delta S^0_{\text{s}} = \int d^{d+1}\sigma\,e \left[ -T^a\delta\tau_a + \frac{1}{2}\mathcal T^{ab}\delta h_{ab} + \mathcal{O}_\Omega\delta\Omega \right]\,,
\end{equation}
where we included the variation of the compensator field. The responses are given by
\begin{equation}
\label{eq:spont-zero-T-responses}
    \begin{split}
        T^a &= \Omega^{-1} P_{\text{s}} v^a\,,\qquad \mathcal{T}^{ab} = \Omega^{-1} P_{\text{s}} h^{ab} + \Omega^{-1} m_{\text{s}} k^a k^b\,,\qquad \mathcal{O}_\Omega = -\Omega^{-2} P_{\text{s}}\,,
    \end{split}
\end{equation}
where we defined 
\begin{equation}
\label{eq:ms-def}
     m_{\text{s}} := 2\pd{P_{\text{s}}}{\vec k^2}\,.
\end{equation}
The energy density is $\varepsilon_{\mathrm{s}}:=\tau_aT^a=-\Omega^{-1}P_{\mathrm{s}}$ so the ideal enthalpy vanishes for both interpretations of $\Omega$. Only a dynamical compensator also requires $P_{\mathrm{s}}=0$, as we discuss below. The Carroll boost and temporal Weyl Ward identities are
\begin{subequations}
\label{eq:spont-zero-T-Weyl-WIs}
\begin{align}
    T^b h_{ba} &= 0\,,\label{eq:vanilla-boost-WI}\\
    \tau_a T^a - \Omega\mathcal{O}_\Omega &= 0\,,\label{eq:vanilla-Weyl-WI}
\end{align}
\end{subequations}
which are manifestly satisfied by the ideal-order responses in~\eqref{eq:spont-zero-T-responses}. If $\Omega$ is taken to be dynamical, the response $\mathcal{O}_\Omega$ is the equation of motion for the Goldstone-like field $\Omega$, which must be satisfied by on-shell fluid configurations
\begin{equation}
\label{eq:zero-spont-P}
	\mathcal O_\Omega = 0 \implies P_{\text{s}} = 0\,,
\end{equation}
though $m_{\text{s}}$ defined in~\eqref{eq:ms-def} may remain nonzero. The mechanism behind the vanishing of $P_{\text{s}}$ when the temporal Weyl symmetry is spontaneously broken is analogous to the vanishing of the dilaton potential discussed in~\cite{Bellazzini:2013fga} (see also~\cite{Afxonidis:2025tns}). If instead $\Omega$ is a Stueckelberg field, $\mathcal{O}_\Omega$ is a response rather than an equation of motion. In this case, one may fix temporal Weyl transformations by setting $\Omega = 1$.

An important special case, which will be of particular interest later, is when $\vec k^2 = 0$, in which case $h_{ab}k^b = 0$, implying that $k^a$ is proportional to $v^a$ with proportionality $\nu$ (cf.~\eqref{eq:aligned-case}), and since $k^a$ is invariant under temporal Weyl transformations, $\nu$ must transform according to
\begin{equation}
\label{eq:nu-trafo}
    \delta_\alpha \nu = -\alpha \nu\,,
\end{equation}
which is the same transformation as the one for the compensator field $\Omega$ in~\eqref{eq:Omega-Weyl-trafo}, though we emphasise that they should not be conflated: $\nu$ is just the time component of $k^a$, $\nu = -\tau_a k^a$ and its temporal Weyl transformation is inherited from that of $\tau_a$. As we will see shortly, $\nu$ contains no physical information. In this special case both $P_s = P_s(0)$ and $m_s = m_s(0)$ become constants, while
\begin{equation}
\label{eq:aligned-case}
    T^a = \Omega^{-1} P_s(0) v^a\,,\qquad \mathcal{T}^{ab} = \Omega^{-1}P_{\text{s}}(0) h^{ab} + \Omega^{-1} m_s(0) \nu^2 v^a v^b\,,
\end{equation}
where, in particular, the term in the momentum-stress tensor proportional to $v^a v^b$ is redundant and may be removed: this removes all reference to $\nu$, which is the sense in which it is unphysical. This leads to the energy-momentum tensor
\begin{equation}
\label{eq:spont-EMT}
    T^a{_b} = \Omega^{-1} P_{\text{s}}(0) \delta^a_b\,.
\end{equation}
In particular, this also shows that if $k^a$ is aligned with the Carrollian vector field $v^a$, and $\Omega$ is interpreted as a dynamical field, the phase becomes trivial. When $\Omega$ is instead a Stueckelberg field, the responses need not vanish, but the intrinsic dynamics is nevertheless trivial at ideal order when $k^a$ is aligned with $v^a$. In this case, the diffeomorphism Ward identity is
\begin{equation}
\label{eq:zero-T-compensated-diffeo}
    e^{-1}\D_b(e T^b{_a}) + T^b\D_a\tau_b - \frac{1}{2}\mathcal{T}^{bc}\D_a h_{bc} - \mathcal{O}_\Omega\D_a\Omega = 0\,.
\end{equation}
Using the identity for the measure
\begin{equation}
\label{eq:measure-derivative}
    e^{-1}\D_a e = -v^b\D_a\tau_b + \frac{1}{2}h^{bc}\D_a h_{bc}\,,
\end{equation}
the left-hand side reduces to $\Omega^{-1}\D_a P_{\text{s}}(0)$, which vanishes identically since $P_{\text{s}}(0)$ is constant. This conclusion is only for the aligned phase and does not necessarily apply to the general functional~\eqref{eq:spont-zero-T-HPF}.

\paragraph{Explicit breaking.} Again in analogy with null fluids~\cite{Armas:2025uyv,Armas:2026nwa}, we may also explicitly break the temporal Weyl symmetry by introducing a physical local scale $\kappa$. We regard $\kappa$ as an independent hydrodynamic scalar rather than a background source or a Goldstone field, and it is held fixed when varying the hydrostatic functional to obtain the geometric responses. As we will show, its spacetime dependence is constrained by the fluid conservation equations. In that case, the most general ideal hydrostatic partition function is
\begin{equation}
\label{eq:explicit-zero-T-HPF}
    S^0_{\text{e}} =  \int d^{d+1}\sigma\,e P_{\text{e}}(\vec k^2,\kappa)\,,
\end{equation}
where now the subscript ``e'' stands for explicit. In particular, the scalar $\kappa$ should not be interpreted as a Carrollian temperature: it is a local scale arising from explicit breaking of the temporal Weyl symmetry, and equilibrium configurations must satisfy
\begin{equation}
\label{eq:kappa-rel}
    \pounds_k \kappa = 0\,.
\end{equation}
Obtaining the currents themselves follows from a calculation identical to what we did above, leading to
\begin{equation}
\label{eq:explicit-spont-zero-T-responses}
    \begin{split}
        T^a &=  P_{\text{e}} v^a\,,\qquad \mathcal{T}^{ab} =  P_{\text{e}} h^{ab} +  m_{\text{e}} k^a k^b\,,
    \end{split}
\end{equation}
where we defined 
\begin{equation}
    m_{\text{e}} := 2\pd{P_{\text{e}}}{\vec k^2}\,.
\end{equation}
The boost Ward identity~\eqref{eq:vanilla-boost-WI} is trivially satisfied at ideal order, since $T^a$ has no spatial legs. The intrinsic Carrollian energy density is\footnote{Note that the Carroll boost Ward identity implies that $T^a \propto v^a$ to all orders in a suitable derivative expansion, which means that the combination $\tau_a T^a$ is Carroll boost-invariant.}
\begin{equation}
    \varepsilon_{\text{e}} := \tau_a T^a = -P_{\text{e}}\,,
\end{equation}
and hence the enthalpy vanishes at ideal order in the zero-temperature phase 
\begin{equation}
\label{eq:vanishing-enthalpy}
    \varepsilon_{\text{e}} + P_{\text{e}} = 0\,,
\end{equation}
as in the compensated phase above. This vanishing of the enthalpy was first observed in the general theory of Carroll perfect fluids in~\cite{deBoer:2017ing} and subsequently discussed in the context of dark energy and inflation in~\cite{deBoer:2021jej}.

We may again consider the case where $k^a$ is aligned with the Carrollian vector field. When the temporal Weyl symmetry is explicitly broken, this implies that 
\begin{equation}
    k^a = \tilde\nu v^a\,,
\end{equation}
for some function $\tilde\nu$ that, unlike $\nu$ in~\eqref{eq:nu-trafo}, does not transform under temporal Weyl transformations. The currents now become
\begin{equation}
    T^a =  P_{\text{e}}(0,\kappa) v^a\,,\qquad \mathcal{T}^{ab} =  P_{\text{e}}(0,\kappa) h^{ab} +  \tilde\nu^2 m_{\text{e}}(0,\kappa) v^a v^b\,.
\end{equation}
As above, the term proportional to $v^a v^b$ is redundant, and hence the scale $\tilde\nu$ drops out entirely, leaving the energy-momentum tensor
\begin{equation}
    T^a{_b} = P_{\text{e}}(0,\kappa) \delta^a_b\,.
\end{equation}
The diffeomorphism Ward identity is
\begin{equation}
\label{eq:zero-T-explicit-diffeo}
    e^{-1}\D_b(e T^b{_a}) + T^b\D_a\tau_b - \frac{1}{2}\mathcal T^{bc}\D_a h_{bc} = 0\,,
\end{equation}
which, using~\eqref{eq:measure-derivative}, reduces to $\D_a P_{\text{e}}(0,\kappa) = 0$. This is a condition on the thermodynamics whenever the pressure depends nontrivially on $\kappa$.

\subsection{Breaking both Carroll boosts and temporal Weyl symmetry}
\label{sec:breaking-both-syms}

The final possibility is that, in addition to the spontaneously broken Carroll boosts which are required for the existence of a finite temperature, the temporal Weyl symmetry is also broken. Exactly as above, there are two symmetry-breaking patterns for the temporal Weyl symmetry: either it is explicitly broken or it is ``spontaneously'' broken, with the associated compensator field being either dynamical or a Stueckelberg field.

\paragraph{Explicitly broken temporal Weyl symmetry.} When the temporal Weyl symmetry is explicitly broken, we may parametrise this by a local scale $\kappa$, just as we did above. The construction of the ideal-order hydrostatic partition function then proceeds in the same way as in Section~\ref{sec:spont-break-boosts}, with the temperature $T$ and $\vec u^2$ given by~\eqref{eq:hydrostatic-data}; the only difference is that the pressure now also depends on the local scale $\kappa$, meaning that
\begin{equation}
    S_{\text{b,e}} = \int e\,P_{\text{b,e}}(T,\vec u^2,\kappa)\,,
\end{equation}
where $\kappa$ satisfies the relation~\eqref{eq:kappa-rel}. The Gibbs--Duhem relation is
\begin{equation}
    dP = s dT + \frac{1}{2}m d\vec u^2 + \pd{P_{\text{b,e}}}{\kappa} d\kappa\,.
\end{equation}
The resulting currents are again given by the expressions~\eqref{eq:current-broken-boosts}. In particular, explicitly breaking the temporal Weyl symmetry lifts the homogeneity constraint~\eqref{eq:P-expr} on the pressure, and the resulting theory is therefore a generalisation of the generic ideal Carrollian fluid developed in~\cite{Armas:2023dcz}, but now realised intrinsically on a null hypersurface. In particular, the Goldstone equation $K_a = 0$ requires either that
\begin{equation}
    \vec u^{a} = 0\,,\qquad\text{or}\qquad sT + m\vec u^2 = 0\,,
\end{equation}
which are precisely the two branches discussed in~\cite{Armas:2023dcz}. The branch with $\vec u^a = 0$ contains the Carrollian fluids obtained by taking the $c \to 0$ limit of relativistic hydrodynamics~\cite{Ciambelli:2018wre,Ciambelli:2018xat,Petkou:2022bmz} as a special case, but the dependence on the local scale $\kappa$ making the explicitly broken phase more general. 

\paragraph{``Spontaneously'' broken temporal Weyl symmetry.} As above, we may also break the temporal Weyl symmetry by introducing a compensator field $\Omega$ transforming as in~\eqref{eq:Omega-Weyl-trafo}. In this case, the ideal-order hydrostatic partition function takes the form
\begin{equation}
    S_{\text{b,s}} = \int d^{d+1}x\,e \mathscr{P}(T,\vec u^2,\Omega)\,,
\end{equation}
where we defined the Weyl-covariant pressure of weight $+1$
\begin{equation}
    \mathscr{P}(T,\vec u^2,\Omega) := \Omega^{-1} P_{\text{b,s}}(\Omega T,\Omega^2 \vec u^2)\,,
\end{equation}
by dressing the pressure appropriately with the compensator field. The variation of this functional is
\begin{equation}
    \delta S_{\text{b,s}} = \int d^{d+1}x\,e \left[ -T^a \delta \tau_a + \frac{1}{2}\mathcal{T}^{ab}\delta h_{ab} - K_a \delta \theta^a + \mathcal{O}_\Omega \delta\Omega  \right]\,,
\end{equation}
where the currents take the same form as in~\eqref{eq:current-broken-boosts}, but written in terms of the dressed pressure $\mathscr{P}$
\begin{equation}
\label{eq:both-broken-currents}
\begin{split}
T^a &= \mathscr{P} v^a + \left(\tilde s T+\tilde m \vec u^2\right)u^a\,,\\
\mathcal T^{ab} &= \mathscr{P}h^{ab} + \tilde m u^au^b - 2\left(\tilde sT+\tilde m \vec u^2\right)
u^{(a}\theta^{b)}\,,\\
K_a &= \left(\tilde sT+\tilde m\vec u^2 \right) h_{ab}u^b\,,\\
\mathcal{O}_\Omega &= \Omega^{-1}(\tilde s T + \tilde m \vec u^2 - \mathscr{P})\,,
\end{split}
\end{equation}
where now
\begin{equation}
    \tilde s := \pd{\mathscr{P}}{T}\,,\qquad \tilde m := 2\pd{\mathscr{P}}{\vec u^2}\,.
\end{equation}
The Ward identity for the temporal Weyl symmetry is
\begin{equation}
    \tau_a T^a + K_a \theta^a - \Omega \mathcal{O}_\Omega = 0\,,
\end{equation}
which is identically satisfied by the ideal-order expressions for the currents in~\eqref{eq:both-broken-currents}. When $\Omega$ is a Stueckelberg field, we may simply fix the temporal Weyl symmetry by setting $\Omega = 1$. Doing so, the temporal Weyl symmetry has disappeared from the description, and we exactly obtain the Carrollian fluid of~\cite{Armas:2023dcz}. If instead $\Omega$ is dynamical, i.e., Goldstone-like, then we require its response to vanish on-shell
\begin{equation}
\label{eq:compensator-eq}
    \mathcal{O}_\Omega = 0 \implies \mathscr{P} = \tilde s T + \tilde m\vec u^2\,,
\end{equation}
which is precisely the homogeneity condition~\eqref{eq:P-expr}, which now arises as the equation of motion for $\Omega$ rather than as the temporal Weyl Ward identity. This in turn implies that
\begin{equation}
    K_a = \mathscr{P} \vec u_a\,,
\end{equation}
so that a nonzero pressure requires 
\begin{equation}
    \vec u^a = 0\,,
\end{equation}
implying that, just like in~\eqref{eq:collapsed-currents}, the on-shell ideal currents collapse to
\begin{equation}
    T^a = 0\,,\qquad \mathcal{T}^{ab} = \mathscr{P}(h^{ab} + 2v^{(a}\theta^{b)} )\,.
\end{equation}
With nonzero pressure, the compensator equation~\eqref{eq:compensator-eq} can be used to solve for $\Omega$. More explicitly, we note that, with $\vec u^2=0$,~\eqref{eq:compensator-eq} implies that
\begin{equation}
	\frac{d}{dy}\left(\frac{P_{\text{b,s}}(y,0)}{y}\right)\Big\vert_{y = \Omega T} = 0\,,\qquad y := \Omega T\,.
\end{equation}
If this has a positive solution $y = y_0$, then
\begin{equation}
	\Omega = \frac{y_0}{T}\,,	
\end{equation}
and hence
\begin{equation}
	\mathscr P=s_0T\,,\qquad
	s_0:=\frac{P_{\text{b,s}}(y_0,0)}{y_0}\,.
\end{equation}
Thus the aligned on-shell constitutive relations coincide with those of the Weyl-invariant thermal phase discussed in Section~\ref{sec:spont-break-boosts}, although the two constructions differ off shell. The phases constructed in this section are summarised in Fig.~\ref{fig:phase-map}.

\section{Fluids moving at the speed of light on null hypersurfaces}
\label{sec:null-fluids-on-surfs}

In this section, we explore which ambient flows may be confined to null hypersurfaces and which of the intrinsic phases described in Section~\ref{sec:Carroll-hydro} they give rise to. We begin by a discussion of exactly when it makes sense to describe an ambient fluid in terms of an intrinsic phase. Unsurprisingly, such an ambient fluid must move at the speed of light, and is therefore an example of a null fluid which we review in Section~\ref{sec:null-fluids-review}. Finally, in Section~\ref{sec:twistless-null-fluids}, we show exactly how ambient null fluids give rise to the zero-temperature intrinsic Carrollian phase described in Section~\ref{sec:Carroll-zero-temperature}.

\subsection{Causal fluids confined to null hypersurfaces}
\label{sec:causal-confinement}

A nonzero vector $W^\mu$ is causal if 
\begin{equation}
\label{eq:causal-cond}
W^2 \leq 0\,.    
\end{equation}
If $W^\mu$ is also tangent to a null hypersurface $N$, it can be expressed in terms of the basis $u^\mu_a$  of $X_*(T_pN)$ as
\begin{equation}
    W^\mu = W^a u^\mu_a\,,
\end{equation}
so that, using~\eqref{eq:induced-metric}
\begin{equation}
    W^2 = h_{ab}W^a W^b \geq 0\,,
\end{equation}
where the last inequality follows from the fact that the Carrollian ruler $h_{ab}$ is positive semidefinite. Combined with the causal condition~\eqref{eq:causal-cond}, this implies that
\begin{equation}
    W^2 = 0 \implies W^\mu \propto v^\mu\,,
\end{equation}
where the last implication follows from the fact that $v^a$ spans the kernel of $h_{ab}$. Hence, only fluids that move at the speed of light may be confined to a null hypersurface. The systematics of such fluids were developed in~\cite{Armas:2025uyv}, which we briefly review below. Before we do so, we note that there are two levels to the statement that $W^\mu$ restricts to $N$: either $W^\mu$ is tangent to just $N$, which requires only
\begin{equation}
\label{eq:tangent-vector-to-N}
    W^\mu\big\vert_N \propto v^\mu \implies W^\flat \wedge dW^\flat\big\vert_N = 0\,,\qquad W^\flat := g(W,-)\,,
\end{equation}
where the second equation tells us that $W_\mu$ is hypersurface-orthogonal on $N$, or twistless (see also the discussion around~\eqref{eq:HSO-induced-conn}).\footnote{To explicitly see how the twistless condition follows from the first equation, note that there always exists a coordinate near $N$ such that $N = \{u = 0\}$ with a normal $1$-form proportional to $du$. Since $W$ is causal and tangent to $N$, the argument above implies that $W^\flat$ is proportional to $du$ on $N$. Hence, in a neighbourhood around $N$ we have $W^\flat = f du + uA$ for a nowhere-vanishing function $f$ and a $1$-form $A$. From this expression, $W^\flat \wedge dW^\flat\big\vert_{u=0} = 0$ follows immediately.} Alternatively, if the twistless condition holds in some neighbourhood around $N$
\begin{equation}
\label{eq:}
    W^\flat \wedge dW^\flat = 0\qquad \text{near $N$}\,,
\end{equation}
and $W^\mu$ is null throughout that neighbourhood, it defines a null foliation there. We will consider both of these cases in Section~\ref{sec:twistless-null-fluids} below.

Finally, if the causal vector is not tangent to $N$, it necessarily crosses the null hypersurface and its restriction does not define a fluid confined to $N$. Such flows generically give rise to open Carrollian systems that exchange energy and momentum with their surroundings. When $N$ is a leaf of a given null foliation, this exchange couples the systems on neighbouring leaves. The foliation is specified independently of the causal fluid velocity. We will discuss this in Section~\ref{sec:open-Carroll-fluids}.

\subsection{Ideal null fluids and zero-temperature hydrodynamics }
\label{sec:null-fluids-review}

Systematically developed in~\cite{Armas:2025uyv}, null fluids are characterised by a null fluid velocity. As per our discussion above, these are the fluids which may be confined to null hypersurfaces provided the null fluid velocity is twistless. Null matter, for which null fluids provide an effective description, has a long and illustrious history in General Relativity. Null dust sources the Vaidya metric~\cite{Vaidya:1951fdr} as well as the gravitational shockwaves obtained from the Aichelburg--Sexl ultraboost~\cite{Aichelburg:1970dh}, which led~\cite{Alves:2018wku} to conjecture a null fluid/gravity correspondence for pressureless null fluids. A fluid/gravity correspondence for general null fluids was formulated in~\cite{Armas:2026nwa}.

The kinematic data of a null fluid consists of a null congruence generated by a nowhere-vanishing null vector $\mathsf v^\mu$, and the geometric construction is nearly identical to our considerations in Sections~\ref{sec:ind-carrollian-structure}, except that now $\mathsf v^\mu$ is null throughout the spacetime. The null vector $\mathsf v^\mu$ transforms under null boosts as 
\begin{equation}
\label{eq:null-boost-fluid}
    \delta_\alpha \mathsf v^\mu = \alpha \mathsf v^\mu\,.
\end{equation}
As discussed in detail in~\cite{Armas:2025uyv}, null fluids do not have a local rest frame. They arise as an infinite boost limit of a causal fluid, whose enthalpy $w$ vanishes in the limit. For an ordinary neutral fluid, the enthalpy is $w = sT$, where $s$ is the entropy density and $T$ the temperature. If the entropy density is positive for all finite nonzero values of the temperature, the zero-enthalpy limit corresponds to $T \to 0$. Nevertheless, as pointed out in~\cite{Armas:2025uyv}, other limiting temperatures are possible depending on the details of the equation of state. 

The null boost transformation of $\mathsf v^\mu$ in~\eqref{eq:null-boost-fluid} implies that, in order to define an effective theory in terms of invariant quantities, one must introduce a Stueckelberg field $\varpi$ which transforms as
\begin{equation}
\label{eq:null-stueckelberg}
    \delta_\alpha \varpi = -\alpha \varpi\,,
\end{equation}
under null boosts with parameter $\alpha$. Using this Goldstone-like field, we define the null boost-invariant vector
\begin{equation}
    \ell^\mu := \varpi \mathsf v^\mu\,,
\end{equation}
which will play the role of the null fluid velocity. The ideal-order energy-momentum tensor describing a null fluid with spontaneously broken null boosts is
\begin{equation}
\label{eq:null-spont-EMT}
    T_{\text{s}}^{\mu\nu} = \mathcal{E} \ell^\mu \ell^\nu + P g^{\mu\nu}\,,
\end{equation}
where $\mathcal{E}$ and $P$ are constants. 

Alternatively, one may consider the case where null boosts are explicitly broken by choosing a preferred normalisation for $\mathsf v^\mu$, which amounts to introducing a physical local scale $\kappa$. In this case, the ideal-order energy-momentum tensor takes the form
\begin{equation}
    T_{\text{e}}^{\mu\nu} = \mathcal{E}(\kappa) \mathsf v^\mu \mathsf v^\nu + P(\kappa) g^{\mu\nu}\,,
\end{equation}
where now the null energy density $\mathcal{E}$ and the null pressure $P$ are parametrised in terms of the local scale $\kappa$, in contrast to the spontaneously broken case~\eqref{eq:null-spont-EMT}, where they are constants.

In the below, it will be convenient to introduce a unified notation for the physical null fluid velocity 
\begin{equation}
\label{eq:unified-fluid-velocity}
    q^\mu := \begin{cases}
        \ell^\mu = \varpi \mathsf v^\mu\,,&\text{spontaneously broken null boosts}\,,\\
        \mathsf v^\mu\,,& \text{explicitly broken null boosts}\,,
    \end{cases}
\end{equation}
as well as 
\begin{equation}
    (\widehat{\mathcal{E}},\widehat{P}) :=
        \begin{cases}
        (\mathcal{E},P)\,, & \text{spontaneously broken null boosts}\,,\\
        (\mathcal{E}(\kappa),P(\kappa))\,, & \text{explicitly broken null boosts}\,.
        \end{cases}
\end{equation}
so that the two ideal-order energy-momentum tensors may be written uniformly as
\begin{equation}
\label{eq:universal-null-EMT}
    T^{\mu\nu} = \widehat{\mathcal{E}} q^\mu q^\nu + \widehat{P}g^{\mu\nu}\,.
\end{equation}
The fluid equations follow from the covariant conservation of the energy-momentum tensor
\begin{equation}
\label{eq:null-fluid-conservation}
    \nabla_\mu T^{\mu\nu} = 0\,,
\end{equation}
which is dictated by diffeomorphism invariance of the theory.

\subsection{Confined null fluids and intrinsic Carrollian hydrodynamics}
\label{sec:twistless-null-fluids}

We now have all the ingredients required to determine the intrinsic Carrollian fluid induced by a null fluid confined to a null hypersurface. As demonstrated in~\eqref{eq:tangent-vector-to-N}, if the null fluid velocity $q^\mu$ is tangent to a particular null hypersurface $N$, it must satisfy
\begin{equation}
    q^\mu\big\vert_N = \Omega v^\mu\,,
\end{equation}
where $\Omega$ is nowhere vanishing on $N$. As discussed in Section~\ref{sec:causal-confinement}, tangency on $N$ implies twistlessness on that leaf. We first impose the stronger condition that $q^\mu$ is twistless in a neighbourhood of $N$
\begin{equation}
\label{eq:null-foliation}
    q^\flat \wedge d q^\flat = 0\,,
\end{equation}
in which case $q^\mu$ defines a null foliation in that neighbourhood. We will distinguish below which parts of the construction also apply when $q^\mu$ is tangent only to a single prescribed null hypersurface. By Frobenius' theorem there exist local functions $u$ and $\Lambda$ such that
\begin{equation}
    \label{eq:Frobenius}
    q_\mu = \Lambda\D_\mu u\,,
\end{equation}
where the level sets 
\begin{equation}
    N_u := \{ u = \text{constant}\} \,,
\end{equation}
define a foliation by null hypersurfaces. As usual, on each leaf $N_u$, the normal form is defined only up to normalisation, and we choose a representative given by
\begin{equation}
\label{eq:leaf-normal}
    v_\mu = e^\sigma \D_\mu u\,,
\end{equation}
for some $\sigma$. A change of normalisation is just the null boost~\eqref{eq:trafos}, under which $\delta_\alpha \sigma = \alpha$. Together,~\eqref{eq:Frobenius} and~\eqref{eq:leaf-normal} imply that
\begin{equation}
    q_\mu = \Omega v_\mu\,,\qquad \Omega := e^{-\sigma}\Lambda\,.
\end{equation}
Since the null fluid velocity $q_\mu$ is invariant under null boosts, $\Omega$ compensates the null boost of the normal $1$-form
\begin{equation}
\label{eq:Omega-trafo}
    \delta_\alpha \Omega = -\alpha \Omega\,.
\end{equation}
Using the submanifold calculus developed in Section~\ref{sec:submanifold-calculus}, we find that
\begin{equation}
    q_a = 0\,, \qquad q^a = \Omega v^a\,,
\end{equation}
implying that $q_\mu$ is normal to each leaf, while $q^\mu$ is tangent with $q^a$ spanning the kernel of $h_{ab}$. We will identify the induced theory with the aligned zero-temperature phase of Section~\ref{sec:Carroll-zero-temperature} by matching its responses and symmetry realisation. On its own, the the fact that $q^a$ is aligned with $v^a$ is not sufficient to determine whether Carroll boosts are broken, since the phase with spontaneously broken Carroll boosts discussed Section~\ref{sec:spont-break-boosts} also may have vanishing spatial fluid velocity on-shell.

\paragraph{Induced responses.}

We now turn our attention to the responses induced on each leaf by the ambient null fluid. These responses may be obtained from the variational framework developed in Section~\ref{sec:variational-calculus}. On a given null hypersurface, the projected responses are given in~\eqref{eq:currents-on-surface} and for the ideal null fluid~\eqref{eq:universal-null-EMT} they become
\begin{equation}
\label{eq:null-projected-responses}
    \widetilde T^a = \widehat{P} v^a\,,\qquad \widetilde{\mathcal{T}}^{ab} = \widehat{\mathcal{E}}\Omega^2 v^a v^b + \widehat{P}h^{ab}\,,\qquad \widetilde{\mathcal{F}}_\mu = \widehat{P}v_\mu\,,
\end{equation}
where we note that the null traction is purely in the normal direction so that
\begin{equation}
\widetilde{\mathcal{F}}_a = u^\mu_a \widetilde{\mathcal{F}}_\mu = 0\,,\qquad v^\mu \widetilde{\mathcal{F}}_\mu = 0\,,
\end{equation}
so that there is no local null tangential traction. Below we will verify from the conservation equations that the intrinsic system has no sources at ideal order.

To identify the corresponding intrinsic Carrollian responses on each leaf of the null foliation~\eqref{eq:null-foliation}, we use the decomposition of the ambient volume form on each leaf given in~\eqref{eq:restriction-of-volume-form}. We may then factorise the ambient volume form in terms of the (dual of the) null fluid velocity as
\begin{equation}
    \boldsymbol{\epsilon}_M = q^\flat\wedge (\Omega^{-1}\boldsymbol{\epsilon}_N)\,,
\end{equation}
where $\Omega^{-1}\boldsymbol{\epsilon}_N$ is the volume form on each leaf normalised with respect to the null fluid velocity, which is precisely the combination that is scale invariant, and where $q^\flat = \Lambda du$ (cf.~\eqref{eq:Frobenius}). In adapted coordinates this is equivalent to
\begin{equation}
    \sqrt{-g}\,d^{d+2}x = \frac{\Lambda}{\Omega}du\, e d^{d+1}x\,.
\end{equation}
Hence we find that
\begin{equation}
\label{eq:foliation-variational-complex}
\begin{split}
    \frac{1}{2}\boldsymbol{\epsilon}_MT^{\mu\nu}\delta g_{\mu\nu} &= q^\flat\wedge \boldsymbol{\epsilon}_N \Omega^{-1}\left[ -\widetilde T^a\delta \tau_a + \frac{1}{2}\widetilde{\mathcal{T}}^{ab}\delta h_{ab} + \widetilde{\mathcal{F}}_\mu \delta \tau^\mu  \right]\\
    &= q^\flat\wedge \boldsymbol{\epsilon}_N \left[ - T^a\delta \tau_a + \frac{1}{2}{\mathcal{T}}^{ab}\delta h_{ab} + \Omega^{-1}\widetilde{\mathcal{F}}_\mu \delta \tau^\mu  \right]\,,
\end{split}
\end{equation}
where we defined the intrinsic responses relative to the transverse factor $q^\flat$
\begin{equation}
\label{eq:intri-responses-1}
    T^a := \Omega^{-1}\widetilde{T}^a = \Omega^{-1}\widehat P v^a \,,\qquad \mathcal{T}^{ab} := \Omega^{-1}\widetilde{\mathcal{T}}^{ab} = \Omega^{-1} \widehat P h^{ab} + \Omega\widehat{\mathcal E}v^av^b\,,
\end{equation}
The factor $\Omega^{-1}$ is fixed by the ambient volume density with this transverse factor, while, as usual, the term proportional to $v^av^b$ belongs to the intrinsic ambiguity~\eqref{eq:intrinsic-ambiguity} and may be removed. Thus the induced Carrollian fluid is insensitive to $\widehat{\mathcal{E}}$ and, in particular, the ambient null energy density therefore cannot be reconstructed from the intrinsic Carrollian responses.

These currents lead to the Carrollian energy-momentum tensor
\begin{equation}
    T^a{_b} = \Omega^{-1}\widehat{P}\delta^a_b\,,
\end{equation}
from which we find that the intrinsic Carrollian energy density and pressure are given by
\begin{equation}
    \varepsilon_{\text{C}} := \tau_a T^a = -\Omega^{-1}\widehat P\,, \qquad \mathcal{P}_{\text{C}} := \Omega^{-1}\widehat P\,,
\end{equation}
which in particular satisfy 
\begin{equation}
\label{eq:zero-enthalpy}
    \varepsilon_{\text{C}} + \mathcal{P}_{\text{C}} = 0\,,
\end{equation}
consistent with our identification of the confined twistless null fluid with the zero-temperature Carrollian phase described in Section~\ref{sec:Carroll-zero-temperature}, which satisfies the same relation (cf.~\eqref{eq:vanishing-enthalpy}).

To make the variational interpretation precise, we consider variations at fixed embedding and fixed $q_\mu$. Now, since $q_\mu\tau^\mu = -\Omega$, we have
\begin{equation}
    v_\mu\delta\tau^\mu = -\Omega^{-1}\delta\Omega \implies    \Omega^{-1}\widetilde{\mathcal F}_\mu\delta\tau^\mu = -\Omega^{-2}\widehat P\delta\Omega\,,
\end{equation}
where we used~\eqref{eq:null-projected-responses}. Therefore the response term involving the null traction in~\eqref{eq:foliation-variational-complex} precisely gives
\begin{equation}
	\mathcal O_\Omega = -\Omega^{-2}\widehat P\,,
\end{equation}
in the intrinsic description involving the compensator field $\Omega$, in precise agreement with~\eqref{eq:spont-zero-T-responses}. Here the compensator $\Omega$ is varied as a background Stueckelberg field rather than as an independent hydrodynamic variable, which would impose $\widehat{P} = 0$ as discussed around~\eqref{eq:zero-spont-P}. Note also that this formally identifies the pairing between sources and responses, which holds also if we consider a single prescribed null hypersurface rather than a null foliation. Indeed,  on a single null hypersurface tangency of $q^\mu$ still defines $\Omega$ and the normalised measure $\Omega^{-1}\boldsymbol{\epsilon}_N$ along $N$. The local response matching above and the tangential conservation equation which we consider below therefore remain valid. Only the interpretation of the ambient fluid functional as an integral over neighbouring leaves requires a null foliation in a neighbourhood.

\paragraph{Induced intrinsic phases.}

The intrinsic Carrollian phase realised by confining the null fluid to a null hypersurface depends on whether the null fluid has spontaneously broken or explicitly broken null boosts. When the null boosts are spontaneously broken, the null fluid comes with the Stueckelberg field $\varpi$ introduced in~\eqref{eq:null-stueckelberg}, which has the same transformation law as $\Omega$ in~\eqref{eq:Omega-trafo}. If
\begin{equation}
	\mathsf v^\mu = f v^\mu\,,
\end{equation}
along $N$ for some function $f$, then it holds in general that
\begin{equation}
	\Omega = f\varpi\,.
\end{equation}
The direct identification $\Omega = \varpi$ applies only when the representatives are chosen such that $\mathsf v^\mu = v^\mu$. Notably, $\Omega$ is not dynamical, leading us to conclude that confining a null fluid with spontaneously broken null boosts to a null hypersurface leads to the aligned zero-temperature Carrollian fluid with a Stueckelberg field for temporal Weyl transformations, whose energy-momentum tensor is given in~\eqref{eq:spont-EMT} with the identification
\begin{equation}
\label{eq:spont-identification}
    P_{\text{s}}(0) = \widehat P\,.
\end{equation}
Here $\widehat P$ is constant, so this realisation has nonzero intrinsic responses but trivial intrinsic ideal dynamics, as discussed in Section~\ref{sec:Carroll-zero-temperature}.
If instead null boosts are explicitly broken in the ambient fluid, the null fluid velocity comes with a preferred normalisation. Since the temporal Weyl symmetry of the induced Carrollian geometry is inherited from null boosts, this preferred normalisation also explicitly breaks the temporal Weyl symmetry on the null hypersurface. We take this preferred normalisation to be
\begin{equation}
    q^\mu = v^\mu\,,
\end{equation}
corresponding to
\begin{equation}
    \Omega = 1\,.
\end{equation}
The physical local scale $\kappa$ of the ambient fluid is inherited in the Carrollian theory, and the induced pressure becomes
\begin{equation}
    \mathcal{P}_{\text{C}}(\kappa) = \widehat P(\kappa) = P_{\text{e}}(0,\kappa)\,.
\end{equation}
Hence, the explicitly broken ambient null fluid confined to a null hypersurface realises the zero-temperature Carrollian fluid with explicitly broken temporal Weyl symmetry discussed in Section~\ref{sec:Carroll-zero-temperature}. 

\paragraph{Conservation laws.}

In addition to matching the constitutive relations, we now match the ideal-order fluid equations by showing how the ambient conservation equation~\eqref{eq:null-fluid-conservation} reduces to Carrollian conservation laws upon restriction to $N$, which is either a leaf of a null foliation or a single prescribed null hypersurface. This is a specific example of the formalism developed in Sec.~\ref{sec:variational-calculus}.

As we saw in~\eqref{eq:geodesic}, the null normal is geodesic implying that the null fluid velocity obeys
\begin{equation}
    q^\mu \nabla_\mu q_\nu = c_q q_\nu\,,
\end{equation}
along $N$ for some inaffinity $c_q$. If $q^\mu$ is twistless, this relation holds for each leaf of the associated null foliation. Plugging the null fluid energy-momentum tensor at ideal order~\eqref{eq:universal-null-EMT} into the ambient conservation equation~\eqref{eq:null-fluid-conservation} gives
\begin{equation}
\label{eq:ambient-conv-law}
    0 = \nabla_\mu T^\mu{_\nu} = \D_\nu \widehat P + q_\nu(\nabla_\mu (q^\mu \widehat{\mathcal{E}}) + \widehat{\mathcal{E}}c_q )\,,
\end{equation}
which when pulled back to $N$ gives simply
\begin{equation}
\label{eq:confined-eom}
    \D_a \widehat{P} = 0\,.
\end{equation}
In particular, the ambient null energy density is absent from this equation, mirroring what happened for the restriction of the constitutive relations above. 

We now demonstrate that~\eqref{eq:confined-eom} matches the intrinsic Carrollian diffeomorphism Ward identities both when the temporal Weyl symmetry is compensated by a Stueckelberg field and when it is explicitly broken. In the former case, the Ward identity is~\eqref{eq:zero-T-compensated-diffeo} with (cf.~\eqref{eq:intri-responses-1})
\begin{equation}
    T^a{_b} = \Omega^{-1}\widehat P\delta^a_b\,,\qquad T^a = \Omega^{-1}\widehat{P} v^a\,,\qquad \mathcal{T}^{ab} = \Omega^{-1}\widehat{P} h^{ab}\,,\qquad \mathcal O_\Omega = -\Omega^{-2}\widehat P\,,
\end{equation}
where we used the identification~\eqref{eq:spont-identification} and removed the intrinsic ambiguity proportional to $v^av^b$ in the expression for $\mathcal{T}^{ab}$. Substituting these expressions and using~\eqref{eq:measure-derivative}, we find
\begin{equation*}
    e^{-1}\D_b(e T^b{_a}) + T^b\D_a\tau_b - \frac{1}{2}\mathcal T^{bc}\D_a h_{bc} - \mathcal O_\Omega\D_a\Omega
     = \D_a(\Omega^{-1}\widehat P) + \Omega^{-2}\widehat P\D_a\Omega = \Omega^{-1}\D_a\widehat P \equiv 0\,,
\end{equation*}
which identically satisfied since $\widehat P$ is constant in this phase.

In the explicitly broken phase, the preferred representative has $\Omega = 1$, and the diffeomorphism Ward identity~\eqref{eq:zero-T-explicit-diffeo} instead gives
\begin{equation}
    e^{-1}\D_b(e T^b{_a}) + T^b\D_a\tau_b - \frac{1}{2}\mathcal T^{bc}\D_a h_{bc} = \D_a\widehat P(\kappa) = 0\,.
\end{equation}
Here $\kappa$ is the physical scale inherited from the ambient fluid, so this is a genuine intrinsic condition when $\widehat P$ depends on $\kappa$. Whenever $\pd{\widehat P}{\kappa} \neq 0$, it requires $\D_a\kappa = 0$ on $N$. Both realisations of the intrinsic phases obtained from confining an ambient twistless null fluid therefore satisfy the intrinsic ideal equations without extra source terms in the diffeomorphism Ward identity, which is the sense in which the Carrollian fluid on each leaf is closed.

As discussed in Section~\ref{sec:variational-calculus}, the ambient fluid nevertheless obeys an additional equation that is not visible from the intrinsic perspective: contracting its conservation law~\eqref{eq:ambient-conv-law} with $\tau^\nu$ gives
\begin{equation}
    \nabla_\mu(q^\mu\widehat{\mathcal E}) + \widehat{\mathcal E}c_q = \Omega^{-1}\tau^\nu\D_\nu\widehat P\,.
\end{equation}
In the Stueckelberg phase the right-hand side vanishes, but the remaining equation still constrains the null congruence. In the explicitly broken phase it also involves the transverse derivative of the pressure. This ambient equation depends on $\widehat{\mathcal E}$, which is absent from the intrinsic responses and their conservation laws, and hence closure of the intrinsic ideal fluid does not imply that the full ambient fluid is captured by the theory induced on the leaves.

\section{Open Carrollian fluids on null foliations}
\label{sec:open-Carroll-fluids}

In the previous section, we considered causal fluids whose velocity is either surface forming or tangent to a prescribed null hypersurface. In this section we consider flows that cross a null foliation, and determine how the ambient conservation equations organise into sourced Carrollian conservation laws on the leaves. The extra terms present in the conservation equations due to transport between leaves bear a resemblance to source terms that appear in sourced Carrollian hydrodynamics as discussed in~\cite{Husnugil:2025edm}. We first derive the exact balance equation on the leaves for an arbitrary ambient energy-momentum tensor and then evaluate its responses for a timelike perfect fluid and for a null limit. We next take a decoupling limit in which the source term vanishes and the Carroll boost Goldstone appears in the finite intrinsic responses, while the ambient energy-momentum tensor diverges. This leads to a realisation of the Carrollian fluid discussed in Section~\ref{sec:breaking-both-syms}.

\subsection{Null foliations as open Carrollian systems}
\label{sec:foliation-balance}

\begin{figure}[t]
	\centering
		
	\begin{tikzpicture}[scale=1.2,
		x={(0.797cm,0.389cm)}, y={(-1.501cm,0.207cm)}, z={(0cm,1.642cm)},
		>={Stealth[length=2.4mm,width=1.8mm]},
		vec/.style={->,very thick,line cap=round},
		lab/.style={font=\small},
		leaf/.style={fill=blue!6, draw=blue!40!black!55, line join=round},
		gen/.style={blue!22},
		rig/.style={orange!85!black},
		flow/.style={very thick, orange!85!black, line cap=round},
		]
		\def\du{1.8}
		\def\sA{-1.0} \def\sB{1.6} \def\yA{-1.5} \def\yB{1.35}
		\def\NTleaf#1#2{%
			\begin{scope}[shift={({0.5*#1*\du},0,{-0.5*#1*\du})}]
				\filldraw[leaf, #2] (\sA,\yA,\sA) -- (\sB,\yA,\sB) -- (\sB,\yB,\sB) -- (\sA,\yB,\sA) -- cycle;
				\foreach \yy in {-1.1,-0.6,-0.1,0.4,0.9}
				\draw[gen] (\sA,\yy,\sA) -- (\sB,\yy,\sB);
		\end{scope}}
		
		\def\pS{-0.3} \def\pT{0.9} \def\pY{-0.8} \def\pZ{0.5}
		
		\def\Uh{1.5}
		\coordinate (Ustart) at ({0.10+0.5*\du},0.25,{0.10-0.5*\du});
		\coordinate (Uend) at (0,-0.05,\Uh);
		\pgfmathsetmacro{\lamcross}{\du/(\du+\Uh)}
		\coordinate (Ucross) at ($(Ustart)!\lamcross!(Uend)$);
		
		\NTleaf{2}{fill opacity=0.5, draw opacity=0.5}
		\node[lab, black!45, anchor=west] at ({\sB+\du},\yA,{\sB-\du}) {$N_{u+2\delta u}$};
		
		\NTleaf{1}{fill opacity=0.9}
		\node[lab, black!75, anchor=west] at ({\sB+0.5*\du},\yA,{\sB-0.5*\du}) {$N_{u+\delta u}$};
		
		\filldraw[fill=orange!16, draw=orange!70!black]
		({\pS+0.5*\du},\pY,{\pS-0.5*\du}) -- ({\pT+0.5*\du},\pY,{\pT-0.5*\du})
		-- ({\pT+0.5*\du},\pZ,{\pT-0.5*\du}) -- ({\pS+0.5*\du},\pZ,{\pS-0.5*\du}) -- cycle;
		\node[lab, anchor=west] at ({\pT+0.5*\du+0.05},{\pY+0.15},{\pT-0.5*\du-0.25}) {$\widetilde{\mathcal F}_a\big|_{u+\delta u}$};
		
		\draw[->, thick, black!75, densely dashed] (\pS,\pY,\pS) -- ({\pS+0.5*\du},{\pY-0.7},{\pS-0.5*\du});
		\draw[->, thick, black!75] ({\pS+0.5*\du},{\pY-0.7},{\pS-0.5*\du}) -- ({\pS+0.5*\du},{\pY},{\pS-0.5*\du});
		\node[lab, black!85, anchor=north west] at ({\pS+0.5*\du+0.02},{\pY-0.72},{\pS-0.5*\du-0.02}) {$\D_u$};
		\node[lab, black!85, anchor=north, align=center] at ({\pS+0.5*\du},{\pY-0.32},{\pS-0.5*\du-0.12}) {$\zeta^a\partial_a$};
		
		\foreach \s/\y in {\pT/\pY, \pT/\pZ, \pS/\pZ}
		\draw[rig, densely dashed] (\s,\y,\s) -- ++({0.5*\du},0,{-0.5*\du});
		
		\fill[orange!85!black] (Ustart) circle (1.3pt);
		\draw[flow] (Ustart) -- (Ucross);
		
		\NTleaf{0}{fill opacity=0.55}
		\node[lab, anchor=west] at (\sB,\yA,\sB) {$N_u$};
		\draw[vec,blue!65!black] (-0.85,1.05,-0.85) -- (0.35,1.05,0.35);
		\node[lab,blue!65!black]  at (2.1,1.9) {$v^\mu$};
		
		\filldraw[fill=orange!24, fill opacity=0.75, draw=orange!70!black]
		(\pS,\pY,\pS) -- (\pT,\pY,\pT) -- (\pT,\pZ,\pT) -- (\pS,\pZ,\pS) -- cycle;
		\node[lab] at (0.35,-0.45,0.35) {$\widetilde{\mathcal F}_a\big|_u$};
		\draw[vec, rig,opacity=0.7] (\pS,\pY,\pS) -- ++({0.5*\du},0,{-0.5*\du}) node[lab, rig, pos=0.5, left=1pt] {$R$};
		\fill (\pS,\pY,\pS) circle (1.3pt);
		
		\draw[flow, ->] (Ucross) -- (Uend)
		node[lab, above=1pt] {crossing flow};
		\fill[orange!60!black] (Ucross) circle (1.5pt);
	\end{tikzpicture}
	
	\caption{A null foliation as a stack of coupled Carrollian systems. The blue hypersurfaces $N_u$ are generated by the tangent null vector $v^\mu$. The auxiliary null vector $\tau^\mu$ defines the transverse evolution vector $R=-\mathcal N\tau=\D_u+\zeta^a\D_a$ (cf.~\eqref{eq:foliation-lapse}), whose integral curves provide an identification between points on neighbouring leaves. The orange patches represent corresponding regions on $N_u$ and $N_{u+\delta u}$ which are related by transport along $R$, on which the tangential null traction $\widetilde{\mathcal F}a$ induced by the ambient fluid is evaluated. The black arrows illustrate the decomposition of $R$ into $\D_u$ and a tangential shift $\zeta^a\D_a$. The orange arrow represents an ambient causal fluid velocity that crosses the leaves. The transport of this traction between leaves sources the intrinsic diffeomorphism Ward identity, $\mathcal W_a = -e^{-1}\mathfrak D_u (e\widetilde{\mathcal F})_a$ (cf.~\eqref{eq:sourced-foliation-diffeo-WI}), where $\mathcal{W}_a = 0$ is the unsourced diffeomorphism Ward identity and $\mathfrak D_u = \D_u+\pounds_\zeta$.}
\label{fig:null-foliation}
\end{figure}
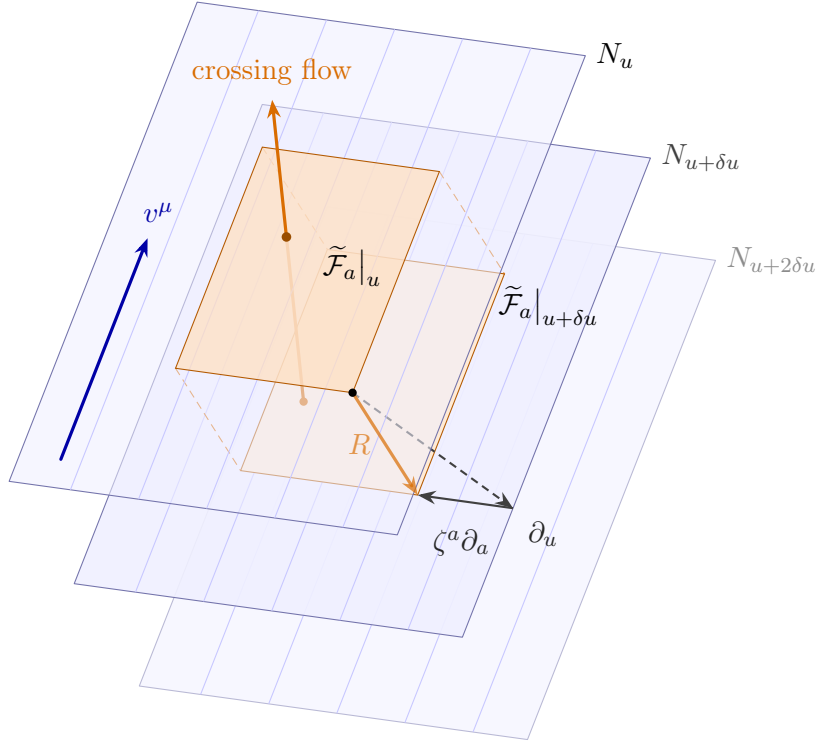

Consider an open region $\mathcal{U}\subset M$ that admits a null foliation $\mathcal{U} = \bigcup_u N_u$, where $N_u$ are the null leaves of the foliation as depicted in Figure~\ref{fig:null-foliation}. This is the same setup we considered in Section~\ref{sec:the-M-vector} (cf.~\eqref{eq:null-foliation-sec-2}). Since all statements are local, we suppress the restriction to $\mathcal{U}$ in what follows. As in Section~\ref{sec:the-M-vector}, it is useful to use adapted coordinates $x^\mu = (u,x^a)$, in terms of which we may write the dual of the normal vector $v^\flat$ and the auxiliary null vector $\tau$ as
\begin{equation}
\label{eq:foliation-lapse}
	v^\flat = \mathcal N du\,,\qquad  \tau^\mu\D_\mu = -\mathcal N^{-1}R\,,\qquad  R := \D_u + \zeta^a\D_a \,,
\end{equation}
where $\mathcal{N}$ is the lapse (cf.~\eqref{eq:foliation-r}), which satisfies $\mathcal N>0$ and which transforms under null boosts as
\begin{equation}
\label{eq:trafo-cal-N}
	\delta_\alpha \mathcal N = \alpha\mathcal N\,.
\end{equation}
The vector $R$ captures transport between the leaves and the shift $\zeta^a$ specifies its tangential part in adapted coordinates. Following $R$ from one leaf to the next changes the leaf coordinates by $dx^a=\zeta^a du$, and so the shift measures the tangential change relative to the integral curves of $\D_u$ as depicted in Figure~\ref{fig:null-foliation}. In particular, $\zeta^a$ is closely related to the $M^a$-field discussed in Section~\ref{sec:the-M-vector}; indeed, from~\eqref{eq:null-foliation-M} we immediately find that
\begin{equation}
\label{eq:zeta-M-rel}
	\zeta^a = -\mathcal{N} M^a\,.
\end{equation}
Just like $M^a$, $\zeta^a$ transforms under Carroll boosts:
\begin{equation}
\label{eq:zeta-boost-trafo}
	\delta_\lambda \zeta^a = - \mathcal{N}\lambda^a\,,
\end{equation}
which follows from~\eqref{eq:trafos}, but unlike $M^a$ it is inert under temporal Weyl transformations. It is useful to introduce the $1$-forms
\begin{equation*}
	\vartheta^a := dx^a - \zeta^a du\,,
\end{equation*}
which annihilate $R$
\begin{equation}
	\vartheta^a(R) = 0\,,
\end{equation}
 and transform under Carroll boosts as
\begin{equation}
\label{eq:vartheta-trafo}
	\delta_\lambda \vartheta^a = \lambda^a v^\flat\,.
\end{equation}
In adapted coordinates, the tangent maps are simply
\begin{equation}
\label{eq:adapted-tangent-maps}
	u^\mu_a\D_\mu = \D_a\,.
\end{equation}
The inverse projections $u^a_\mu$ defined~\eqref{eq:explicit-inverse} are uniquely determined by the conditions
\begin{equation}
	u^a_\mu u^\mu_b = \delta^a_b\,,\qquad \tau^\mu u_\mu^a = 0\,,
\end{equation}
which allows us to identify
\begin{equation}
	u^a_\mu dx^\mu = \vartheta^a\,,
\end{equation}
which, using~\eqref{eq:vartheta-trafo}, reproduces the transformation property of $u^a_\mu$ in~\eqref{eq:projector-trafo}. In terms of these objects, the metric~\eqref{eq:ambient-metric-decomp} takes the form
\begin{equation}
\label{eq:foliation-metric}
    g = h_{ab}\vartheta^a\vartheta^b - 2\mathcal N du \tau_a\vartheta^a\,, 
\end{equation}
with the measure given by
\begin{equation}
\label{eq:drastic-measure}
	\sqrt{-g}\, d^{d+2}x = \mathcal N e \,du\, d^{d+1}x \,,
\end{equation}
where the transformation property~\eqref{eq:trafo-cal-N} of $\mathcal N$ under null boosts ensures that the product $\mathcal N e$ remains invariant.

\paragraph{Induced responses on leaves.}

We next work out the responses induced on the leaves of the foliation following the same procedure that lead to~\eqref{eq:foliation-variational-complex}. The starting point is again the decomposition in~\eqref{eq:variation-decomp} of $T^{\mu\nu}\delta g_{\mu\nu}$ on the surface, which leads to the universal projections in~\eqref{eq:currents-on-surface}. Using the measure in adapted coordinates~\eqref{eq:drastic-measure} as well as the identity
\begin{equation}
	\delta\tau^\mu = -\mathcal N^{-1}u^\mu_a\delta\zeta^a - \mathcal N^{-1}\tau^\mu\delta\mathcal N\,,
\end{equation}
which follows from varying $\tau^\mu$ in~\eqref{eq:foliation-lapse} at fixed embedding and foliation, we get
\begin{equation}
\label{eq:foliation-variation}
\begin{split}
	\frac{1}{2}\sqrt{-g}\,d^{d+2}x\,T^{\mu\nu}\delta g_{\mu\nu} &= e\,du\,d^{d+1}x\left[-T^a\delta\tau_a + \frac{1}{2}\mathcal T^{ab}\delta h_{ab}
	+ \mathcal O_{\mathcal N}\delta\mathcal N - \widetilde{\mathcal F}_a\delta\zeta^a\right]\\
	&= e\,du\,d^{d+1}x\Big[-T^a\delta\tau_a + \frac12\mathcal T^{ab}\delta h_{ab}\\
	&\hspace{3.4cm} + (\mathcal O_{\mathcal N} + M^a\widetilde{\mathcal F}_a)\delta\mathcal N
	+\mathcal N\widetilde{\mathcal F}_a\delta M^a\Big]\,,
\end{split}
\end{equation}
where we used~\eqref{eq:zeta-M-rel} to rewrite the variation in terms of $M^a$, and where we defined the responses
\begin{equation}
\label{eq:foliation-responses}
    T^a := \mathcal N\widetilde T^a\,,\qquad   \mathcal T^{ab} := \mathcal N\widetilde{\mathcal T}^{ab}\,,\qquad  \widetilde{\mathcal F}_a := u^\mu_a \widetilde{\mathcal F}_\mu\,, \qquad   \mathcal O_{\mathcal N} := -\tau^\mu \widetilde{\mathcal F}_\mu\,,
\end{equation}
written in terms of the tilded responses in~\eqref{eq:currents-on-surface}. These should not be confused with the similarly named responses in~\eqref{eq:foliation-variational-complex}, which were defined relative to the null fluid velocity. Here, the existence of a foliation is independent of the fluid itself, and the responses are defined relative to this foliation. The tangential projection of the null traction is conjugate to $\zeta^a$, while $\mathcal N$ has the response $\mathcal O_{\mathcal N}$. The response identities~\eqref{eq:embedded-response-identity} and~\eqref{eq:embedded-Ward-identities} immediately give
\begin{equation}
\label{eq:foliations-WIs}
\begin{split}
    T^a = \mathcal N\left(h^{ab}\widetilde{\mathcal F}_b+v^a\mathcal O_{\mathcal N}\right)\,,\quad T^b h_{ba} = \mathcal N h^b_a\widetilde{\mathcal F}_b\,, \quad \tau_aT^a+\mathcal N\mathcal O_{\mathcal N}=0\,.
\end{split}
\end{equation}
Thus the temporal projection of the traction is already fixed by the induced energy response. These are precisely the Carroll boost and temporal Weyl Ward identities induced by the embedding as discussed in Section~\ref{sec:variational-calculus} with the transverse measure given by~\eqref{eq:drastic-measure}.

\paragraph{Conservation laws.}

We now turn our attention to the diffeomorphism Ward identity. Recall that the tangential projection to $N$ of the ambient diffeomorphism Ward identity~\eqref{eq:ambient-cons} gives rise to the sourced intrinsic-looking diffeomorphism Ward identity~\eqref{eq:single-surf-sourced-WI}, where the source involves transverse ambient data. The null foliation gives a particularly nice  realisation of the source term in terms of transport between leaves.  As mentioned in Section~\ref{sec:variational-calculus}, the precise Carrollian avatars of their tilded progenitors depend on the details of how the theory is realised on the hypersurface(s); here, using~\eqref{eq:foliation-variation}, the energy-momentum tensor and the expression for the diffeomorphism Ward identity become
\begin{equation}
\label{eq:foliation-WI-guy}
\begin{split}
    T^a{_b} &= -\tau_bT^a + \mathcal T^{ac}h_{cb}=\mathcal N\widetilde T^a{_b}\,,\\
    \mathcal W_a &:= e^{-1}\D_b(eT^b{_a}) + T^b\D_a\tau_b
    - \frac{1}{2}\mathcal T^{bc}\D_a h_{bc}
    - \mathcal O_{\mathcal N}\D_a\mathcal N\,,
\end{split}
\end{equation}
where $\widetilde T^a{_b}$ is defined in~\eqref{eq:embedded-EMT}, and where $ \mathcal W_a$ is related to its tilded parent~\eqref{eq:intrinsic-looking-diffeo-WI} via
\begin{equation}
\label{eq:foliation-intrinsic-Ward}
    \mathcal W_a=\mathcal N\widetilde{\mathcal W}_a
       +\widetilde T^b{_a}\D_b\mathcal N
       -\mathcal O_{\mathcal N}\D_a\mathcal N\,.
\end{equation}
Now, to derive the diffeomorphism Ward identity on the leaves, we may formally vary~\eqref{eq:foliation-variation} under $u$-dependent tangential diffeomorphisms that preserve the leaves
\begin{equation}
	\label{eq:tangential-diffeo}
	\xi = \xi^a(u,x^b)\D_a\,.
\end{equation}
Let $X$ denote a tensor intrinsic to the leaves, so that
\begin{equation}
	\delta_\xi X = \pounds^{N_u}_\xi X\,,
\end{equation}
where, as in~\eqref{eq:has-intrinsic-Lie-derivative}, $\pounds^{N_u}$ is the intrinsic Lie derivative on $N_u$. The sources $(\tau_a,h_{ab},\mathcal{N})$ in~\eqref{eq:foliation-variation} are all of this kind, while the shift $\zeta^a$ is not; to derive its transformation, we first note that the transverse vector $R = \D_u + \zeta^a\D_a$ defined in~\eqref{eq:foliation-lapse} transforms under this $\xi$ as
\begin{equation}
	\delta_\xi  R = \pounds_\xi R = [\xi,R] = -\D_u \xi + [\xi,\zeta]\,,
\end{equation}
where the first Lie bracket is on ambient vector fields, while the second is on vector fields on the leaves. Then, since $\delta _\xi R = (\delta_\xi \zeta^a)\D_a$ (the coordinate vector $\D_u$ is inert), we find that
\begin{equation}
\label{eq:cov-D-u}
	\delta_\xi \zeta = -\D_u \xi - [\zeta,\xi] =: -\mathfrak D_u \xi\,,
\end{equation}
where we defined the covariant transverse derivative
\begin{equation}
	\mathfrak D_u := \D_u + \pounds_\zeta\,,
\end{equation}
which satisfies
\begin{equation}
	\delta_\xi \big( \mathfrak D_u X \big) = \pounds^{N_u}_\xi\big( \mathfrak D_u X \big)\,,
\end{equation}
where we used that $\pounds^{N_u}_{[\xi,\zeta]} = \pounds^{N_u}_\xi \pounds^{N_u}_\zeta - \pounds^{N_u}_\zeta \pounds^{N_u}_\xi$. Hence, $\zeta^a$ plays the role of a connection for $u$-dependent diffeomorphisms of the leaves. Hence, varying the right hand side~\eqref{eq:foliation-variation} under~\eqref{eq:tangential-diffeo} and using~\eqref{eq:cov-D-u} gives, after integrating by parts,
\begin{equation}
	-e\mathcal W_a-\mathfrak D_u(e\widetilde{\mathcal F}_a)\,.
\end{equation}
On the other hand, transforming the left-hand side of~\eqref{eq:foliation-variation} under~\eqref{eq:tangential-diffeo} with $\xi^\mu = u^\mu_a \xi^a$ gives
\begin{equation}
	-\mathcal N e\,u^\nu_a\nabla_\mu T^\mu{_\nu}\,,
\end{equation}
and thus we get
\begin{equation}
	\mathcal N u^\nu_a\nabla_\mu T^\mu{_\nu} = \mathcal W_a +e^{-1}\mathfrak D_u(e\widetilde{\mathcal F}_a)\,,
\end{equation}
and ambient conservation then implies the Ward identity under leaf-preserving diffeomorphisms~\eqref{eq:tangential-diffeo}
\begin{equation}
\label{eq:sourced-foliation-diffeo-WI}
	\mathcal W_a = -e^{-1}\mathfrak D_u(e\widetilde{\mathcal F}_a)\,.
\end{equation}
Thus, on a null foliation, the source term in the tangentially projected diffeomorphism Ward identity is provided by the transverse transport of the traction between neighbouring leaves.

From the intrinsic perspective, each leaf therefore behaves as an open Carrollian system, and the failure of the intrinsic Carrollian energy-momentum to be conserved is due to exchange between adjacent leaves. Note in particular that having vanishing traction on a single leaf is not enough for the intrinsic diffeomorphism Ward identity to close, since its transverse derivative does not necessarily vanish. In Section~\ref{sec:decouling-limit}, we will explore a decoupling limit in which the source term decouples throughout the foliation.

Finally, we work out the explicit form of the transverse projection of the ambient conservation~\eqref{eq:single-surface-trans-eq}, and since
\begin{equation}
	R^\mu = -\mathcal N\tau^\mu\,,
\end{equation}
the $\tau^\mu$-projection is the same as the projection along $R^\mu$ up to a factor. Hence ambient conservation implies
\begin{equation}
\label{eq:transverse-explicit}
	R^\nu\nabla_\mu T^\mu{_\nu} = \nabla_\mu \big( T^\mu{_\nu}R^\nu \big) -T^\mu{_\nu}\nabla_\mu R^\nu = \frac{1}{\mathcal N e}\D_\mu \big(\mathcal N e T^\mu{_\nu}R^\nu\big) - \frac{1}{2}T^{\mu\nu}\pounds_R g_{\mu\nu} = 0\,,
\end{equation}
where in the second line we used $\sqrt{-g}=\mathcal N e$ and symmetry of $T^{\mu\nu}$. Using~\eqref{eq:foliation-responses}, we can write the transverse projection of the ambient energy-momentum tensor as
\begin{equation}
\label{eq:transverse-timelike-cond}
	T^\mu{_\nu}R^\nu = \mathcal O_{\mathcal N}R^\mu -u^\mu_a\mathcal T^{ab}\tau_b\,.
\end{equation}
where we used that $\tau_\mu = u^a_\mu\tau_a$. We note that, unlike the tangential energy-momentum tensor $T^a{_b}$, this expression is sensitive to the $v^av^b$ component of the projected ambient energy-momentum tensor. In Section~\ref{sec:Carroll-recovery}, we will see that there exists a thermal Carrollian limit where this transverse equation constrains only the ambient fluid data without imposing any additional constraints on the intrinsic fluid.

\subsection{Crossing fluids and the Carroll boost field}
\label{sec:timelike-Carroll}

Recall from Section~\ref{sec:causal-confinement} that a causal ambient fluid velocity $W^\mu$ crosses the leaves of the foliation whenever $v_\mu W^\mu\neq0$, cf.~Figure~\ref{fig:null-foliation}. In particular, every timelike velocity crosses, as does every null velocity that is not aligned with $v^\mu$. Crucially, a crossing fluid velocity determines a preferred transverse null direction given by
\begin{equation}
\label{eq:r-W}
    r_W^\mu := -\frac{W^\mu}{v_\nu W^\nu} + \frac{W^2}{2(v_\nu W^\nu)^2}v^\mu\,,
\end{equation}
which satisfies the conditions
\begin{equation}
	r_W^2=0\,,\qquad\text{and}\qquad v_\mu r_W^\mu=-1\,.
\end{equation}
We first consider the case where $W^\mu = U^\mu$ is a unit timelike velocity, i.e.,
\begin{equation}
\label{eq:timelike-normalisation}
	U^2 = -1\,,
\end{equation}
and we furthermore choose the orientation of $U^\mu$ such that
\begin{equation}
\label{eq:B-def}
	B:=-v_\mu U^\mu>0\,.
\end{equation}
We may decompose $U^\mu$ in terms of ambient frame fields as
\begin{equation}
    U^\mu = A v^\mu + B\tau^\mu + s^\mu\,,\qquad v_\mu s^\mu = \tau_\mu s^\mu = 0\,,
\end{equation}
for a vector $s^\mu$ and functions $A,B$, though these are not independent since the normalisation~\eqref{eq:timelike-normalisation} implies that
\begin{equation}
	A = \frac{1 + s^2}{2B}\,.
\end{equation}
We then suggestively define
\begin{equation}
\label{eq:theta-mu-def}
	\tilde\theta^\mu := \frac{s^\mu}{B}\,,
\end{equation}
which we call the Carroll boost field. In terms of $\tilde\theta^\mu$, the timelike fluid velocity $U^\mu$ can be written as
\begin{equation}
\label{eq:disting-vector}
    U^\mu = \frac{1}{2B}v^\mu + B r_{\tilde\theta}^\mu\,,\qquad r_{\tilde\theta}^\mu := \tau^\mu + \tilde\theta^\mu + \frac{1}{2}\tilde\theta^2v^\mu\,,
\end{equation}
where $r_W^\mu$ with $W = U$ in~\eqref{eq:r-W}  can be written in terms of $r_{\tilde\theta}^\mu$ as
\begin{equation}
    r_U^\mu := \frac{U^\mu}{B} - \frac{v^\mu}{2B^2} = r_{\tilde\theta}^\mu\,.
\end{equation}
In this way, $\tilde\theta^\mu$ measures the difference between the auxiliary null vector $\tau^\mu$ and the preferred null vector~\eqref{eq:r-W} provided by the ambient fluid velocity. Its tangential projection is
\begin{equation}
\label{eq:theta-a-sec5}
	\tilde\theta^a := u^a_\mu\tilde\theta^\mu\,,	
\end{equation}
and satisfies $\tau_a\tilde\theta^a = 0$. This construction of $\tilde\theta^a$ is very similar to what we did with the $M^a$-field discussed in Section~\ref{sec:the-M-vector}, but its origin is different. Whereas $M^a$ arises by choosing an independent reference splitting (cf.~\eqref{eq:r-a}), $\tilde\theta^a$ is fixed by the physical ambient fluid velocity $U^\mu$. Then,  since, $\delta U^\mu=0$ implies $\delta B=\alpha B$ and $\delta r_U^\mu=-\alpha r_U^\mu$, the transformation of $\tilde\theta^\mu$ defined in~\eqref{eq:theta-mu-def} is
\begin{equation*}
    \delta\tilde\theta^\mu=-\alpha\tilde\theta^\mu-\lambda^\mu  + (\lambda_\nu\tilde\theta^\nu)v^\mu\,.
\end{equation*}
This in turn implies that the transformation of $\tilde\theta^a$ is 
\begin{equation}
\label{eq:trafo-boost-field}
	\delta\tilde\theta^a = -\alpha\tilde\theta^a-\lambda^a + (\lambda_b\tilde\theta^b)v^a\,,
\end{equation}
where we used that  $\delta u^a_\mu\tilde\theta^\mu=\lambda^a v_\mu\tilde\theta^\mu = 0$. This is very similar to the transformation of the Carroll boost Goldstone $\theta^a$ in~\eqref{eq:trafo-2}, but unlike the Carroll boost Goldstone, the Carroll boost field $\tilde\theta^a$ considered here satisfies the condition $\tau_a\tilde\theta^a = 0$. Therefore, to compare with the Goldstone law~\eqref{eq:trafo-2}, we must impose the condition $\tau_a\theta^a = 0$ on the Goldstone. Preserving this condition for the Goldstone $\theta^a$ (which does not have a tilde) requires
\begin{equation}
	\delta(\tau_a\theta^a)=\lambda_a\theta^a-\chi = 0 \implies 	\chi = \lambda_b\theta^b\,.
\end{equation}
In this sense, the transformation of the Carroll boost field $\tilde\theta^a$ in~\eqref{eq:trafo-boost-field} is the same as the tranformation of the gauge-fixed Carroll boost Goldstone with residual Stueckelberg transformations given by $\chi = \lambda_a \tilde\theta^a$. 

However, unlike the Carroll boost Goldstone $\theta^a$, the Carroll boost field $\tilde\theta^a$ is not an independent fluid degree of freedom since it is fixed by the ambient fluid velocity $U^\mu$ via~\eqref{eq:disting-vector}. In particular, there is no Stueckelberg symmetry acting on $\tilde\theta^a$, and the condition $\tau_a \tilde\theta^a = 0$ follows from its definition rather than from fixing a Stueckelberg symmetry. Nevertheless, we will see in the decoupling limit of Section~\ref{sec:decouling-limit} that the intrinsic responses retain $\tilde\theta^a$ while the ambient fluid velocity becomes singular. In that case $\tilde\theta^a$ does become an independent field which may be identified with the Carroll boost Goldstone of an autonomous Carrollian fluid, as we discuss in Section~\ref{sec:Carroll-recovery}.

\paragraph{Projected ideal energy-momentum tensor.}

We now work out the explicit induced Carrollian responses for an ambient timelike ideal fluid. The energy-momentum tensor for an ideal timelike fluid is
\begin{equation}
\label{eq:timelike-ideal-EMT}
    T^{\mu\nu} = w_{\text{T}}U^\mu U^\nu + p_{\text{T}} g^{\mu\nu}\,,
\end{equation}
where we defined the enthalpy
\begin{equation}
	w_{\text{T}} := \varepsilon_{\text{T}} + p_{\text{T}}\,,
\end{equation}
where the subscript ``T'' stands for timelike.  It is furthermore useful to introduce the quantities
\begin{equation}
\label{eq:crossing-fluid-data}
    Q := w_{\text{T}}B^2\,,\qquad \rho := \frac{w_{\text{T}}}{2} - p_{\text{T}}\,,\qquad \tilde\tau_a := \tau_a + \tilde\theta_a\,,\qquad \chi^a := \tilde\theta^a + \frac{1}{2}\tilde\theta^2v^a\,,
\end{equation}
where $\tilde\theta_a := h_{ab}\tilde\theta^b$. The boost-invariant $1$-form $\tilde\tau_a$ is analogous to $\check\tau_a$ defined in~\eqref{eq:check-objets} except it is written in terms of the Carroll boost vector rather than the Carroll boost Goldstone. It follows from~\eqref{eq:disting-vector} that 
\begin{equation}
	u^a_\mu U^\mu = \frac{v^a}{2B}+B\chi^a\,,
\end{equation}
which implies that the projections~\eqref{eq:currents-on-surface} take the form
\begin{equation}
\label{eq:timelike-projected-responses}
\begin{split}
    \widetilde T^a &= -\rho v^a - Q\chi^a\,,\\
    \widetilde{\mathcal T}^{ab}  &= p_{\text{T}} h^{ab} + \frac{w_{\text{T}}}{4B^2}v^av^b + w_{\text{T}} v^{(a}\chi^{b)} + Q \chi^a\chi^b\,,\\
    \widetilde{\mathcal F}_\mu &= -w_{\text{T}}B U_\mu + p_{\text{T}} v_\mu  \implies \widetilde{\mathcal F}_a = -Q\tilde\tau_a\,,
\end{split}
\end{equation}
where we used that $u^\nu_a \tilde\theta_\nu = h_{ab}\tilde\theta^b$, which follows from $h_{\mu\nu} = u^a_\mu u^b_\nu h_{ab}$, which also implies that $\tilde\theta^2 = h_{ab}\tilde\theta^a\tilde\theta^b$. We note that the momentum-stress has a term proportional to $v^a v^b$, which is precisely the ambiguity~\eqref{eq:intrinsic-ambiguity}, so $\widetilde{\mathcal T}^{ab}$ in~\eqref{eq:timelike-projected-responses} is equivalent to
\begin{equation}
\label{eq:timelike-mom-stress}
    \widetilde{\mathcal T}^{ab} =   p_{\text{T}} h^{ab}+w_{\text{T}}v^{(a}\tilde\theta^{b)} + Q\left(\tilde\theta^a\tilde\theta^b + \tilde\theta^2v^{(a}\tilde\theta^{b)}\right)\,.
\end{equation}
The Carrollian energy-momentum tensor $T^a{_b}$ in~\eqref{eq:foliation-WI-guy} and $\mathcal{O}_{\mathcal N}$ from~\eqref{eq:foliation-responses} are then given by
\begin{equation}
\label{eq:timelike-Car-EMT}
    T^a{_b} = \mathcal{N} p_{\text{T}}\delta^a_b + \mathcal N\left(\frac{w_{\text{T}}}{2}v^a + Q\chi^a\right)\tilde\tau_b\,, \qquad \mathcal O_{\mathcal N} = -\rho - \frac{1}{2}Q\tilde\theta^2\,.
\end{equation}
In terms of these objects, the tangential diffeomorphism Ward identity~\eqref{eq:sourced-foliation-diffeo-WI} becomes
\begin{equation}
\label{eq:sourced-explicit-WI-timelike}
    \mathcal W_a=e^{-1}\mathfrak D_u(eQ\tilde\tau_a)\,,
\end{equation}
while  the Carroll boost Ward identity~\eqref{eq:foliations-WIs} reads
\begin{equation}
\label{eq:anomaly-boost}
    T^b h_{ba}=\mathcal N h^b_a\widetilde{\mathcal F}_b=-\mathcal N Q\tilde\theta_a\,.
\end{equation}
This very explicitly shows how a generic perfect timelike fluid fails to give rise to a closed Carrollian fluid on a leaf, with $Q = w_{\text{T}}B^2$ defined in~\eqref{eq:crossing-fluid-data} controlling this failure.

\paragraph{Crossing null fluids.}

We next consider null fluids as discussed in Section~\ref{sec:null-fluids-on-surfs}, except they now are allowed to cross the leaves of the foliation. The calculations are nearly identical to what we did above, so we will be brief. A crossing null fluid satisfies 
\begin{equation}
	B_q := -v_\mu q^\mu \neq 0\,,
\end{equation}
where $q^\mu$ is the ambient null fluid velocity (cf.~\eqref{eq:unified-fluid-velocity}). The calculation that led to~\eqref{eq:disting-vector} gives, mutatis mutandis,
\begin{equation}
	q^\mu = B_qr_{\tilde\theta_q}^\mu\,,\qquad r_{\tilde\theta_q}^\mu := \tau^\mu + \tilde\theta_q^\mu + \frac{1}{2}\tilde\theta_q^2v^\mu\,,
\end{equation}
with no extra term proportional to $v^\mu$ since $q^2 = 0$. We then define the null counterparts of $Q$, $\tilde\tau_a$ and $\chi^a$ in~\eqref{eq:crossing-fluid-data} as
\begin{equation}
	Q_q := \widehat{\mathcal E}B_q^2\,,\qquad \tilde\tau^q_a = \tau_a + h_{ab}\tilde\theta^b_q\,,\qquad \chi^a_q := \tilde\theta_q^a + \frac{1}{2}\tilde\theta^2_qv^a\,.
\end{equation}
Recall that the ideal null energy-momentum tensor~\eqref{eq:universal-null-EMT} reads
\begin{equation}
	T^{\mu\nu} = \widehat{\mathcal{E}}q^\mu q^\nu + \widehat{P}g^{\mu\nu}\,,
\end{equation}
so that the projections to the leaves~\eqref{eq:currents-on-surface} take the form
\begin{align}
\label{eq:crossing-null-responses}
\begin{aligned}
    \widetilde T^a &= \widehat P v^a - Q_q\chi_q^a\,,&
    \widetilde{\mathcal T}^{ab} &= \widehat P h^{ab} + Q_q\chi_q^a\chi_q^b\,,\\
    \widetilde{\mathcal F}_a &= -Q_q\tilde\tau^q_a\,,&
    \mathcal O_{\mathcal N} &= \widehat P-\frac{1}{2}Q_q\theta_q^2\,.
\end{aligned}
\end{align}
As above, $\widetilde{\mathcal T}^{ab}$ contains a term proportional to $v^a v^b$, and so it is equivalent to
\begin{equation}
	\widetilde{\mathcal T}^{ab} = \widehat P h^{ab} + Q_q( \tilde\theta^a_q \tilde\theta^b_q +v^{(a} \tilde\theta^{b)}_q\tilde\theta^2_q )\,. 
\end{equation}
The tangential diffeomorphism Ward identity~\eqref{eq:sourced-explicit-WI-timelike} now reads 
\begin{equation}
	\mathcal{W}_a = e^{-1}\mathfrak D_u(e Q_q \tilde\tau^q_a)\,.
\end{equation}
Thus, just as for the timelike fluid, a generic crossing null fluid induces open Carrollian fluids on the leaves. We note in particular that the energy density
\begin{equation}
	\tau_aT^a = \mathcal N\left(-\widehat P + \frac{1}{2}Q_q\tilde\theta_q^2\right)\,,
\end{equation}
no longer gives rise to vanishing enthalpy, unlike for the confined null fluid~\eqref{eq:zero-enthalpy}.

In the limit $Q_q \to 0$ all terms responsible for crossing vanish and produces the confined null fluid of~\eqref{eq:null-projected-responses}, i.e.,
\begin{equation}
	\widetilde T^a = \widehat P v^a\,,\qquad \widetilde{\mathcal T}^{ab} = \widehat P h^{ab}\,,\qquad \widetilde{\mathcal F}_a = 0\,,
\end{equation}
where we removed the intrinsic ambiguity proportional to $v^av^b$ in $\widetilde{\mathcal T}^{ab}$.

\subsection{Decoupling limit}
\label{sec:decouling-limit}

\begin{figure}[t]
	\centering
	\begin{tikzpicture}[scale=1.2,
		>={Stealth[length=2.1mm,width=1.6mm]},
		lab/.style={font=\small,align=center},
		path1/.style={very thick,orange!85!black},
		midarrow/.style={postaction={decorate},decoration={markings,
				mark=at position 0.55 with {\arrow{Stealth}}}}]
		\begin{scope}
			\clip (0,0) rectangle (5.5,3.6);
			\foreach \c in {0.06,0.24,0.6,1.5,4.0}
			\draw[black!18] plot[domain=0:5.5,samples=70] (\x,{\c*\x*\x});
		\end{scope}
		\draw[->,thick] (0,0) -- (5.7,0) node[right,lab] {$B$};
		\draw[->,thick] (0,0) -- (0,3.8) node[above,lab] {$w_{\text{T}}$};
		\draw[path1,midarrow] (4.6,2.8) -- (0,2.8);
		\draw[path1,densely dashed,midarrow] plot[domain=4.6:0,samples=70] (\x,{2.8*\x*\x/(4.6*4.6)});
		\node[lab,fill=white,inner sep=2pt] at (2.7,3.12) {$w_{\text{T}}\to w_{\text{T},0}\neq0$};
		\node[lab,fill=white,inner sep=2pt] at (3.7,0.75) {$w_{\text{T}}/B^2$ fixed};
		\fill[orange!85!black] (4.6,2.8) circle (1.8pt);
		\node[lab,anchor=west] at (4.72,2.75) {crossing parent fluid};
		\fill[blue!65!black] (0,2.8) circle (2pt);
		\fill[orange!85!black] (0,0) circle (2pt);
		\draw[->,black!60] (4.6,-0.45) -- (1.0,-0.45)
		node[midway,below,lab] {large boost: $B\to0$};
		\node[lab,text=black!60] at (3.4,4.0) {Grey contours: constant $w_{\text{T}}/(4B^2)$};
		\node[lab,anchor=east,text width=4.5cm] at (-0.45,2.8)
		{\textcolor{blue!65!black}{Thermal Carrollian fluid}\\
			finite intrinsic response\\singular ambient EMT\\
			\hyperref[sec:decouling-limit]{\S\ref*{sec:decouling-limit}}--\hyperref[sec:Carroll-recovery]{\ref*{sec:Carroll-recovery}}};
		\node[lab,anchor=east,text width=4.5cm] at (-0.45,0.1)
		{\textcolor{orange!85!black}{Null fluid}\\finite ambient EMT\\
			\hyperref[sec:null-fluids-on-surfs]{\S\ref*{sec:null-fluids-on-surfs}}};
	\end{tikzpicture}
	\caption{The two large boost limits of a crossing timelike fluid, illustrated here for $w_{\text{T}}>0$. The two paths correspond to different scalings of the enthalpy as $B \to 0$ in which the direction of the ambient fluid velocity approaches the null direction $v^\mu$. The grey curves are contours of $w_{\text{T}}/(4B^2)$, which is the coefficient of $v^\mu v^\nu$ in the ambient energy-momentum tensor~\eqref{eq:singular-ambient-EMT}. Along the dashed line, $w_{\text{T}} = \mathcal O(B^2)$, so this coefficient remains finite and the limit is a ambient null fluid. Along the solid line, $w_{\text{T}}\to w_{\text{T},0}\neq0$, so the ambient energy-momentum tensor becomes singular, while the intrinsic Carrollian responses remain finite (up to the intrinsic ambiguity~\eqref{eq:intrinsic-ambiguity}) and retain the Carroll boost field $\tilde\theta^a$, which, as we show below, becomes a bona fide Carroll boost Goldstone in the limit.}
	\label{fig:Bw-plot}
\end{figure}
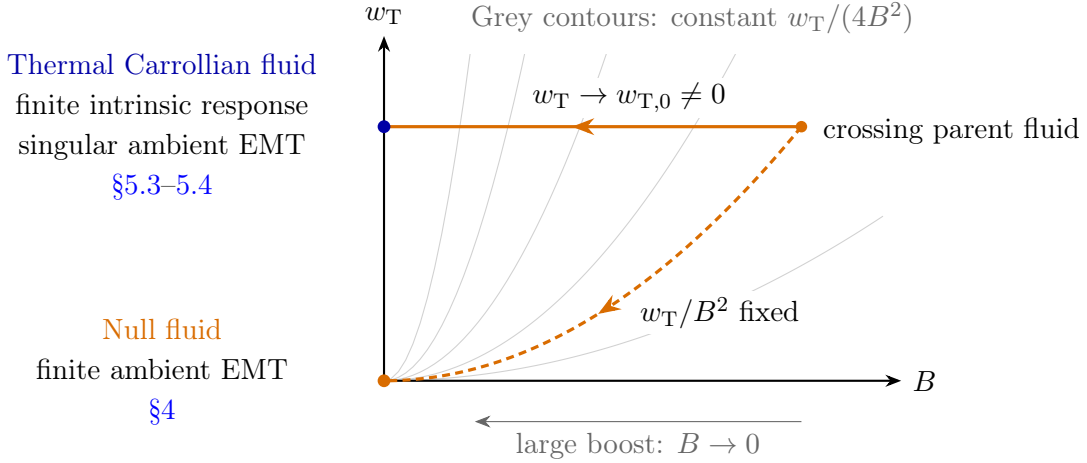

We now introduce a decoupling limit of the open Carrollian system induced by the timelike ambient fluid, where the source term in the tangential Ward identity~\eqref{eq:sourced-explicit-WI-timelike} vanishes, while the Carroll boost field remains part of the description. The decoupling limit is defined by (cf.~Figure~\ref{fig:Bw-plot})
\begin{equation}
\label{eq:decoupling-limit}
    B = \epsilon b\,, \qquad \epsilon \to 0\,,
\end{equation}
where $b>0$, $p_{\text{T}}$, $w_{\text{T}}$ and $\tilde\theta^a$ and their derivatives are kept finite. Then
\begin{equation}
	Q = \epsilon^2w_{\text{T}}b^2\,,
\end{equation}
 and hence also $\widetilde{\mathcal F}_a = \mathcal O(\epsilon^2)$ throughout the neighbourhood where the foliation is defined, implying that the transverse source in~\eqref{eq:sourced-foliation-diffeo-WI} vanishes in the limit. The limits of $T^a$, $\mathcal T^{ab}$ and $\mathcal{O}_{\mathcal{N}}$  in~\eqref{eq:timelike-projected-responses} and~\eqref{eq:timelike-Car-EMT} are
\begin{equation}
\label{eq:limiting-thermal-responses}
    T^a = -\varepsilon_{\text{C}}v^a\,, \qquad  \mathcal T^{ab} = p_{\text{C}}h^{ab} + 2(\varepsilon_{\text{C}} + p_{\text{C}})v^{(a}\tilde\theta^{b)}\,, \qquad \mathcal O_{\mathcal N} = -\varepsilon_{\text{C}}/\mathcal N\,,
\end{equation}
where we defined
\begin{equation}
\label{eq:limiting-thermodynamics}
    p_{\text{C}}:=\mathcal N p_{\text{T},0}\,,\qquad
    \varepsilon_{\text{C}} := \mathcal N\left(\frac{w_{\text{T},0}}{2} - p_{\text{T},0}\right)\,,\qquad     \varepsilon_{\text{C}} + p_{\text{C}} = \frac{\mathcal N w_{\text{T},0}}{2}\,,
\end{equation}
where the subscript ``$0$'' indicates the limiting value, e.g., 
\begin{equation}
	p_{\text{T},0} := \lim\limits_{\epsilon\to 0}p_{\text{T}}\,.
\end{equation}
In what follows, we will drop this additional subscript to avoid clutter. These are precisely the currents in~\eqref{eq:both-broken-currents} on the branch with zero spatial fluid velocity, i.e.,
\begin{equation}
	u^a = -v^a \iff \vec u^a = 0 \,,
\end{equation}
and with the identifications
\begin{equation}
\label{eq:identifications}
	p_{\text{C}}\ = \mathscr{P}\,, \qquad  \varepsilon_{\text{C}} = \tilde s T - \mathscr{P}\,,\qquad  \Omega = \mathcal{N}^{-1} \,.
\end{equation}
Hence, the induced responses precisely take the form of the phase discussed  in Section~\ref{sec:breaking-both-syms} where both the Carroll boosts and the temporal Weyl symmetry are broken, and where $\mathcal{N}^{-1}$ plays the role of a Stueckelberg field for temporal Weyl transformation rather than a dynamical field, which would have imposed $\varepsilon_{\text{C}} = 0$. This matching alone does not yet establish that the limiting theory realises that intrinsic phase, for that we must also show that the Ward identities close and that the transverse equation imposes no additional constraint on the intrinsic data. We return to these questions in Section~\ref{sec:Carroll-recovery}.

Although the induced responses remain finite in the limit, the full ambient energy-momentum tensor becomes singular
\begin{equation}
\label{eq:singular-ambient-EMT}
    T^{\mu\nu}   = \frac{w_{\text{T}}}{4\epsilon^2b^2}v^\mu v^\nu + w_{\text{T}} v^{(\mu} r_{\tilde\theta}^{\nu)} + p_{\text{T}}g^{\mu\nu} + \epsilon^2 w_{\text{T}} b^2r_{\tilde\theta}^\mu r_{\tilde\theta}^\nu\,,
\end{equation}
since the first term diverges in the decoupling limit~\eqref{eq:decoupling-limit}. But, crucially, from the intrinsic perspective, this divergent term only contributes a term proportional to $v^a v^b$ to the Carrollian momentum-stress tensor $\widetilde{\mathcal{T}}^{ab}$, which is precisely the intrinsic ambiguity~\eqref{eq:intrinsic-ambiguity} that we may ignore. Thus the decoupling limit is not a regular limit of the ambient stress tensor, and it becomes finite only after passing to the intrinsic description which is insensitive to the $v^av^b$ contribution.

\paragraph{Relation to the null fluid limit.}

The decoupling limit~\eqref{eq:decoupling-limit} is very similar to the large-boost limit that leads to the null fluids developed in~\cite{Armas:2025uyv,Armas:2026nwa}, which we briefly discussed in Section~\ref{sec:null-fluids-review}. They both send the projective direction of the ambient timelike fluid velocity to the null direction given by $v^\mu$, corresponding to $B \to 0$ in~\eqref{eq:disting-vector}, but they differ in the way the enthalpy scales as we now explain (both limits are illustrated in Figure~\ref{fig:Bw-plot}).

The null fluid limit corresponds to taking $B\to 0$ while keeping $w_{\text{T}}/B^2$ fixed, which for the energy-momentum tensor in~\eqref{eq:singular-ambient-EMT} leads to
\begin{equation}
	T^{\mu\nu} \to \widehat{\mathcal E} v^\mu v^\nu	+ \widehat P g^{\mu\nu}\,, \qquad \widehat{\mathcal E}:=\lim_{B\to0}\frac{w_{\text{T}}}{4B^2}\,,\qquad \widehat P:=\lim_{B\to0}p_{\text{T}}\,,
\end{equation}
which is precisely the ideal null fluid energy-momentum tensor~\eqref{eq:universal-null-EMT}. Since the ambient enthalpy $w_{\text{T}} \to 0$ under this scaling, all dependence on $\tilde\theta^\mu$ drops out. 

The decoupling limit considered here instead keeps $w_{\text{T}}\to w_{\text{T},0} \neq 0$ finite as $B\to0$. In that case, the mixed term in~\eqref{eq:singular-ambient-EMT} remains finite, though the coefficient of $v^\mu v^\nu$ diverges, but, as discussed above, it precisely leads to the intrinsic ambiguity~\eqref{eq:intrinsic-ambiguity} and therefore does not contribute to the Carrollian responses. In terms of the parameter $\epsilon$ introduced in~\eqref{eq:decoupling-limit}, the two scalings of the enthalpy are therefore
\begin{equation}
\label{eq:two-large-boost-limits}
    \epsilon\to0\,,\quad B = \epsilon b:\qquad
\begin{cases}
	w_{\text{T}}/\epsilon^2 \ \text{finite} & \Rightarrow\text{finite ambient null fluid}\,,\\[1mm]
	w_{\text{T}}\to w_{\text{T},0}\neq0 & \Rightarrow\text{finite intrinsic Carrollian responses}\,.
\end{cases}
\end{equation}
In the next section we show that, in the decoupling limit, the Carroll boost field $\tilde\theta^a$ may be interpreted as a genuine Carroll boost Goldstone and that the limiting responses satisfy intrinsic thermal Carrollian dynamics.

\subsection{Thermal fluid and the Carroll boost Goldstone}
\label{sec:Carroll-recovery}

We now show that the Carrollian responses obtained in the decoupling limit indeed define an autonomous thermal Carrollian fluid with spontaneously broken Carroll boosts. Since the traction is $\mathcal O(\epsilon^2)$ throughout the foliation, the source term in~\eqref{eq:sourced-foliation-diffeo-WI} disappears as $\epsilon\to0$, and the tangential projection of the ambient diffeomorphism Ward identity reduces to an intrinsic Carrollian conservation law. This is not yet sufficient to identify the limiting theory with the Carrollian fluid of Section~\ref{sec:breaking-both-syms}. To complete the identification, we must provide a consistent thermodynamic interpretation of the limiting responses and show that the surviving Carroll boost field becomes an independent Carroll boost Goldstone. Lastly, we must also verify that the transverse equation~\eqref{eq:transverse-timelike-cond} imposes no additional constraint on the intrinsic Carrollian dynamics.

\paragraph{Identification of thermodynamic variables.}

We first show that the limiting responses admit a consistent thermodynamic interpretation. We take the timelike ambient fluid to have temperature $T_{\text{T}}$ and an equation of state of the form
\begin{equation}
	p_{\text{T}}=p_{\text{T}}(T_{\text{T}})\,,
\end{equation}
in terms of which the entropy is 
\begin{equation}
	s_{\text{T}} = \pd{p_{\text{T}}}{T_{\text{T}}}\,,
\end{equation}
which we assume to be nonzero, while the enthalpy is given by
\begin{equation}
		w_{\text{T}} = T_{\text{T}}s_{\text{T}}\,.
\end{equation}
To identify the Carrollian temperature $T_{\text{C}}$, we note that the identification of the Carrollian pressure in~\eqref{eq:limiting-thermodynamics} implies that
\begin{equation}
	p_{\text{C}} = \mathcal{N} p_{\text{T}}(T_{\text{T}})\,,\qquad \varepsilon_{\text{C}} + p_{\text{C}} = \frac{\mathcal N s_{\text{T}}T_{\text{T}}}{2}\,.
\end{equation}
Then, writing
\begin{equation}
	T_{\text{C}} = T_{\text{C}}(T_{\text{T}},\mathcal{N})\,,
\end{equation}
the Carrollian entropy density is given by
\begin{equation}
	s_{\text{C}} = \pd{p_{\text{C}}}{T_{\text{C}}}\bigg\vert_{\mathcal{N}} = \mathcal{N} s_{\text{T}} \pd{T_{\text{T}}}{T_{\text{C}}}\bigg\vert_{\mathcal{N}}\,.
\end{equation}
Then, using the Euler relation
\begin{equation}
\label{eq:Euler-rel}
	T_{\text{C}} s_{\text{C}} = \varepsilon_{\text{C}} + p_{\text{C}}\,,
\end{equation}
we get
\begin{equation}
	T_{\text{C}} \pd{T_{\text{T}}}{T_{\text{C}}}\bigg\vert_{\mathcal{N}} = \frac{T_{\text{T}}}{2} \implies \pd{T_{\text{C}}}{T_{\text{T}}}\bigg\vert_{\mathcal{N}} = \frac{2T_{\text{C}}}{T_{\text{T}}} \,,
\end{equation}
which, by separating variables, gives
\begin{equation}
	T_{\text{C}} = F(\mathcal{N})T_{\text{T}}^2\,,
\end{equation}
where $F(\mathcal{N})$ is the integration function. Since, from the intrinsic perspective, the Carrollian temperature has temporal Weyl weight $+1$ (cf.~\eqref{eq:hydrostatic-data}), while $T_{\text{T}}$ has weight zero, we can write
\begin{equation}
	F(\mathcal{N}) = \frac{\mathcal{N}}{T_\star}\,,
\end{equation}
for some positive constant reference temperature $T_\star$ that provides the normalisation of the Carrollian temperature relative to the ambient temperature, and hence
\begin{equation}
	T_{\text{C}} = \frac{\mathcal N T_{\text{T}}^2}{T_\star}\,.
\end{equation}
With this identification, we get the intrinsic equation of state
\begin{equation}
\label{eq:limit-EOS}	
	p_{\text{C}}(T_{\text{C}},\mathcal N) = \mathcal N
	p_{\text{T}}\left(\sqrt{\frac{T_\star T_{\text{C}}}{\mathcal N}}\right)\,.
\end{equation}
This agrees precisely with the projected response~\eqref{eq:limiting-thermodynamics}. Furthermore, the Carrollian pressure also gives rise to the correct response to $\mathcal N$, since
\begin{equation}
\label{eq:O-N}
	\pd{p_{\text{C}}}{\mathcal N}\bigg\vert_{T_{\text{C}}} = p_{\text{T}} - \frac{w_{\text{T}}}{2} =
-\frac{\varepsilon_{\text{C}}}{\mathcal N} = \mathcal O_{\mathcal N}\,,
\end{equation}
so that the Gibbs--Duhem relation becomes
\begin{equation}
	dp_{\text{C}} = s_{\text{C}} dT_{\text{C}} + \mathcal{O}_{\mathcal N}d\mathcal{N}\,.
\end{equation}
Hence, writing $\Omega=\mathcal N^{-1}$ (cf.~\eqref{eq:identifications}) for the Stueckelberg field precisely recovers the phase described in~\ref{sec:breaking-both-syms}, and the constitutive relations obtained here in the decoupling limit admit the full thermodynamic interpretation of that phase.

Interestingly, the background temporal Weyl compensator $\Omega = \mathcal N^{-1}$ drops out of the description when $\mathcal O_{\mathcal N} = 0$. For the ambient fluid this requires, using~\eqref{eq:O-N}, that $w_{\text{T}} = 2 p_{\text{T}}$, corresponding to the quadratic equation of state
\begin{equation}
	p_{\text{T}} = a T_{\text{T}}^2\,, \qquad a > 0\,.	
\end{equation}
In this case
\begin{equation}
	p_{\text{C}} = aT_\star T_{\text{C}}\,,\qquad \varepsilon_{\text{C}} = 0 = \mathcal O_{\mathcal N}\,.	
\end{equation}
Hence, in this special case temporal Weyl symmetry is restored in the limiting intrinsic theory and the constitutive relations coincide with those of the aligned phase with spontaneously broken Carroll boosts discussed in Section~\ref{sec:spont-break-boosts} with $s_0 = aT_\star$ in~\eqref{eq:const-ent}.

We now consider the decoupling limit of the tangential diffeomorphism Ward identity~\eqref{eq:sourced-explicit-WI-timelike}, which, as mentioned above, becomes sourceless. Contracting it with $v^a$ and using the relation~\eqref{eq:measure-derivative} for the derivative of the measure gives
\begin{equation}
	v^a\D_a\varepsilon_{\text{C}} + w_{\text{C}} e^{-1}\D_a(ev^a) + \mathcal O_{\mathcal N}v^a\D_a\mathcal N = 0\,.
\end{equation}
Using the first law
\begin{equation}
	d\varepsilon_{\text{C}} = T_{\text{C}}ds_{\text{C}} -\mathcal O_{\mathcal N}d\mathcal N\,,
\end{equation}
and the Euler relation~\eqref{eq:Euler-rel}, this turns into the intrinsic entropy equation
\begin{equation}
\label{eq:intrinsic-entropy-conservation}
	e^{-1}\D_a(es_{\text{C}}v^a) = 0\,,
\end{equation}
where $-s_{\text{C}} v^a$ is the ideal Carrollian entropy current. Thus the temporal component of the tangential diffeomorphism Ward identity in the decoupling limit is precisely the ideal entropy conservation equation of the intrinsic Carrollian fluid. We will see below that the transverse ambient equation is also related to this same equation.

\paragraph{The transverse equation and intrinsic entropy.}

Even though the $v^\mu v^\nu$ term of the ambient energy-momentum tensor~\eqref{eq:singular-ambient-EMT} diverges in the decoupling limit, the conservation of the divergent part of the energy-momentum could still lead to constraints. As we now demonstrate, the divergent term  precisely enters the transverse equation~\eqref{eq:transverse-explicit} and gives a conservation law closely related to the entropy conservation equation~\eqref{eq:intrinsic-entropy-conservation}. The energy-momentum tensor~\eqref{eq:singular-ambient-EMT} after taking the decoupling limit can be written to leading order as
\begin{equation}
    T^{\mu\nu} = \epsilon^{-2} A_{(-2)} v^\mu v^\nu + \mathcal{O}(1)\,,\qquad A_{(-2)} := \frac{w_{\text{T}}}{4b^2}\,,
\end{equation}
where $A_{(-2)}$ is the leading coefficient of the divergent term, with $w_{\text{T}}$ and $b$ taking their limiting values. This then implies that
\begin{equation}
	T^\mu{_\nu}R^\nu = \epsilon^{-2}\mathcal{N}A_{(-2)}v^\mu + \mathcal{O}(1)\,.
\end{equation}
We furthermore have that
\begin{equation}
	v^\mu v^\nu \pounds_R g_{\mu\nu} = -2v_\mu([R,v])^\mu = 0\,,
\end{equation}
which follows from the fact that $R(u) = 1$ and $v(u) = 0$, implying that $[R,v](u) = 0$ so that $[R,v]$ is tangent to the leaves. Combining these, we find that the transverse equation~\eqref{eq:transverse-explicit} at order $\epsilon^{-2}$ becomes
\begin{equation}
\label{eq:transverse-LO}
	0 = e^{-1}\D_\mu(e \mathcal{N}^2 A_{(-2)}v^\mu ) = e^{-1}\D_a (e \mathcal{N}^2 A_{(-2)}v^a)\,, 
\end{equation}
since in the adapted coordinates we use here, we have $v^\mu = (0,v^a)$. The thermodynamic identities derived above imply that
\begin{equation}
	\mathcal{N} = \frac{T_{\text{C}} T_\star}{T_{\text{T}}^2}\,,
\end{equation}
so that
\begin{equation}
	\mathcal{N}^2 A_{(-2)} = \frac{T_\star}{2}s_{\text{C}}\left(\frac{T_{\text{C}}}{bT_{\text{T}}}\right)^2\,.
\end{equation}
Plugging this into the leading-order transverse equation~\eqref{eq:transverse-LO}, we get
\begin{equation}
	0 = \frac{T_\star}{2}\left(\frac{T_{\text{C}}}{b T_{\text{T}}}\right)^2 e^{-1}\D_a(e s_{\text{C}} v^a) + \frac{T_\star s_{\text{C}}}{2}\pounds^{N_u}_v \left(\frac{T_{\text{C}}}{b T_{\text{T}}}\right)^2\,.
\end{equation}
The first equation is proportional to the entropy equation~\eqref{eq:intrinsic-entropy-conservation}, which vanishes on its own, and hence the above reduces to the condition
\begin{equation}
	\pounds^{N_u}_v \left(\frac{T_{\text{C}}}{b T_{\text{T}}}\right) = 0\,.
\end{equation} 
In this way, the transverse equation imposes that $T_{\text{C}}$ is proportional to $b T_{\text{T}}$ along $v^a$. A special case of this is if
\begin{equation}
	T_{\text{C}} = b T_{\text{T}}\,,
\end{equation}
in which case 
\begin{equation}
	b = \frac{\mathcal N T_{\text{T}}}{T_\star}\,,
\end{equation}
so that the transverse equation~\eqref{eq:transverse-LO} exactly reduces to the entropy conservation equation~\eqref{eq:intrinsic-entropy-conservation}.

We still must consider the finite piece of the transverse equation. This will determine the subleading component $A_{(0)}$ in the $\epsilon$-expansion\footnote{We assume for simplicity that the coefficient of $v^\mu v^\nu$ admits an expansion in even powers $\epsilon$.} of the $v^\mu v^\nu$ term in~\eqref{eq:singular-ambient-EMT}. More precisely, this means that~\eqref{eq:singular-ambient-EMT} up to $\mathcal{O}(1)$ can be written as
\begin{equation}
	T^{\mu\nu} = \epsilon^{-2}A_{(-2)}v^\mu v^\nu + \left[ A_{(0)} v^\mu v^\nu + w_{\text{T}} v^{(\mu}r^{\nu)}_{\tilde\theta} + p_{\text{T}}g^{\mu\nu} \right] + \mathcal O(\epsilon^2)\,.
\end{equation} 
Therefore, the subleading transverse equation obtained from~\eqref{eq:transverse-explicit} takes the form
\begin{equation}
\label{eq:finite-complementary-transport}
    e^{-1}\D_a(\mathcal N^2e A_{(0)} v^a) = -\mathcal N R_\nu\nabla_\mu (w_{\text{T}}v^{(\mu}r_{\tilde\theta}^{\nu)}+p_{\text{T}}g^{\mu\nu})\,.
\end{equation}
Using the leading-order transverse equation~\eqref{eq:transverse-LO}, we can write the equation above as
\begin{equation}
	\pounds^{N_u}_v \left( \frac{A_{(0)}}{A_{(-2)}} \right) = -\frac{1}{\mathcal N A_{(-2)}} R_\nu\nabla_\mu (w_{\text{T}}v^{(\mu}r_{\tilde\theta}^{\nu)}+p_{\text{T}}g^{\mu\nu})\,.
\end{equation}
Importantly, this does not impose a further constraint on the intrinsic Carrollian fluid and only affects ambient data. In this way, it becomes a condition on the possible ambient fluids that have a well-defined decoupling limit.

\paragraph{Emergence of the boost Goldstone.}

Finally, we demonstrate how the Carroll boost field $\tilde\theta^a$ becomes a bona fide Carroll boost Goldstone in the decoupling limit. With $B=\epsilon b$, the ambient fluid velocity~\eqref{eq:disting-vector} takes the form
\begin{equation}
\label{eq:subleading-boost-data}
	U^\mu = \frac{1}{2\epsilon b}v^\mu + \epsilon b\left(\tau^\mu + \tilde\theta^\mu + \frac{1}{2}\tilde\theta^2v^\mu\right)\,.
\end{equation}
Before taking the decoupling limit, the Carroll boost field $\tilde\theta^\mu$ is determined by the ambient fluid velocity. As we saw above, even though the fluid velocity $U^\mu$ diverges, $\tilde\theta^\mu$ combines with the divergent leading term to produce a finite term in the ambient energy-momentum tensor~\eqref{eq:singular-ambient-EMT}. Consequently, $\tilde\theta^a$ survives in the intrinsic responses~\eqref{eq:limiting-thermal-responses} in the decoupling limit. Since the diffeomorphism Ward identity closes on the intrinsic fields and does not involve the singular ambient fluid velocity, we may interpret $\tilde\theta^a$ as the Carroll boost Goldstone introduced in Section~\ref{sec:spont-break-boosts}, i.e.,
\begin{equation}
\label{eq:Goldstone-identification}
	\tilde\theta^a \xrightarrow{\text{decoupl.~lim.}} \theta^a\,.
\end{equation}
Moreover, the Stueckelberg symmetry with parameter $\chi$ (cf.~\eqref{eq:trafo-2}) emerges in the decoupling limit: if we perform a transformation of the form
\begin{equation}
\label{eq:stueckelberg-emerg}
	\delta_\chi\theta^a=\chi v^a\,,
\end{equation}
the responses~\eqref{eq:limiting-thermal-responses} transform as
\begin{equation}
	\begin{split}
		\delta_\chi T^a &= 0\,,\qquad \delta_\chi\mathcal T^{ab} = 2(\varepsilon_{\text{C}} + p_{\text{C}})\chi v^av^b\,,\\
		\delta_\chi T^a{_b} &= 0\,,\qquad \delta_\chi\mathcal O_{\mathcal N} = 0\,.
	\end{split}
\end{equation}
Hence, since the transformation of the momentum-stress tensor involves the intrinsic ambiguity~\eqref{eq:intrinsic-ambiguity}, all responses and their Ward identities are invariant under the emergent Stueckelberg symmetry~\eqref{eq:stueckelberg-emerg}. Together with the induced Carroll boost transformation, this precisely leads to the identification~\eqref{eq:Goldstone-identification}, while to recover the exact representative inherited from $\tilde\theta^a$ one must impose $\tau_a \theta^a = 0$. It is interesting to note that the boost Goldstone also appears as a subleading field in the small-$c$ expansion of the fluid velocity in the construction of Carrollian fluids in~\cite{Armas:2023dcz}. Here, the decoupling limit corresponds to an infinite boost and gives rise to a Carrollian fluid in one dimension lower.

We have thus shown that the decoupling limit~\eqref{eq:decoupling-limit} produces an autonomous thermal Carrollian fluid, even though the ambient energy-momentum tensor diverges. The divergent part is invisible from the intrinsic perspective, while the finite part gives rise to the thermodynamic data and an independent Carroll boost Goldstone.

\section{Discussion}
\label{sec:discussion}

In this work, we have developed theories for fluids on a null hypersurface and shown how they may be realised by restricting ambient fluids to an embedded null hypersurface. The null embedding gives rise to a Carrollian structure with an additional temporal Weyl symmetry, and both symmetries enter the classification of ideal fluid phases. Thermal fluids on a null hypersurface require spontaneously broken Carroll boosts, and the temporal Weyl symmetry further constrains these theories in different ways depending on whether a compensator is a background Stueckelberg field or a dynamical Goldstone field.

The null submanifold calculus we developed makes the relation to ambient physics explicit by identifying the Carrollian responses induced by projections of the ambient energy-momentum tensor. Similarly, we showed that the intrinsic diffeomorphism Ward identity obtained by projecting the ambient diffeomorphism Ward identity along the null hypersurface is generally sourced by transverse ambient data, and there is an additional independent transverse equation originating from the transverse projection of the ambient diffeomorphism Ward identity. The same geometric framework also clarifies the role of the vector $M^a$ introduced in~\cite{Hartong:2015xda}: it parametrises the difference between the auxiliary null vector $\tau^\mu$ and a chosen reference transverse splitting. 

Demanding that a causal fluid be confined to a null hypersurface forces the fluid velocity to be null and aligned with $v^\mu$. The resulting intrinsic responses realise the aligned zero-temperature phases. If the ambient fluid flows cross the null hypersurface, we instead get open Carrollian fluids. A null foliation organises such open fluids as a stack of systems coupled by transverse transport between the null leaves. In particular, we have shown that there exists a decoupling limit of a timelike ambient fluid that recovers an autonomous thermal Carrollian fluid. In this limit, the ambient energy-momentum tensor is singular, but the corresponding intrinsic responses remain finite since the divergent term lies in the intrinsic ambiguity of the Carrollian momentum-stress tensor, while the Carroll boost field survives as an independent Goldstone with an emergent intrinsic Stueckelberg redundancy. 

These results suggest several interesting directions for further study, and we list a few of these below:
\begin{description}
	\item[Flat space holography.] Another natural application of our framework is to null infinity $\scri$ which is a null hypersurface in the conformally completed spacetime. Since massive matter ends up at timelike infinity, null matter is the relevant ambient theory at $\scri$, and it would be interesting to compare our Ward identities with the Bondi flux-balance laws. In particular, a precise application at $\scri$ would need to take into account changes of conformal completion which rescale the spatial metric and are therefore distinct from the temporal Weyl transformations considered here. Another worthwhile direction would be to investigate gravitational radiation using Ashtekar's formulation~\cite{Ashtekar:1981hw}, where the radiative degrees of freedom are encoded in equivalence classes of connections on $\scri$. It would be interesting to relate these to our induced connection, which should be related to the shear and the news. A formulation using Cartan geometry could provide a natural formalism for this investigation, building on the gauging of Carrollian geometries~\cite{Hartong:2015xda,Figueroa-OFarrill:2022mcy}, the Cartan geometry of null infinity~\cite{Herfray:2021qmp}, and the very recent work on the geometry of gravitational radiation~\cite{Hartong:2026bvh, Fiorucci:2026nyg}, as well as existing constructions for lightlike manifolds using Cartan geometry~\cite{Papadopoulos:2018mma,palomo2020lightlikemanifoldscartangeometries,moron2025cartangeometriesmodelfuture}.
	
	\item[Black hole horizons and the membrane paradigm.] Probably the most important null hypersurfaces are black hole horizons. It would be interesting to investigate the relation between the formalism developed in this paper and the membrane paradigm~\cite{Damour:1978cg,Price:1986yy,Thorne:1986iy,Parikh:1997ma}, building on work that has already connected Carrollian geometry and the membrane paradigm~\cite{Donnay:2019jiz,Freidel:2022vjq,Redondo-Yuste:2022czg}. In particular, it would be interesting to understand how the intrinsic Carrollian responses and the ambient null traction are related to the horizon fluid equations and their associated fluxes.
	
    \item[Derivative corrections and entropy production.] In this work, we considered perfect fluids on null hypersurfaces. The variational calculus and the intrinsic response identities obtained by projection in~\ref{sec:variational-calculus} remain valid at all derivative orders, providing a framework for studying which first-order transport coefficients can be inherited from ambient fluids. It would also be useful to determine which derivative corrections survive the thermal decoupling limit, and whether they exactly reproduce the first-order Carrollian fluid of~\cite{Armas:2023dcz}. Along these same lines, it would be interesting how exchange between leaves contributes to entropy production, and how the second law behaves in the decoupling limit.

	\item[Dynamical null hypersurfaces.] Throughout this work, we have kept the embedding fixed. Allowing it to vary would require a variational principle where transverse deformations are constrained by the null condition in contrast to non-null submanifolds~\cite{Armas:2013hsa,Armas:2013goa}. The resulting shape equations should relate the intrinsic Carrollian responses, the extrinsic geometry and the ambient null traction identified in our variational calculus. It would be particularly interesting to investigate how these equations combine with the sourced Carrollian Ward identities to describe the coupled evolution of the fluid and the null hypersurface where it lives. This could provide a natural framework for studying deformations of null surfaces, including perturbed black hole horizons.
\end{description}

\section*{Acknowledgments}
I am grateful to Jay Armas, Jos\'e Figueroa-O'Farrill, Jelle Hartong, Niels Obers and Gianbattista-Piero Nicosia for useful discussions, and in particular to Jay Armas, Jos\'e Figueroa-O'Farrill, Jelle Hartong and Niels Obers for comments on an early draft. LLMs from OpenAI and Anthropic were used as adversarial referees for a near-final draft of the paper. The work of EH is supported by Carlsberg Foundation grant CF24-1656. The Center of Gravity is a Center of Excellence funded by the Danish National Research Foundation under grant No.~184.

\providecommand{\href}[2]{#2}\begingroup\raggedright\endgroup
	

\begin{thebibliography}{10}
	
	\bibitem{Hawking:1973uf}
	S.~W. Hawking and G.~F.~R. Ellis, {\em {The Large Scale Structure of
			Space-Time}}.
	\newblock Cambridge Monographs on Mathematical Physics. Cambridge University
	Press,
	2011.
	\newblock
	
	\bibitem{Gourgoulhon:2005ng}
	E.~Gourgoulhon and J.~L. Jaramillo, ``{A 3+1 perspective on null hypersurfaces
		and isolated horizons},''
	\href{http://dx.doi.org/10.1016/j.physrep.2005.10.005}{{\em Phys. Rept.}
		{\bfseries 423} (2006) 159--294},
	\href{http://arxiv.org/abs/gr-qc/0503113}{{\ttfamily arXiv:gr-qc/0503113}}.
	
	\bibitem{Penrose:1962ij}
	R.~Penrose, ``{Asymptotic properties of fields and space-times},''
	\href{http://dx.doi.org/10.1103/PhysRevLett.10.66}{{\em Phys. Rev. Lett.}
		{\bfseries 10} (1963) 66--68}.
	
	\bibitem{Penrose:1965am}
	R.~Penrose, ``{Zero rest mass fields including gravitation: Asymptotic
		behavior},'' \href{http://dx.doi.org/10.1098/rspa.1965.0058}{{\em Proc. Roy.
			Soc. Lond. A} {\bfseries 284} (1965) 159}.
	
	\bibitem{Duval:2014uoa}
	C.~Duval, G.~W. Gibbons, P.~A. Horvathy, and P.~M. Zhang, ``{Carroll versus
		Newton and Galilei: two dual non-Einsteinian concepts of time},''
	\href{http://dx.doi.org/10.1088/0264-9381/31/8/085016}{{\em Class. Quant.
			Grav.} {\bfseries 31} (2014) 085016},
	\href{http://arxiv.org/abs/1402.0657}{{\ttfamily arXiv:1402.0657 [gr-qc]}}.
	
	\bibitem{Duval:2014uva}
	C.~Duval, G.~Gibbons, and P.~Horvathy, ``{Conformal Carroll groups and BMS
		symmetry},'' \href{http://dx.doi.org/10.1088/0264-9381/31/9/092001}{{\em
			Class.Quant.Grav.} {\bfseries 31} (2014) 092001},
	\href{http://arxiv.org/abs/1402.5894}{{\ttfamily arXiv:1402.5894 [gr-qc]}}.
	
	\bibitem{Hartong:2015xda}
	J.~Hartong, ``{Gauging the Carroll Algebra and Ultra-Relativistic Gravity},''
	\href{http://dx.doi.org/10.1007/JHEP08(2015)069}{{\em JHEP} {\bfseries 08}
		(2015) 069},
	\href{http://arxiv.org/abs/1505.05011}{{\ttfamily arXiv:1505.05011 [hep-th]}}.
	
	\bibitem{Penna:2018gfx}
	R.~F. Penna, ``{Near-horizon Carroll symmetry and black hole Love numbers},''
	\href{http://arxiv.org/abs/1812.05643}{{\ttfamily arXiv:1812.05643
			[hep-th]}}.
	
	\bibitem{Donnay:2019jiz}
	L.~Donnay and C.~Marteau, ``{Carrollian Physics at the Black Hole Horizon},''
	\href{http://dx.doi.org/10.1088/1361-6382/ab2fd5}{{\em Class. Quant. Grav.}
		{\bfseries 36} no.~16, (2019) 165002},
	\href{http://arxiv.org/abs/1903.09654}{{\ttfamily arXiv:1903.09654 [hep-th]}}.
	
	\bibitem{Ciambelli:2019lap}
	L.~Ciambelli, R.~G. Leigh, C.~Marteau, and P.~M. Petropoulos, ``{Carroll
		Structures, Null Geometry and Conformal Isometries},''
	\href{http://dx.doi.org/10.1103/PhysRevD.100.046010}{{\em Phys. Rev. D}
		{\bfseries 100} no.~4, (2019) 046010},
	\href{http://arxiv.org/abs/1905.02221}{{\ttfamily arXiv:1905.02221
			[hep-th]}}.
	
	\bibitem{deBoer:2017ing}
	J.~de~Boer, J.~Hartong, N.~A. Obers, W.~Sybesma, and S.~Vandoren, ``{Perfect
		Fluids},'' \href{http://dx.doi.org/10.21468/SciPostPhys.5.1.003}{{\em SciPost
			Phys.} {\bfseries 5} no.~1, (2018) 003},
	\href{http://arxiv.org/abs/1710.04708}{{\ttfamily arXiv:1710.04708 [hep-th]}}.
	
	\bibitem{Ciambelli:2018wre}
	L.~Ciambelli, C.~Marteau, A.~C. Petkou, P.~M. Petropoulos, and K.~Siampos,
	``{Flat holography and Carrollian fluids},''
	\href{http://dx.doi.org/10.1007/JHEP07(2018)165}{{\em JHEP} {\bfseries 07}
		(2018) 165}, \href{http://arxiv.org/abs/1802.06809}{{\ttfamily
			arXiv:1802.06809 [hep-th]}}.
	
	\bibitem{Ciambelli:2018xat}
	L.~Ciambelli, C.~Marteau, A.~C. Petkou, P.~M. Petropoulos, and K.~Siampos,
	``{Covariant Galilean versus Carrollian hydrodynamics from relativistic
		fluids},'' \href{http://dx.doi.org/10.1088/1361-6382/aacf1a}{{\em Class.
			Quant. Grav.} {\bfseries 35} no.~16, (2018) 165001},
	\href{http://arxiv.org/abs/1802.05286}{{\ttfamily arXiv:1802.05286
			[hep-th]}}.
	
	\bibitem{Petkou:2022bmz}
	A.~C. Petkou, P.~M. Petropoulos, D.~R. Betancour, and K.~Siampos,
	``{Relativistic fluids, hydrodynamic frames and their Galilean versus
		Carrollian avatars},'' \href{http://dx.doi.org/10.1007/JHEP09(2022)162}{{\em
			JHEP} {\bfseries 09} (2022) 162},
	\href{http://arxiv.org/abs/2205.09142}{{\ttfamily arXiv:2205.09142
			[hep-th]}}.
	
	\bibitem{deBoer:2021jej}
	J.~de~Boer, J.~Hartong, N.~A. Obers, W.~Sybesma, and S.~Vandoren, ``{Carroll
		Symmetry, Dark Energy and Inflation},''
	\href{http://dx.doi.org/10.3389/fphy.2022.810405}{{\em Front. in Phys.}
		{\bfseries 10} (2022) 810405},
	\href{http://arxiv.org/abs/2110.02319}{{\ttfamily arXiv:2110.02319
			[hep-th]}}.
	
	\bibitem{Freidel:2022bai}
	L.~Freidel and P.~Jai-akson, ``{Carrollian hydrodynamics from symmetries},''
	\href{http://dx.doi.org/10.1088/1361-6382/acb194}{{\em Class. Quant. Grav.}
		{\bfseries 40} no.~5, (2023) 055009},
	\href{http://arxiv.org/abs/2209.03328}{{\ttfamily arXiv:2209.03328
			[hep-th]}}.
	
	\bibitem{Freidel:2022vjq}
	L.~Freidel and P.~Jai-akson, ``{Carrollian hydrodynamics and symplectic
		structure on stretched horizons},''
	\href{http://dx.doi.org/10.1007/JHEP05(2024)135}{{\em JHEP} {\bfseries 05}
		(11, 2024) 135}, \href{http://arxiv.org/abs/2211.06415}{{\ttfamily
			arXiv:2211.06415 [gr-qc]}}.
	
	\bibitem{Redondo-Yuste:2022czg}
	J.~Redondo-Yuste and L.~Lehner, ``{Non-linear black hole dynamics and
		Carrollian fluids},'' \href{http://dx.doi.org/10.1007/JHEP02(2023)240}{{\em
			JHEP} {\bfseries 02} (2023) 240},
	\href{http://arxiv.org/abs/2212.06175}{{\ttfamily arXiv:2212.06175 [gr-qc]}}.
	
	\bibitem{deBoer:2023fnj}
	J.~de~Boer, J.~Hartong, N.~A. Obers, W.~Sybesma, and S.~Vandoren, ``{Carroll
		stories},'' \href{http://dx.doi.org/10.1007/JHEP09(2023)148}{{\em JHEP}
		{\bfseries 09} (2023) 148}, \href{http://arxiv.org/abs/2307.06827}{{\ttfamily
			arXiv:2307.06827 [hep-th]}}.
	
	\bibitem{Armas:2023dcz}
	J.~Armas and E.~Have, ``{Carrollian Fluids and Spontaneous Breaking of Boost
		Symmetry},'' \href{http://dx.doi.org/10.1103/PhysRevLett.132.161606}{{\em
			Phys. Rev. Lett.} {\bfseries 132} no.~16, (2024) 161606},
	\href{http://arxiv.org/abs/2308.10594}{{\ttfamily arXiv:2308.10594
			[hep-th]}}.
	
	\bibitem{Chabirand:2026sxe}
	V.~Chabirand, A.~Fiorucci, P.~M. Petropoulos, and M.~Vilatte, ``{Kinetic Theory
		of Carroll Hydrodynamics},''
	\href{http://arxiv.org/abs/2605.06786}{{\ttfamily arXiv:2605.06786
			[hep-th]}}.
	
	\bibitem{Levy1965}
	J.-M. Lévy-Leblond, ``Une nouvelle limite non-relativiste du groupe de
	Poincaré,'' {\em Annales de l'I.H.P. Physique théorique} {\bfseries 3}
	no.~1, (1965) 1--12. \url{http://eudml.org/doc/75509}.
	
	\bibitem{SenGupta1966OnAA}
	N.~D.~S. Gupta, ``On an analogue of the Galilei group,'' {\em Il Nuovo Cimento
		A (1965-1970)} {\bfseries 44} (1966) 512--517.
	
	\bibitem{Henneaux:1979vn}
	M.~Henneaux, ``{Geometry of Zero Signature Space-times},''
	{\em Bull.Soc.Math.Belg.} {\bfseries 31} (1979) 47--63.
	
	\bibitem{Katsuno_1980}
	K.~Katsuno, ``Null hypersurfaces in Lorentzian manifolds: I,''
	\href{http://dx.doi.org/10.1017/S0305004100057479}{{\em Mathematical
			Proceedings of the Cambridge Philosophical Society} {\bfseries 88} no.~1,
		(1980) 175–182}.
	
	\bibitem{Katsuno_1981}
	K.~Katsuno, ``Null hypersurfaces in Lorentzian manifolds II,''
	\href{http://dx.doi.org/10.1017/S0305004100058412}{{\em Mathematical
			Proceedings of the Cambridge Philosophical Society} {\bfseries 89} no.~3,
		(1981) 525–532}.
	
	\bibitem{Mars:1993mj}
	M.~Mars and J.~M.~M. Senovilla, ``{Geometry of general hypersurfaces in
		space-time: Junction conditions},''
	\href{http://dx.doi.org/10.1088/0264-9381/10/9/026}{{\em Class. Quant. Grav.}
		{\bfseries 10} (1993) 1865--1897},
	\href{http://arxiv.org/abs/gr-qc/0201054}{{\ttfamily arXiv:gr-qc/0201054}}.
	
	\bibitem{Guti_rrez_2015}
	M.~Gutiérrez and B.~Olea, ``Induced Riemannian structures on null
	hypersurfaces,'' \href{http://dx.doi.org/10.1002/mana.201400355}{{\em
			Mathematische Nachrichten} {\bfseries 289} no.~10, (Dec., 2015) 1219–1236},
	\href{http://arxiv.org/abs/1207.1030}{{\ttfamily arXiv:1207.1030 [math.DG]}}.
	
	\bibitem{Freidel:2024emv}
	L.~Freidel and P.~Jai-akson, ``{Geometry of Carrollian Stretched Horizons},''
	\href{http://dx.doi.org/10.1088/1361-6382/adaf6e}{{\em Class. Quant. Grav.}
		{\bfseries 42} no.~6, (6, 2024) 065010},
	\href{http://arxiv.org/abs/2406.06709}{{\ttfamily arXiv:2406.06709 [gr-qc]}}.
	
	\bibitem{Ciambelli:2025unn}
	L.~Ciambelli and P.~Jai-akson, ``{Foundations of Carrollian geometry},''
	\href{http://dx.doi.org/10.1016/j.physrep.2026.05.005}{{\em Phys. Rept.}
		{\bfseries 1188} (2026) 1--51},
	\href{http://arxiv.org/abs/2510.21651}{{\ttfamily arXiv:2510.21651
			[hep-th]}}.
	
	\bibitem{Carter:1997pb}
	B.~Carter, ``{Brane dynamics for treatment of cosmic strings and vortons},'' in
	{\em {Recent developments in gravitation and mathematical physics.
			Proceedings, 2nd Mexican School, Tlaxcala, Mexico, December 1-7, 1996}}.
	\newblock 1997.
	\newblock
	\href{http://arxiv.org/abs/hep-th/9705172}{{\ttfamily arXiv:hep-th/9705172
			[hep-th]}}.
	\newblock
	
	\bibitem{Armas:2013hsa}
	J.~Armas, ``{How Fluids Bend: the Elastic Expansion for Higher-Dimensional
		Black Holes},'' \href{http://dx.doi.org/10.1007/JHEP09(2013)073}{{\em JHEP}
		{\bfseries 09} (2013) 073},
	\href{http://arxiv.org/abs/1304.7773}{{\ttfamily arXiv:1304.7773 [hep-th]}}.
	
	\bibitem{Armas:2013goa}
	J.~Armas, ``{(Non)-Dissipative Hydrodynamics on Embedded Surfaces},''
	\href{http://dx.doi.org/10.1007/JHEP09(2014)047}{{\em JHEP} {\bfseries 09}
		(2014) 047},
	\href{http://arxiv.org/abs/1312.0597}{{\ttfamily arXiv:1312.0597 [hep-th]}}.
	
	\bibitem{Armas:2017pvj}
	J.~Armas and J.~Tarrio, ``{On actions for (entangling) surfaces and DCFTs},''
	\href{http://dx.doi.org/10.1007/JHEP04(2018)100}{{\em JHEP} {\bfseries 04}
		(2018) 100},
	\href{http://arxiv.org/abs/1709.06766}{{\ttfamily arXiv:1709.06766 [hep-th]}}.
	
	\bibitem{Armas:2019gnb}
	J.~Armas, J.~Hartong, E.~Have, B.~F. Nielsen, and N.~A. Obers, ``{Newton-Cartan
		Submanifolds and Fluid Membranes},''
	\href{http://dx.doi.org/10.1103/PhysRevE.101.062803}{{\em Phys. Rev. E}
		{\bfseries 101} no.~6, (2020) 062803},
	\href{http://arxiv.org/abs/1912.01613}{{\ttfamily arXiv:1912.01613
			[hep-th]}}.
	
	\bibitem{Kapec:2016jld}
	D.~Kapec, P.~Mitra, A.-M. Raclariu, and A.~Strominger, ``{2D Stress Tensor for
		4D Gravity},'' \href{http://dx.doi.org/10.1103/PhysRevLett.119.121601}{{\em
			Phys. Rev. Lett.} {\bfseries 119} no.~12, (2017) 121601},
	\href{http://arxiv.org/abs/1609.00282}{{\ttfamily arXiv:1609.00282
			[hep-th]}}.
	
	\bibitem{Pasterski:2016qvg}
	S.~Pasterski, S.-H. Shao, and A.~Strominger, ``{Flat Space Amplitudes and
		Conformal Symmetry of the Celestial Sphere},''
	\href{http://dx.doi.org/10.1103/PhysRevD.96.065026}{{\em Phys. Rev. D}
		{\bfseries 96} no.~6, (2017) 065026},
	\href{http://arxiv.org/abs/1701.00049}{{\ttfamily arXiv:1701.00049
			[hep-th]}}.
	
	\bibitem{Pasterski:2017kqt}
	S.~Pasterski and S.-H. Shao, ``{Conformal basis for flat space amplitudes},''
	\href{http://dx.doi.org/10.1103/PhysRevD.96.065022}{{\em Phys. Rev.}
		{\bfseries D96} no.~6, (2017) 065022},
	\href{http://arxiv.org/abs/1705.01027}{{\ttfamily arXiv:1705.01027 [hep-th]}}.
	
	\bibitem{Pasterski:2021raf}
	S.~Pasterski, M.~Pate, and A.-M. Raclariu, ``{Celestial Holography},'' in {\em
		{Snowmass 2021}}.
	\newblock 11, 2021.
	\newblock \href{http://arxiv.org/abs/2111.11392}{{\ttfamily arXiv:2111.11392
			[hep-th]}}.
	
	\bibitem{Raclariu:2021zjz}
	A.-M. Raclariu, ``{Lectures on Celestial Holography},''
	\href{http://arxiv.org/abs/2107.02075}{{\ttfamily arXiv:2107.02075
			[hep-th]}}.
	
	\bibitem{Pasterski:2021rjz}
	S.~Pasterski, ``{Lectures on celestial amplitudes},''
	\href{http://dx.doi.org/10.1140/epjc/s10052-021-09846-7}{{\em Eur. Phys. J.
			C} {\bfseries 81} no.~12, (2021) 1062},
	\href{http://arxiv.org/abs/2108.04801}{{\ttfamily arXiv:2108.04801
			[hep-th]}}.
	
	\bibitem{McLoughlin:2022ljp}
	T.~McLoughlin, A.~Puhm, and A.-M. Raclariu, ``{The SAGEX review on scattering
		amplitudes chapter 11: soft theorems and celestial amplitudes},''
	\href{http://dx.doi.org/10.1088/1751-8121/ac9a40}{{\em J. Phys. A} {\bfseries
			55} no.~44, (2022) 443012}, \href{http://arxiv.org/abs/2203.13022}{{\ttfamily
			arXiv:2203.13022 [hep-th]}}.
	
	\bibitem{Donnay:2023mrd}
	L.~Donnay, ``{Celestial holography: An asymptotic symmetry perspective},''
	\href{http://dx.doi.org/10.1016/j.physrep.2024.04.003}{{\em Phys. Rept.}
		{\bfseries 1073} (2024) 1--41},
	\href{http://arxiv.org/abs/2310.12922}{{\ttfamily arXiv:2310.12922
			[hep-th]}}.
	
	\bibitem{Zhu:2026ofh}
	B.~Zhu, ``{Topics in Celestial holography: A bottom-up perspective},''
	\href{http://arxiv.org/abs/2606.24285}{{\ttfamily arXiv:2606.24285
			[hep-th]}}.
	
	\bibitem{Hartong:2015usd}
	J.~Hartong, ``{Holographic Reconstruction of 3D Flat Space-Time},''
	\href{http://dx.doi.org/10.1007/JHEP10(2016)104}{{\em JHEP} {\bfseries 10}
		(2016) 104},
	\href{http://arxiv.org/abs/1511.01387}{{\ttfamily arXiv:1511.01387 [hep-th]}}.
	
	\bibitem{Fiorucci:2025twa}
	A.~Fiorucci, S.~Pekar, P.~Marios~Petropoulos, and M.~Vilatte,
	``{Carrollian-Holographic Derivation of Gravitational Flux-Balance Laws},''
	\href{http://dx.doi.org/10.1103/qv17-ks32}{{\em Phys. Rev. Lett.} {\bfseries
			135} no.~26, (2025) 261602},
	\href{http://arxiv.org/abs/2505.00077}{{\ttfamily arXiv:2505.00077
			[hep-th]}}.
	
	\bibitem{Hartong:2025jpp}
	J.~Hartong, E.~Have, V.~Nenmeli, and G.~Oling, ``{Boundary Energy-Momentum
		Tensors for Asymptotically Flat Spacetimes},''
	\href{http://arxiv.org/abs/2505.05432}{{\ttfamily arXiv:2505.05432
			[hep-th]}}.
	
	\bibitem{Hartong:2026rbr}
	J.~Hartong, E.~Have, V.~Nenmeli, and G.~Oling, ``{The Energy-Momentum-News
		Complex near Future Null Infinity},''
	\href{http://arxiv.org/abs/2607.07872}{{\ttfamily arXiv:2607.07872
			[hep-th]}}.
	
	\bibitem{Fiorucci:2026nyg}
	A.~Fiorucci, S.~Pekar, P.~M. Petropoulos, and M.~Vilatte, ``{Carroll-Cotton
		Tensors and Gravitational Radiation at Null Infinity},''
	\href{http://arxiv.org/abs/2608.28741}{{\ttfamily arXiv:2608.28741
			[hep-th]}}.
	
	\bibitem{Hartong:2026bvh}
	J.~Hartong, ``{The Geometry of Gravitational Radiation},''
	\href{http://arxiv.org/abs/2609.09954}{{\ttfamily arXiv:2609.09954
			[hep-th]}}.
	
	\bibitem{Bagchi:2025vri}
	A.~Bagchi, A.~Banerjee, P.~Dhivakar, S.~Mondal, and A.~Shukla, ``{The
		Carrollian kaleidoscope},''
	\href{http://dx.doi.org/10.1140/epjc/s10052-026-15437-1}{{\em Eur. Phys. J.
			C} {\bfseries 86} no.~4, (2026) 429},
	\href{http://arxiv.org/abs/2506.16164}{{\ttfamily arXiv:2506.16164
			[hep-th]}}.
	
	\bibitem{Nguyen:2025zhg}
	K.~Nguyen, ``{Lectures on Carrollian Holography},''
	\href{http://arxiv.org/abs/2511.10162}{{\ttfamily arXiv:2511.10162
			[hep-th]}}.
	
	\bibitem{Ruzziconi:2026bix}
	R.~Ruzziconi, ``{Carrollian Physics and Holography},''
	\href{http://arxiv.org/abs/2602.02644}{{\ttfamily arXiv:2602.02644
			[hep-th]}}.
	
	\bibitem{Donnay:2022aba}
	L.~Donnay, A.~Fiorucci, Y.~Herfray, and R.~Ruzziconi, ``{Carrollian Perspective
		on Celestial Holography},''
	\href{http://dx.doi.org/10.1103/PhysRevLett.129.071602}{{\em Phys. Rev.
			Lett.} {\bfseries 129} no.~7, (2022) 071602},
	\href{http://arxiv.org/abs/2202.04702}{{\ttfamily arXiv:2202.04702
			[hep-th]}}.
	
	\bibitem{Bagchi:2022emh}
	A.~Bagchi, S.~Banerjee, R.~Basu, and S.~Dutta, ``{Scattering Amplitudes:
		Celestial and Carrollian},''
	\href{http://dx.doi.org/10.1103/PhysRevLett.128.241601}{{\em Phys. Rev.
			Lett.} {\bfseries 128} no.~24, (2022) 241601},
	\href{http://arxiv.org/abs/2202.08438}{{\ttfamily arXiv:2202.08438
			[hep-th]}}.
	
	\bibitem{Donnay:2022wvx}
	L.~Donnay, A.~Fiorucci, Y.~Herfray, and R.~Ruzziconi, ``{Bridging Carrollian
		and celestial holography},''
	\href{http://dx.doi.org/10.1103/PhysRevD.107.126027}{{\em Phys. Rev. D}
		{\bfseries 107} no.~12, (2023) 126027},
	\href{http://arxiv.org/abs/2212.12553}{{\ttfamily arXiv:2212.12553
			[hep-th]}}.
	
	\bibitem{Bagchi:2023ysc}
	A.~Bagchi, K.~S. Kolekar, and A.~Shukla, ``{Carrollian Origins of Bjorken
		Flow},'' \href{http://dx.doi.org/10.1103/PhysRevLett.130.241601}{{\em Phys.
			Rev. Lett.} {\bfseries 130} no.~24, (2023) 241601},
	\href{http://arxiv.org/abs/2302.03053}{{\ttfamily arXiv:2302.03053
			[hep-th]}}.
	
	\bibitem{Bagchi:2023rwd}
	A.~Bagchi, K.~S. Kolekar, T.~Mandal, and A.~Shukla, ``{Heavy-ion collisions,
		Gubser flow, and Carroll hydrodynamics},''
	\href{http://dx.doi.org/10.1103/PhysRevD.109.056004}{{\em Phys. Rev. D}
		{\bfseries 109} no.~5, (2024) 056004},
	\href{http://arxiv.org/abs/2310.03167}{{\ttfamily arXiv:2310.03167
			[hep-th]}}.
	
	\bibitem{Armas:2025uyv}
	J.~Armas, E.~Have, and G.-P. Nicosia, ``{Null matter and the ultrarelativistic
		origin of hydrodynamics at zero temperature},''
	\href{http://arxiv.org/abs/2509.25320}{{\ttfamily arXiv:2509.25320
			[hep-th]}}.
	
	\bibitem{Armas:2026nwa}
	J.~Armas, E.~Have, and G.-P. Nicosia, ``{Null fluid/gravity correspondence},''
	\href{http://dx.doi.org/10.1007/JHEP08(2026)220}{{\em JHEP} {\bfseries 08}
		(2026) 220}, \href{http://arxiv.org/abs/2602.20268}{{\ttfamily
			arXiv:2602.20268 [hep-th]}}.
	
	\bibitem{Figueroa-OFarrill:2020gpr}
	J.~Figueroa-O'Farrill, ``{On the intrinsic torsion of spacetime structures},''
	\href{http://arxiv.org/abs/2009.01948}{{\ttfamily arXiv:2009.01948
			[hep-th]}}.
	
	\bibitem{Hajicek:1973hof}
	P.~{H{\'a}j\'i{\v{c}}ek}, ``{Exact models of charged black holes: I. Geometry
		of totally geodesic null hypersurface},''
	\href{http://dx.doi.org/10.1007/BF01646541}{{\em Communications in
			Mathematical Physics} {\bfseries 34} no.~1, (Mar., 1973) 37--52}.
	
	\bibitem{Andringa:2010it}
	R.~Andringa, E.~Bergshoeff, S.~Panda, and M.~de~Roo, ``{Newtonian Gravity and
		the Bargmann Algebra},''
	\href{http://dx.doi.org/10.1088/0264-9381/28/10/105011}{{\em
			Class.Quant.Grav.} {\bfseries 28} (2011) 105011},
	\href{http://arxiv.org/abs/1011.1145}{{\ttfamily arXiv:1011.1145 [hep-th]}}.
	
	\bibitem{Hansen:2021fxi}
	D.~Hansen, N.~A. Obers, G.~Oling, and B.~T. S{\o}gaard, ``{Carroll Expansion of
		General Relativity},''
	\href{http://dx.doi.org/10.21468/SciPostPhys.13.3.055}{{\em SciPost Phys.}
		{\bfseries 13} no.~3, (2022) 055},
	\href{http://arxiv.org/abs/2112.12684}{{\ttfamily arXiv:2112.12684
			[hep-th]}}.
	
	\bibitem{Bagchi:2024rje}
	A.~Bagchi, A.~Banerjee, J.~Hartong, E.~Have, and K.~S. Kolekar, ``{Strings near
		black holes are Carrollian. Part II},''
	\href{http://dx.doi.org/10.1007/JHEP11(2024)024}{{\em JHEP} {\bfseries 11}
		(2024) 024}, \href{http://arxiv.org/abs/2407.12911}{{\ttfamily
			arXiv:2407.12911 [hep-th]}}.
	
	\bibitem{Blair:2025nno}
	C.~D.~A. Blair, N.~A. Obers, and Z.~Yan, ``{Carroll geometry meets de Sitter
		space via holography},''
	\href{http://dx.doi.org/10.1016/j.physletb.2026.140641}{{\em Phys. Lett. B}
		{\bfseries 879} (2026) 140641},
	\href{http://arxiv.org/abs/2506.19720}{{\ttfamily arXiv:2506.19720
			[hep-th]}}.
	
	\bibitem{Armas:2023ouk}
	J.~Armas and E.~Have, ``{Ideal fracton superfluids},''
	\href{http://dx.doi.org/10.21468/SciPostPhys.16.1.039}{{\em SciPost Phys.}
		{\bfseries 16} no.~1, (2024) 039},
	\href{http://arxiv.org/abs/2304.09596}{{\ttfamily arXiv:2304.09596
			[hep-th]}}.
	
	\bibitem{Jain:2023nbf}
	A.~Jain, K.~Jensen, R.~Liu, and E.~Mefford, ``{Dipole superfluid
		hydrodynamics},'' \href{http://dx.doi.org/10.1007/JHEP09(2023)184}{{\em JHEP}
		{\bfseries 09} (2023) 184}, \href{http://arxiv.org/abs/2304.09852}{{\ttfamily
			arXiv:2304.09852 [hep-th]}}.
	
	\bibitem{deBoer:2020xlc}
	J.~de~Boer, J.~Hartong, E.~Have, N.~A. Obers, and W.~Sybesma, ``{Non-Boost
		Invariant Fluid Dynamics},''
	\href{http://dx.doi.org/10.21468/SciPostPhys.9.2.018}{{\em SciPost Phys.}
		{\bfseries 9} no.~2, (2020) 018},
	\href{http://arxiv.org/abs/2004.10759}{{\ttfamily arXiv:2004.10759
			[hep-th]}}.
	
	\bibitem{Figueroa-OFarrill:2022kcd}
	J.~Figueroa-O'Farrill, R.~Grassie, and S.~Prohazka, ``{Lifshitz symmetry: Lie
		algebras, spacetimes and particles},''
	\href{http://dx.doi.org/10.21468/SciPostPhys.14.3.035}{{\em SciPost Phys.}
		{\bfseries 14} no.~3, (2023) 035},
	\href{http://arxiv.org/abs/2206.11806}{{\ttfamily arXiv:2206.11806
			[hep-th]}}.
	
	\bibitem{Hartong:2015zia}
	J.~Hartong and N.~A. Obers, ``{Ho\v{r}ava-Lifshitz gravity from dynamical
		Newton-Cartan geometry},''
	\href{http://dx.doi.org/10.1007/JHEP07(2015)155}{{\em JHEP} {\bfseries 07}
		(2015) 155},
	\href{http://arxiv.org/abs/1504.07461}{{\ttfamily arXiv:1504.07461 [hep-th]}}.
	
	\bibitem{PartitionFunction2}
	K.~Jensen, M.~Kaminski, P.~Kovtun, R.~Meyer, A.~Ritz, and A.~Yarom, ``{Towards
		hydrodynamics without an entropy current},''
	\href{http://dx.doi.org/10.1103/PhysRevLett.109.101601}{{\em Phys. Rev.
			Lett.} {\bfseries 109} (2012) 101601},
	\href{http://arxiv.org/abs/1203.3556}{{\ttfamily arXiv:1203.3556 [hep-th]}}.
	
	\bibitem{Banerjee:2012iz}
	N.~Banerjee, J.~Bhattacharya, S.~Bhattacharyya, S.~Jain, S.~Minwalla, and
	T.~Sharma, ``{Constraints on Fluid Dynamics from Equilibrium Partition
		Functions},'' \href{http://dx.doi.org/10.1007/JHEP09(2012)046}{{\em JHEP}
		{\bfseries 09} (2012) 046}, \href{http://arxiv.org/abs/1203.3544}{{\ttfamily
			arXiv:1203.3544 [hep-th]}}.
	
	\bibitem{Jensen:2014ama}
	K.~Jensen, ``{Aspects of hot Galilean field theory},''
	\href{http://dx.doi.org/10.1007/JHEP04(2015)123}{{\em JHEP} {\bfseries 04}
		(2015) 123}, \href{http://arxiv.org/abs/1411.7024}{{\ttfamily arXiv:1411.7024
			[hep-th]}}.
	
	\bibitem{Banerjee:2015uta}
	N.~Banerjee, S.~Dutta, and A.~Jain, ``{Equilibrium partition function for
		nonrelativistic fluids},''
	\href{http://dx.doi.org/10.1103/PhysRevD.92.081701}{{\em Phys. Rev. D}
		{\bfseries 92} (2015) 081701},
	\href{http://arxiv.org/abs/1505.05677}{{\ttfamily arXiv:1505.05677 [hep-th]}}.
	
	\bibitem{deBoer:2017abi}
	J.~de~Boer, J.~Hartong, N.~A. Obers, W.~Sybesma, and S.~Vandoren,
	``{Hydrodynamic Modes of Homogeneous and Isotropic Fluids},''
	\href{http://dx.doi.org/10.21468/SciPostPhys.5.2.014}{{\em SciPost Phys.}
		{\bfseries 5} no.~2, (2018) 014},
	\href{http://arxiv.org/abs/1710.06885}{{\ttfamily arXiv:1710.06885 [hep-th]}}.
	
	\bibitem{Armas:2020mpr}
	J.~Armas and A.~Jain, ``{Effective field theory for hydrodynamics without
		boosts},'' \href{http://dx.doi.org/10.21468/SciPostPhys.11.3.054}{{\em
			SciPost Phys.} {\bfseries 11} no.~3, (2021) 054},
	\href{http://arxiv.org/abs/2010.15782}{{\ttfamily arXiv:2010.15782
			[hep-th]}}.
	
	\bibitem{Bellazzini:2013fga}
	B.~Bellazzini, C.~Csaki, J.~Hubisz, J.~Serra, and J.~Terning, ``{A Naturally
		Light Dilaton and a Small Cosmological Constant},''
	\href{http://dx.doi.org/10.1140/epjc/s10052-014-2790-x}{{\em Eur. Phys. J. C}
		{\bfseries 74} (2014) 2790}, \href{http://arxiv.org/abs/1305.3919}{{\ttfamily
			arXiv:1305.3919 [hep-th]}}.
	
	\bibitem{Afxonidis:2025tns}
	E.~Afxonidis, J.~K. Ghosh, D.~Musso, D.~Naegels, and I.~Salazar~Landea,
	``{Scale without conformal symmetry in hydrodynamics},''
	\href{http://dx.doi.org/10.1140/epjc/s10052-025-14685-x}{{\em Eur. Phys. J.
			C} {\bfseries 85} no.~9, (2025) 976},
	\href{http://arxiv.org/abs/2507.22247}{{\ttfamily arXiv:2507.22247
			[hep-th]}}.
	
	\bibitem{Vaidya:1951fdr}
	P.~C. Vaidya, ``{The gravitational field of a radiating star},''
	\href{http://dx.doi.org/10.1007/BF03173260}{{\em Proc. Indian Acad. Sci. A}
		{\bfseries 33} no.~5, (1951) 264}.
	
	\bibitem{Aichelburg:1970dh}
	P.~C. Aichelburg and R.~U. Sexl, ``{On the Gravitational field of a massless
		particle},'' \href{http://dx.doi.org/10.1007/BF00758149}{{\em Gen. Rel.
			Grav.} {\bfseries 2} (1971) 303--312}.
	
	\bibitem{Alves:2018wku}
	D.~W.~F. Alves and H.~Nastase, ``{Hopfion solutions in gravity and a null
		fluid/gravity conjecture},''
	\href{http://arxiv.org/abs/1812.08630}{{\ttfamily arXiv:1812.08630
			[hep-th]}}.
	
	\bibitem{Husnugil:2025edm}
	S.~H{\"u}sn{\"u}gil and L.~Lehner, ``{Sourced Carrollian fluids dual to black
		hole horizons},'' \href{http://dx.doi.org/10.1103/lc67-5ytb}{{\em Phys. Rev.
			D} {\bfseries 112} no.~10, (2025) 104043},
	\href{http://arxiv.org/abs/2508.20284}{{\ttfamily arXiv:2508.20284 [gr-qc]}}.
	
	\bibitem{Ashtekar:1981hw}
	A.~Ashtekar, ``{Radiative Degrees of Freedom of the Gravitational Field in
		Exact General Relativity},'' \href{http://dx.doi.org/10.1063/1.525169}{{\em
			J. Math. Phys.} {\bfseries 22} (1981) 2885--2895}.
	
	\bibitem{Figueroa-OFarrill:2022mcy}
	J.~Figueroa-O'Farrill, E.~Have, S.~Prohazka, and J.~Salzer, ``{The gauging
		procedure and carrollian gravity},''
	\href{http://dx.doi.org/10.1007/JHEP09(2022)243}{{\em JHEP} {\bfseries 09}
		(2022) 243}, \href{http://arxiv.org/abs/2206.14178}{{\ttfamily
			arXiv:2206.14178 [hep-th]}}.
	
	\bibitem{Herfray:2021qmp}
	Y.~Herfray, ``{Carrollian manifolds and null infinity: a view from Cartan
		geometry},'' \href{http://dx.doi.org/10.1088/1361-6382/ac635f}{{\em Class.
			Quant. Grav.} {\bfseries 39} no.~21, (2022) 215005},
	\href{http://arxiv.org/abs/2112.09048}{{\ttfamily arXiv:2112.09048 [gr-qc]}}.
	
	\bibitem{Papadopoulos:2018mma}
	G.~Papadopoulos, ``{Geometry and symmetries of null G-structures},''
	\href{http://dx.doi.org/10.1088/1361-6382/ab1ee3}{{\em Class. Quant. Grav.}
		{\bfseries 36} no.~12, (2019) 125006},
	\href{http://arxiv.org/abs/1811.03500}{{\ttfamily arXiv:1811.03500
			[hep-th]}}.
	
	\bibitem{palomo2020lightlikemanifoldscartangeometries}
	F.~J. Palomo, ``Lightlike manifolds and Cartan geometries,''
	\href{http://arxiv.org/abs/2003.09448}{{\ttfamily arXiv:2003.09448
			[math.DG]}}. \url{https://arxiv.org/abs/2003.09448}.
	
	\bibitem{moron2025cartangeometriesmodelfuture}
	R.~Morón and F.~J. Palomo, ``Cartan geometries with model the future lightlike
	cone of Lorentz-Minkowski spacetime,''
	\href{http://arxiv.org/abs/2508.20202}{{\ttfamily arXiv:2508.20202
			[math.DG]}}. \url{https://arxiv.org/abs/2508.20202}.
	
	\bibitem{Damour:1978cg}
	T.~Damour, ``{Black Hole Eddy Currents},''
	\href{http://dx.doi.org/10.1103/PhysRevD.18.3598}{{\em Phys. Rev. D}
		{\bfseries 18} (1978) 3598--3604}.
	
	\bibitem{Price:1986yy}
	R.~H. Price and K.~S. Thorne, ``{Membrane Viewpoint on Black Holes: Properties
		and Evolution of the Stretched Horizon},''
	\href{http://dx.doi.org/10.1103/PhysRevD.33.915}{{\em Phys. Rev. D}
		{\bfseries 33} (1986) 915--941}.
	
	\bibitem{Thorne:1986iy}
	K.~S. Thorne, R.~H. Price, and D.~A. Macdonald, eds., {\em {BLACK HOLES: THE
			MEMBRANE PARADIGM}}.
	\newblock 1986.
	
	\bibitem{Parikh:1997ma}
	M.~Parikh and F.~Wilczek, ``{An Action for black hole membranes},''
	\href{http://dx.doi.org/10.1103/PhysRevD.58.064011}{{\em Phys. Rev. D}
		{\bfseries 58} (1998) 064011},
	\href{http://arxiv.org/abs/gr-qc/9712077}{{\ttfamily arXiv:gr-qc/9712077}}.
	
\end{thebibliography}
\end{document}